\documentclass[]{emulateapj}
\usepackage{natbib}
\newcommand{\onecolumn}{}

\newcommand{\figurepath}{}

\usepackage{amsmath}
\usepackage{amssymb}
\usepackage{xspace}
\usepackage{color}
\usepackage{ifthen}
\usepackage{xspace}

\definecolor{gray}{rgb}{0.5,0.5,0.5} 

\newcommand{\powersep}{{\ensuremath{\times}}}

\newcommand{\s}{{\ensuremath{\mathrm{s}}}\xspace}
\newcommand{\g}{{\ensuremath{\mathrm{g}}}\xspace}
\newcommand{\K}{{\ensuremath{\mathrm{K}}}\xspace}
\newcommand{\cm}{{\ensuremath{\mathrm{cm}}}\xspace}
\newcommand{\yr}{{\ensuremath{\mathrm{yr}}}\xspace}

\newcommand{\Msun}{{\ensuremath{\mathrm{M}_{\odot}}}\xspace}
\newcommand{\Rsun}{{\ensuremath{\mathrm{R}_{\odot}}}\xspace}

\newcommand{\kLsun}{{\ensuremath{\mathrm{kL}_{\odot}}}\xspace}

\newcommand{\erg}{{\ensuremath{\mathrm{erg}}}\xspace}
\newcommand{\ergs}{{\ensuremath{\mathrm{erg}\,\mathrm{s}^{-1}}}\xspace}
\newcommand{\erggs}{{\ensuremath{\mathrm{erg}\,\mathrm{g}^{-1}\,\mathrm{s}^{-1}}}\xspace}

\newcommand{\kms}{{\ensuremath{\mathrm{km}\,\mathrm{s}^{-1}}}\xspace}

\newcommand{\B}{{\ensuremath{\mathrm{B}}}\xspace}

\newcommand{\kt}{{\ensuremath{\mathrm{kt}}}\xspace}
\newcommand{\ccm}{{\ensuremath{\mathrm{cm}^{3}}}\xspace}
\newcommand{\dex}{{\ensuremath{\mathrm{dex}}}\xspace}

\newcommand{\kB}{{\ensuremath{k_{\mathrm{B}}}}\xspace}
\newcommand{\NA}{{\ensuremath{N_{\mathrm{\!A}}}}\xspace}
\newcommand{\eV}{{\ensuremath{\mathrm{e}\!\mathrm{V}}}\xspace}

\newcommand{\MeV}{{\ensuremath{\mathrm{MeV}}}\xspace}

\newcommand{\Zsun}{{\ensuremath{\mathrm{Z}_{\odot}}}\xspace}

\newcommand{\amu}{{\ensuremath{\mathrm{u}}}\xspace}

\newcommand{\GK}{{\ensuremath{\mathrm{GK}}}\xspace}
\newcommand{\kK}{{\ensuremath{\mathrm{kK}}}\xspace}

\newcommand{\lSect}[1]{{\label{sec:#1}}}
\newcommand{\lFig}[1]{{\label{fig:#1}}}
\newcommand{\lEq}[1]{{\label{eq:#1}}}
\newcommand{\lTab}[1]{{\label{tab:#1}}}

\newcommand{\Tabff}[1]{{\ref{tab:#1}}}
\newcommand{\Tab}[1]{{Table~\Tabff{#1}}}
\newcommand{\Tabs}[1]{{Tables~\Tabff{#1}}}

\newcommand{\PAN}[1]{{{#1}}}

\newcommand{\FIGFF}[2]{{\ref{fig:#2}\PAN{#1}}}
\newcommand{\Figff}[1]{{\FIGFF{}{#1}}}
\newcommand{\FIG}[2]{{Fig.~\FIGFF{#1}{#2}}}
\newcommand{\Fig}[1]{{\FIG{}{#1}}}
\newcommand{\FIGS}[2]{{Figs.~\FIGFF{#1}{#2}}}
\newcommand{\Figs}[1]{{\FIGS{}{#1}}}

\newcommand{\Sectff}[1]{{\ref{sec:#1}}}
\newcommand{\Sect}[1]{{\S~\Sectff{#1}}}

\newcommand{\Eqref}[1]{{\ref{eq:#1}}}
\newcommand{\Eqff}[1]{{(\Eqref{#1})}}

\newcommand{\Eq}[1]{{Eq.~\Eqff{#1}}}

\newcommand{\KEPLER}[1]{{\textsc{Kepler}}\xspace}

\newcommand{\Solar}[1]{{\ensuremath{\left[\right.}#1\ensuremath{\left.\right]}}}
\newcommand{\I}[2]{{\isotope{}{#1}{#2}}}
\newcommand{\El}[1]{{\I{}{#1}}}

\newcommand{\RatEl}[2]{{\El{#1}/\El{#2}}}
\newcommand{\SolRatEl}[2]{{\Solar{\RatEl{#1}{#2}}}}

\newcommand{\Ep}[1]{{\ensuremath{10^{#1}}}}
\newcommand{\Epp}[2]{{\ensuremath{10^{#1#2}}}}

\newcommand{\E}[1]{{\ensuremath{\powersep\Ep{#1}}}}
\newcommand{\EE}[2]{{\ensuremath{\powersep\Ep{#1#2}}}}

\newcommand{\Ye}{{\ensuremath{Y_{\mathrm{e}}}}\xspace}

\newcommand{\Tc}{{\ensuremath{T_{\mathrm{c}}}}\xspace}
\newcommand{\rhoc}{{\ensuremath{\rho_{\mathrm{c}}}}\xspace}
\newcommand{\Sc}{{\ensuremath{S_{\mathrm{c}}}}\xspace}
\newcommand{\Yec}{{\ensuremath{Y_{\mathrm{e,c}}}}\xspace}
\newcommand{\Reff}{{\ensuremath{R_{\mathrm{eff}}}}\xspace}
\newcommand{\Leff}{{\ensuremath{L_{\mathrm{eff}}}}\xspace}
\newcommand{\Teff}{{\ensuremath{T_{\mathrm{eff}}}}\xspace}

\newcommand{\BES}{{\ensuremath{BE_{\mathrm{S=4}}}}\xspace}
\newcommand{\MS}{{\ensuremath{M_{\mathrm{S=4}}}}\xspace}

\newcommand{\BEYecore}{{\ensuremath{BE_{\mathrm{Ye core}}}}\xspace}
\newcommand{\EExp}{{\ensuremath{E^{\mathrm{exp}}}}\xspace}
\newcommand{\SNAkB}{{\ensuremath{S/\NA\kB}}\xspace}
\newcommand{\Tcol}{{\ensuremath{T_{\mathrm{col}}}}\xspace}
\newcommand{\Ekin}{{\ensuremath{E_{\mathrm{kin}}}}\xspace}

\newcommand{\anue}{{\ensuremath{\bar{\nu}_{\El{e}}}}\xspace}

\newcommand{\ltaprx}{\lesssim}

\slugcomment{Draft for ApJ, \today} 

\shorttitle{Pop III Stars 10--100\,\Msun}
\shortauthors{Heger \& Woosley}

\begin{document}


\title{Nucleosynthesis and Evolution of Massive Metal-Free Stars}


\author{Alexander~Heger}
\affil{Theoretical
Astrophysics Group, MS B227, Los Alamos National Laboratory, Los
Alamos, NM 87545;\\
and 
Department of Astronomy and Astrophysics,
University of California, Santa Cruz, CA 95064}
\email{alex@ucolick.org}

\and

\author{S.~E.\ Woosley}
\affil{Department of Astronomy and Astrophysics,
University of California, Santa Cruz, CA 95064}
\email{woosley@ucolick.org}

\begin{abstract} 
  The evolution and explosion of metal-free stars with masses $10 -
  100\,\Msun$ are followed, and their nucleosynthetic yields, light
  curves, and remnant masses determined. Such stars would have been
  the first to form after the Big Bang and may have left a distinctive
  imprint on the composition of the early universe. When the supernova
  yields are integrated over a Salpeter initial mass function (IMF),
  the resulting elemental abundance pattern is qualitatively solar,
  but with marked deficiencies of odd-$Z$ elements with $7 \le Z \le
  13$.  Neglecting the contribution of the neutrino wind from the
  neutron stars that they make, no appreciable abundances are made for
  elements heavier than germanium. The computed pattern compares
  favorably with what has been observed in metal-deficient stars with
  $[Z] \ltaprx -3$.  The amount of ionizing radiation from this
  generation of stars is $\sim2.16\,\MeV$ per baryon ($4.15\,\B$ per
  \Msun; where $1\,\B = 1\,\mbox{Bethe} = 10^{51}\,\erg$) for a
  Salpeter IMF, and may have played a role in reionizing the universe.
  Most of the stars end their lives as blue supergiants and make
  supernovae with distinctive light curves resembling SN~1987A, but
  some produce primary nitrogen by dredge up and become red
  supergiants. These make brighter supernovae like typical Type
  IIp's. For the lower mass supernovae considered, the distribution of
  remnant masses clusters around typical modern neutron star masses,
  but above $20\,\Msun$ to $30\,\Msun$, with the value depending on
  explosion energy, black holes are copiously formed by fallback, with
  a maximum hole mass of $\sim40\,\Msun$.  A novel automated fitting
  algorithm is developed for determining optimal combinations of
  explosion energy, mixing, and initial mass function in the large
  model data base to agree with specified data sets.  The model is
  applied to the low metallicity sample of \citet{cay04} and the two
  ultra-iron-poor stars HE0107-5240 and HE1327-2326.  Best agreement
  with these low metallicity stars is achieved with very little
  mixing, and none of the metal-deficient data sets considered show
  the need for a high energy explosion component. To the contrary,
  explosion energies somewhat less than $1.2\,\B$ seem to be preferred
  in most cases.

\end{abstract}

\keywords{stars-supernovae, stars-metal poor, nucleosynthesis,
  supernova - light curves, black holes, neutron stars, abundances -
  observations}

\section{Introduction}
\lSect{intro}

There must have been a first generation of stars whose initial
composition, reflecting that of the Big Bang, was essentially metal
free.  The term Population III is often applied to such stars
\citep[e.g.,][]{osh08}.  To the extent that lines of heavy elements
and dust are responsible for mass loss, and in the absence of deep
mixing, such stars would have lived their lives with near constant
mass.  As a result, the average mass at death of such stars would have
been larger, even for a metallicity-independent IMF.  They would also
typically have been bluer stars, especially at death, and their
supernovae might have been fainter.  Their nucleosynthesis would also
differ, at least for the many elements that are in any sense
``secondary''.

Such stars have been of interest for a long time
\citep[e.g.,][]{Eze71}, but recent measurements suggesting the
existence of stars at a very early epoch in the universe
\citep{Spe03,Kog03} have invigorated their study.  So, too, have new
observational data sets that highly constrain the nucleosynthesis of
the first few generations of stars \citep[e.g.,][]{cay04,Lai08} and
the discovery of ultra-iron poor stars
\citep[e.g.,][]{Dep02,Chr02,Fre05} which may be providing clues as to
how the first generation of stars formed and exploded
\citep[e.g.,][]{Ume03}.

Zero metal stars are also a limiting case of low metallicity stars.
While this is the first in a series of papers that will treat the
evolution and explosion of all stars up to solar metallicity and
beyond, it is expected that many of the results will apply equally
well to stars of $\Ep{-4}\,\Zsun$ or even $\Ep{-2}\,\Zsun$, so long as
mass loss remains negligible.  An open issue is whether and when forms
of mass loss not dependent upon atomic lines of heavy elements or
grains will become important.  Will $100\,\Msun$ zero metallicity
stars near the end of their lives resemble Eta Carina?  It is also
uncertain just how fast this first generation of stars rotated and how
rotationally induced mixing may have affected not only mass loss and
nucleosynthesis, but also the presupernova structure
\citep[e.g.,][]{Mey06}.

Here we consider a fiducial set of models whose physics is
comparatively simple. The initial composition is pristine Big Bang
nucleosynthesis and mass loss is neglected at all stages of the
evolution.  Similarly, rotation is ignored and, along with it, any
consequences of rotationally-induced mixing and strong asymmetry in
the explosion.  The mass range studied is $10\,\Msun$ to $100\,\Msun$.
Except between $90\,\Msun$ and $100\,\Msun$, this selection avoids the
complications of the pair instability on the upper end and asymptotic
giant branch mass loss on the lower end.  Each of these approximations
is a potentially serious omission, and we hope to return with these
embellishments in future works.

Other recent surveys of zero metallicity massive stars are
\citet{Woo95}, \citet{Mar01}, \citet{Heg02}, \citet{Chi04},
\citet{Ume05}, \citet{Hir06}, and \citet{Tom07}.  This one differs in
following the evolution of a large number of stars in great physical
detail, from the zero age main sequence through explosion as a
supernovae. A variety of supernova properties (energies, mixing, mass
cut) are considered for each mass.  Nucleosynthesis, supernova light
curves and remnant masses are calculated.  Even this ``simple'' study
took us five years to complete.  As we shall see, our simple models
agree reasonably well with all the available data and make some
interesting predictions.  The yields of all isotopes in all stars and
supernova models calculated (extended versions of \Tabs{0.6-yield-mix}
- \Tabff{1.2-yield-nomix-Ye}) are available in the electronic version
of the paper and more detailed data including all the presupernova
models and the yield data base and search tool is available online.

\section{Initial Models and Procedure}
\lSect{models}

\subsection{Masses Studied}

This study is specifically of non-rotating stars that experience iron
core collapse in the end and avoid the pair instability.  Because the
ignition of multiple shells of nuclear burning can often lead to
dramatic variation in the final stellar structure for stars that
differ only slightly in initial mass \citep{Bar90}, the grid of masses
employed is very fine - 120 different masses in the
range $10\,\Msun$ to $100\,\Msun$.  Frequently, model stars differing
by only $0.1\,\Msun$ had substantially different presupernova
properties, especially towards the lighter end of the range studied.
The masses of the stars studied are given in \Tab{ionize}.

\subsection{The Kepler Code and Adopted Physics}
\lSect{kepler}

All calculations of evolution, nucleosynthesis, and light curves were
done with the \KEPLER  code.  \KEPLER is a one-dimensional implicit
hydrodynamics package adapted to the study of stellar evolution and
explosive astrophysical phenomena \citep{Wea78,WHW02}.  The convective
criterion is Ledoux, but with a substantial semi-convective diffusion
coefficient, about $10\,\%$ of the radiative diffusion coefficient in
regions that are stable by the Ledoux criterion and unstable by
Schwarzschild \citep{Woo88}.  A  small amount of convective
overshooting is included by forcing convective boundary zones to be
semiconvective.  Opacities from the OPAL and Los Alamos tables are
used wherever the helium mass fraction exceeds \Ep{-5} and the
temperature is less than $\Ep9\,\K$. Energy generation is followed
using the small 19 isotope nuclear reaction network prior to oxygen
depletion, and a 128 isotope quasi-equilibrium network thereafter as
described in \citet{Wea78}.  The critical nuclear reaction rates and
other stellar physics are those described in \citet{WHW02} and
\citet{Woo07}.  Compositional information on the neutron excess, which
can affect the structure during late stages of evolution, was taken at
each time step in the calculation from the larger nucleosynthesis
co-processing network (\Sect{network}).  Since this quantity evolves
slowly, this procedure proved a stable and cost effective way of
implementing some of the results of a large network in a large survey.
No rotation was included in the present study and mass loss was taken
to be zero.  Calculations by \cite{Kud02,mey02}, for example, suggest
that the mass loss for non-rotating stars lighter than $100\,\Msun$
and metallicity $\lesssim \Ep{-5}$ \Zsun \ can be neglected, though
see \citet{Mey06}.

\subsection{Nucleosynthesis Network and Initial Composition}
\lSect{network}

Nucleosynthesis was followed with an adaptive reaction network that
included from $\sim250$ isotopes on the main sequence to $\sim900$
isotopes during the explosion \citep{rau02}.  Isotopes were added and
removed as necessary to follow the nuclear reaction flow, with decisions
based upon a conservative set of assumptions regarding abundances and
flows in the neighborhood.

The initial composition of Pop III stars is, by assumption, pristine
Big Bang material.  We adopt a Big Bang composition from
\citet{Cyb01,Cyb02}.  The assumed initial mass fractions of \I1H,
\I2H, \I3{He}, and \I4{He} are $0.7513$, $4.3\E{-5}$, $2.1\E{-5}$ and
$0.2487$, respectively.  A mass fraction of $1.9\E{-9}$ was taken for
\I7{Li}.

\section{Presupernova Evolution}

\subsection{Ionizing Photon Yield on the Main Sequence}

The great majority of the energy released in the evolution of a
massive star does not come out as the kinetic energy of stellar winds
or of the supernova it produces in the end, but as the light it emits
as a stable star.  Most of this light is emitted during the long-lived
main sequence evolution (central hydrogen burning, about $90\,\%$ of
the star's lifetime).  It is thought that this emission, primarily in
the ultraviolet band may have played a key role in ionizing the
universe after the recombination epoch \citep[e.g.,][]{Loe01}.  At the
end of central hydrogen burning the stars expand, making softer
radiation, but many stars here remained very blue compared to
their more metal-rich counterparts.  A few of our more massive stars
became red supergiants during their late evolution (\Tab{pre-SN}) due
to the production of primary nitrogen.

\Tab{ionize} gives, for all models, the total energy, $E$, number
of photons, $N_\gamma$, and the average wavelength, $\bar{\lambda}$,
of all photons emitted during the presupernova evolution that are
capable of causing hydrogen ionization
($\mathrm{H}+\gamma\longmapsto\mathrm{H}^++\mathrm{e^-}$, marked as
``HII''), first helium ionization
($\mathrm{He}+\gamma\longmapsto\mathrm{He}^++\mathrm{e^-}$, marked as
``HeII''), second helium ionization
($\mathrm{He}^++\gamma\longmapsto\mathrm{He}^{2+}+\mathrm{e^-}$,
marked as ``HeIII''), or in the Lyman-Werner (LW) Band
($\mathrm{H}_2+\gamma\longmapsto2\mathrm{H}$, marked as ``LW'').  The
values given for the LW Band only include photon energies between
$11.2\,\eV$ and $13.6\,\eV$, i.e., they exclude the hydrogen ionizing
photons.  The data are obtained by integrating an time-dependent
assumed black body flux and spectrum based on luminosity and effective
stellar temperature over the entire stellar evolution.  \Fig{ionize}
shows the number of photons for these energies normalized to the
stellar mass (photons per baryon).

For convenience, we also give, in \Tab{photon}, the photon numbers in
bins relevant to primordial radiation chemistry so that they can be
added up according to the desired cutoffs.  The corresponding energies 
are given in the Table.  We use the following bin boundaries:

\smallskip
\noindent
\begin{tabular}{l@{+ $\gamma\longmapsto$ }l@{ for h$\nu >$ }r@{\,\eV (}r@{\,\AA)}}
H       & H$^+$ + e$^-$     &  13.5984 & 911.76 \\
He      & He$^+$  + e$^-$   &  24.5874 & 504.26 \\
He$^+$  & He$^{2+}$ + e$^-$ &  54.4161 & 227.8 \\
H$_2$   & 2 H               &  11.18  &  1109 \\
H$_2$   & H$_2^+$  + e$^-$  &  15.42   &   804 \\
H$^-$   & H    + e$^-$      &  0.755  &  16421 \\
H$_2^+$ & H    + H$^+$      &  2.65   &   4678 \\
H$_2^+$ & 2 H$^+$ + e$^-$   &  30     &   413.3 \\
\end{tabular}
\smallskip

\subsection{Energy Generation and Convective History}
\lSect{conv}

The convective and energy generation histories for four representative
stars ($15\,\Msun$, $25\,\Msun$, $40\,\Msun$, and $80\,\Msun$) are
given in \Figs{conv15+25} and \Figff{conv40+80}.

As has been known for a long time, stars with zero metallicity are
different from all other stars \citep{Eze71}, even those with just a
trace of heavy elements.  In order to burn hydrogen they must first
contract and burn a trace of helium to produce the catalyst for the
CNO cycle hydrogen burning (the pp-cycles are never important in such
massive stars).  This need for high temperature and density to obtain
nuclear energy leads to denser, higher entropy cores from the
beginning, a characteristic that affects the evolution throughout all
subsequent stages.  Also, without heavy elements in their outer
envelope, these stars are stable against opacity driven pulsations
that would tear down their modern counterparts as well as nuclear
driven pulsations \citep{Bar01}.  They also have weaker hydrogen
burning shells and that adds to their tendency to be compact blue
stars throughout their entire evolution.

During core hydrogen burning, the central mass fraction of \I{14}N
gradually rises but is typically about \Epp-9.  Only during the very
end of hydrogen burning are much higher values reached, approaching
\Epp-7.  In contrast, during most of hydrogen shell burning, (i.e.,
during core He burning and thereafter) values of \Epp-7 are typical at
the base of the shell.  In the more massive stars, even those that do
not make primary nitrogen (see below), the hydrogen burning shell can
nevertheless become very vigorous (e.g., the $40\,\Msun$ star shown in
\Fig{conv40+80}) and move outwards in mass at a high rate.

Of the stars whose convective histories are plotted in
\Figs{conv15+25} and \Figff{conv40+80}, the $15\,\Msun$, $25\,\Msun$
and $40\,\Msun$ stars all die with envelopes that, at least in their
outer extremities, are radiative. As \Tab{pre-SN} shows, these three
stars are blue supergiants at death. The $15\,\Msun$ and $40\,\Msun$
stars die without having ever made primary nitrogen in excess of
10$^{-6}$ by mass fraction their hydrogen layers.  The $25\,\Msun$
star did make primary nitrogen of about \Epp-3 in the convective part
of its envelope and, given sufficient time, would have expanded to red
giant proportions. But it died in transit while still a very blue star
(\Tab{pre-SN}).  This also happened for the $20.5\,\Msun$,
$21.5\,\Msun$, $27$\,\Msun, $28\,\Msun$, and $30\,\Msun$ models, but
above $30\,\Msun$, primary nitrogen production always led to a red
supergiant progenitor.  An example of the latter behavior is the
$80\,\Msun$ star in \Fig{conv40+80}.

Carbon burning is convective in the center of the $15\,\Msun$ star,
but radiative in the heavier ones - consistent with what has been seen
in solar metallicity stars and showing that the evolution overall is
most sensitive to the helium core mass, not much to the metallicity.

For the $80\,\Msun$ star in \Fig{conv40+80}, a mild nuclear
instability is encountered during the ignition of the second oxygen
burning shell, notable as a series of growing spikes of energy
generation (dark blue) just above a mass coordinate of $7\,\Msun$,
starting at $\Epp-4\,\yr$ prior to core collapse.  These shell flashes
reflect the onset of the ``pulsational pair instability''
\citep{Heg02}.  As the mass of the star becomes larger, these flashes
become more violent and happen earlier, eventually moving to the
center of the star.  They can briefly expand the core and reduce the
neutrino losses, as seen in the reduction in purple coloration
following each flash.  During the peak of the last pulse (note
logarithmic time scale that make later pulses take more space on the
\textsl{x}-axis) burning has proceeded far enough for the stellar core
to collapses at this point.

\subsection{Primary nitrogen production}

As noted in the previous section, some of the models produced primary
nitrogen at the base of the hydrogen envelope, a phenomenon that not
only affected their nucleosynthesis, but greatly altered their
structure.

In the $25\,\Msun$ model nitrogen production commenced late in the
evolution (\Fig{conv15+25}), just following central carbon depletion,
as the convective helium burning \emph{shell} encroached on the base
of the hydrogen envelope. The phenomenon is inhibited in lower mass
stars (like the $15\,\Msun$ model) presumably because the exoergic
nature of carbon burning (including neutrino losses) causes core
expansion and a weaker helium shell.  Because of the weak hydrogen
burning activity, the entropy contrast between the helium convective
shell and the base of the hydrogen envelope was not great at the time
they touched, $S/N_A k = 12.8$ as compared to $13.4$.  As the carbon
and hydrogen were mixed together as a consequence of convective
overshoot mixing, the nitrogen mass fraction at the base of the
hydrogen shell rose rapidly.  This led to increased energy generation,
a rise in entropy, and the onset of hydrogen shell convection.  The
(``undershoot'' and) overshoot mixing of both shells continued to hold
them in contact and the nitrogen abundance rose. Carbon was dredged up
into the hydrogen shell immediately became nitrogen.  This mixing
continued until shortly before the star's death. By that time the
entropy at the hydrogen shell had risen to $27.6$.

We regard this synthesis of nitrogen as unavoidable, but with an
uncertainty in the yield of at least an order of magnitude, especially
if the effects of rotation were to be included.  We repeated the
calculation of the $25\,\Msun$ model with convective overshoot turned
off, expecting the primary nitrogen to be reduced.  Instead a different
phenomenon was observed.  A trace of hydrogen mixed into the helium
shell by the first contact caused a burst of energy generation that
raised the entropy.  Consequently the outer part of the helium shell,
that part into which hydrogen had been mixed became convective and
merged with the hydrogen shell.  The substantial amount of carbon in
this boundary layer then all became nitrogen in the envelope. Almost
the same amount of nitrogen was made as before, but this was a
coincidence and the yield was found to depend upon the zoning and time
step at the moment the two layers touched.

Even above $25\,\Msun$ some stars did \emph{not} make significant
primary nitrogen.  The $40\,\Msun$ model is an example
(\Fig{conv40+80}).  The helium shell at the end of carbon burning in
this star contained little mass, only $1.3\,\Msun$ and burned only a 
few percent of its helium to carbon before the star
died.  Consequently, the helium convective burning shell was weak and
never made it to the base of the hydrogen shell.  The primary nitrogen
that was made survived in the hydrogen shell \emph{from hydrogen
burning} from its earlier, pre-helium burning evolution.  Its mass
fraction, between $\Epp-7$ and $\Epp-6$, was adequate to catalyze
hydrogen burning and produce a convective region that extended almost
to the surface of the star.

In still higher mass stars, of which we may take $80\,\Msun$ as
representative, copious primary nitrogen production occurred at an
earlier stage in the evolution and was more robust.  Here the nitrogen
was made during helium \emph{core} burning, again by the encroachment
of the helium convective zone on the hydrogen shell.  As the hydrogen
convective core receded near the end of central hydrogen burning, a
gradient of hydrogen concentration was left behind that resulted in
several solar masses with hydrogen mass fraction below $10\,\%$. The
helium convection zone first collided with an appreciable hydrogen
abundance when the central helium mass fraction was $0.50$.  The initial
entropy contrast was, as in the case of the $25\,\Msun$ star, quite
small, and the hydrogen envelope was entirely radiative.

In the collision, a trace of hydrogen was initially convected
downwards and combined with carbon to make nitrogen in the outer part
of the helium convective core.  The energy released raised the entropy
and temporarily shut off the convection, truncating nitrogen
production after about $\Epp-4\,\Msun$ had been made.  When the helium
convective core grew again, most of this nitrogen was convected
downwards in the helium core and became \I{22}{Ne} (with interesting
implications for the \textsl{s}-process.  Later, when the central
helium abundance was $0.38$, however, the convective core re-expanded
into the hydrogen layer again, making more nitrogen, this time about
$\Epp-3\,\Msun$.  Enough of this was mixed into the hydrogen envelope
to ignite hydrogen burning and produce a convection zone that extended
out to a mass coordinate of about $55\,\Msun$ (\Fig{conv40+80}). It
also shut off convection in the outer part of the helium convection
zone (see the vertical line at $\Ep{5.35}\,\s$ in \Fig{conv40+80}).
Over time the hydrogen burning replenished a thin shell of helium ash
at its base.  When helium burning was almost over (central mass
fraction $0.02$), convective dredge up destroyed this thin buffer
shell and carbon began to be dredged up directly into the
envelope. The nitrogen abundance rose rapidly at this point
($\Ep{4.2}\,\s$ in \Fig{conv40+80}), eventually reaching
$0.25\,\Msun$, and the star became a fully developed red supergiant.

Rotation, which was neglected here, is likely to have a large effect,
on primary nitrogen production, especially on the more massive stars
\citep{Hir08}.  Shear mixing between the highly differentially rotating
helium core and hydrogen envelope will amplify the effects of
convective overshoot that we have included here and lead to greater
nitrogen production and more frequent red supergiant progenitors.  This
could have a dramatic effect on the mixing during the explosion and we
intend to include these effects in a subsequent publication.

\subsection{Presupernova Models - Structure and Composition}

\Fig{presn} and \Tab{pre-SN} show the composition and density
structure of our presupernova stars.  By and large, these structures
are the same as seen in solar metallicity stars with the same final
helium core mass.  Thus whatever central engine blows up solar
metallicity stars below $40\,\Msun$ should blow up these just as well,
though the degree of rotation and the amount of fallback after the
initial explosion will differ.

The density of the hydrogen envelope in the $15\,\Msun$ stars is much
larger and its radius smaller for reasons we have discussed - chiefly
the weak hydrogen burning shell in stars with low metallicity. The
$25\,\Msun$ star, as previously noted, is somewhat anomalous in having
produced primary nitrogen shortly before its death, but not yet
expanded to red giant proportions.  The primary nitrogen and extent of
the convection zone are apparent in the bottom frame of
\Fig{presn}.  But in general, the helium core structures and especially
the iron cores of the zero metallicity stars resemble their solar
counterparts.  There are greater variations from mass to mass than
systematic differences in the two metallicities.

Presupernova properties of the zero metallicity stars are given in
\Tab{pre-SN}.  For comparison, the central density, temperature,
entropy, and electron mole number of a $15\,\Msun$ solar metallicity
star \citep{Woo07} at the same stage in the collapse would be
$7.25\E9\,\g\,\cm^{-3}$, $7.64\E9\,\K$, $0.698$, and $0.436$.  For
$25\,\Msun$ the equivalent numbers would be $3.34\E9\,\g\,\cm^{-3}$, 
$7.89\E9\,\K$, $0.892$, and $0.443$.

The figures also make clear two possible choices for locating the
piston in the explosion - the edge of the iron core, where the
electron mole number, $\Ye$ has a sudden decrease, moving inwards,
from $\Ye \approx 0.50$ to $\Ye \ltaprx 0.48$, and the location of
entropy per baryon, $\SNAkB = 4.0$.  The latter, typically at the base
of the oxygen burning shell (\Fig{presn}), is where the density begins
to decline rapidly (moving outwards). The physical motivation for
these choices is discussed in the next section.

\section{Explosion and Mixing}

Lacking a robust, first principles model for how core-collapse
supernovae explode, one must parameterize the explosion.  Even if such
a fundamental understanding existed, the number of stellar masses to
be explored is so large that a parameterized approach would be
necessary, but this admittedly adds an element of uncertainty and
ambiguity in all studies of nucleosynthesis in supernovae.  Assuming
that the central engine in these stars initially operates the same as
in their modern counterparts (though fallback may differ), then
present day constraints on nucleosynthesis, supernova light curves,
and remnant masses \citep{Wea78,Woo07} limit our choices of explosion
energy and mass cut.  It is possible, however, that early stars of a
given mass differed in ways other than metallicity.  More rapid
rotation, for example, may have led to a change in the central engine
and hence different explosion energies, symmetries, and mixing.  Thus
we explore here a wide range of explosion energies and make mixing
essentially a free parameter.

\subsection{The Piston Model}
\lSect{pist}

One can simulate an explosion either by moving a piston (essentially
time-dependent momentum deposition) or depositing energy (usually done
instantly, but could, in principle, depend on time and space).  We
choose the former approach because it is easy to implement in a
Lagrangian code and preserves, approximately, the entropy in the
material close to the piston.  Depositing energy is a reasonable
alternative, but brings in additional parameters (e.g., in what mass
to deposit the energy) and may artificially increase the entropy of
the material in which the energy is deposited.

Besides choosing the location of the piston one must also chose how to
move it.  A rapid large amplitude motion gives a powerful explosion; a
short or slow motion gives a weak one.  Here we explore a range of
kinetic energies at infinity of $0.3\E{51}\,\erg$ to $\Ep{52}\,\erg$
with 8 intermediate values.  These are named Models ``zmX'' where
``z'' stand for the zero metallicity models of this paper, ``m'' is
the mass of the star in solar masses, X is a letter denoting the
explosion energy and is one of: A ($0.3\,$Bethe), B ($0.6\,\B$), C
($0.9\,\B$), D ($1.2\,\B$), E ($1.5\,\B$), F ($1.8\,\B$), G
($2.4\,\B$), H, ($3.0\,\B$), I ($5.0\,\B$), and J($10.0\,\B$)
respectively, and the piston mass is situated at $\SNAkB = 4.0$.  We
also followed two additional explosion series with the piston located
deeper in the star at the edge of the iron core.  These were Models P
(explosion energy $1.2\,\B$) and V (explosion energy $10\,\B$)

An important point here is that the energy of the models given is
their \emph{kinetic energy at infinity}.  This is not the same as the
energy often calculated by those who study core collapse, because
our energy automatically includes the gravitational binding energy of
all the mass that was ejected.  That is the energy that must be
provided by the central engine is always larger than the energy with
which our models are labeled, especially for more massive stars. This 
difference is also metallicity dependent.

For example, in our Model z25D, the piston is located at $\SNAkB =
4.0$ which is at $2.17\,\Msun$. In the explosion (plus fallback), all
the material external to a final remnant mass of $4.16\,\Msun$ is
ejected with a kinetic energy of $1.2\,\B$.  The binding energy of
that ejected material in the presupernova star is $0.775\,\B$, so the
central engine, plus any energy generation associated with fallback
actually had to provide $1.98\,\B$.  For a corresponding solar
metallicity star, S25A \citep{Woo07}, the remnant mass was
$2.09\,\Msun$ and the binding energy in the presupernova star at that
point was $0.865\,\B$, so the central engine had to provide
$2.07\,\B$. In general, the low metallicity stars are more tightly
bound and that makes them harder to explode.  But the amount of mass
that falls back is also larger and that means less total energy is
required to provide the assigned kinetic energy to the material that
does escape.

This perhaps cumbersome definition of explosion energy has been used
in the nucleosynthesis community for decades and will probably persist
until the central engine responsible for the explosion is better
understood.

\subsection{Fallback and Remnant Masses}
\lSect{fallback}

The idea of ``fallback'' in supernovae, the incomplete ejection of
matter outside the neutron star, was originally introduced by
\citet{Col66,Col71}.  \citet{Che89} gave analytic estimates for the
time history of fallback and pointed out that fallback would be
enhanced in compact stars (e.g., blue supergiants as opposed to red
ones). This is because the expanding helium core encounters
significant mass at an earlier time in more compact stars, thus
decelerating the deep interior at a time when its density is higher.
The early arrival of the ``reverse shock'' at the neutron star thus
increases its mass, perhaps turning it into a black hole.

Determining this number accurately poses a computational problem for
Lagrangian hydrodynamics codes like \KEPLER.  If the inner boundary
(here ``the piston'') is reflecting and is held fixed at a constant
radius (here $\Ep9\,\cm$), then the matter that falls back can pile up
on it, ultimately distorting the accretion flow.  Later the deposition
of energy by radioactive decay can even cause matter that has fallen
in to move out again.  This problem can be circumvented, in part, by
fine zoning near the piston and by moving the piston into a smaller
radius, beyond the point where the infall becomes supersonic. In
practice, however, attempts to do this resulted in code instability
and small steps.

A different approach was therefore adopted.  Explosions calculated in
\KEPLER were linked at $100\,\s$, a time after nucleosynthesis has
ceased but fallback has not commenced, to an explicit Eulerian
hydrodynamics code.  A transmitting inner boundary condition
\citep[see also][]{Mac01} was assumed and the fallback calculated for
the approximately 1200 supernova studied here.  The results, for the
full range of masses, explosion energies and choices for piston mass,
have been determined by \citet{Zha07} and were employed for all
nucleosynthetic yields reported here.  \Fig{remnant} shows these
masses for the explosions of various energies using the
$S/N_{\mathrm{A}}k_{\mathrm{B}}=4$ piston location.  As the figure
shows, fallback is a very important effect in zero metallicity stars
leading, for a standard explosion energy of $1.2\,B$, to delayed black
hole formation in stars over about $25\,\Msun$.  This can have a
dramatic effect on the light curves, nucleosynthesis, and remnant
properties of such massive stars.

\subsection{Mixing}
\lSect{mixing}

Multi-dimensional effects such as mixing cannot be followed in our
one-dimensional code and this adds another source of uncertainty to
our calculations.  This mixing can be quite important, both for the
light curve and the nucleosynthesis.  In the more massive stars,
nuclei that might not have been ejected without mixing can escape when
mixing precedes fallback.  Mixing was studied in SN~1987A by many
groups and should be similar for the compact stars calculated here
(\Tab{pre-SN}).  For stars with radii $\ltaprx 2\E{12}\,\cm$, it may
be less because the perturbations has less time to grow before
freezing out \citep{Her94}.

Here we adopt a consistent, but artificial prescription for mixing
across all masses and explosion energies, the same essentially as used
by \citet{Pin88} in their study of mixing in SN~1987A.  A running
boxcar average of width $\Delta M$ is moved through the star a total
of $n$ times until the desired mixing is obtained.  The default values
$\Delta M$ and $n$ are $10\,\%$ of the mass of the helium core and 4,
respectively.  This gives, for example, the mixed composition for a
$25\,\Msun$ star in \Fig{mixed}.  Since this parametric mixing is
applied after the explosion is all over, other choices of parameters
ranging from no mixing to nearly complete mixing may explored provided
that the distribution of yields with Lagrangian mass in the unmixed
model is known (\Sect{robot}).

\section{Light Curves}

\subsection{UV-Transients from Shock Break Out}

The optical display begins as the supernova shock erupts from the
surface giving rise to a brief, hard ultra-violet flash.  For the
progenitors with radius $R \sim 10^{12}$ cm and explosion energies of
$1.2\,\B$, the flash should resemble that estimated for SN~1987A,
\citep[e.g.,][]{Ens92}, i.e., $L \sim 5\E{44}\,\ergs$ for about two
minutes with a color temperature, $\Tcol\sim\Ep6\,\K$.  The displays
of the more compact stars will be fainter, last a shorter time and
have a somewhat harder spectrum.  Obviously for supernova that may be
happening at high redshift, say $z\sim5-10$, these properties would
need to be scaled to the local frame and corrected for ultraviolet
absorption between here and the source, which may be significant.

\subsection{Typical Light Curves}

For those compact presupernova with radiative envelopes - essentially
those with radii smaller than about $50\,\Rsun$ in \Tab{pre-SN} - the
light curves resemble that of SN~1987A, a brief faint plateau followed
by a broad peak powered by radioactive decay (\Figs{lc}). Depending on
details of the mixing, the dip just before the radioactive peak could
be greater (less mixing) or absent (with a lot of mixing, as in
87A). For supernovae that experience a lot of fallback, i.e., the
lower energy explosions for a given mass, the light curve is much
fainter because most of the \I{56}{Ni} falls into the collapsed
remnant.

\section{Nucleosynthesis}

For a given mass star, piston location, mixing prescription, and
energy, the nucleosynthesis is completely determined.  Mixing is
applied across all the nucleosynthesis, including the part that falls
back.  Otherwise the two operations, mixing and fallback, would not be
commutative.  Mixing ejecta after some has already been removed from
the grid gives a different result than mixing before.  \citet{Woo95}
did not include mixing and, as a result, their ejected compositions
were more sensitive to fallback.

\subsection{Nucleosynthesis in Representative Models}

\Tabs{0.6-yield-mix} through \Tabff{10.0-yield-mix} give a subset of
the nucleosynthesis determined for the supernovae studied here. They
provide the total ejected masses of each stable isotope from hydrogen
through germanium for six different mass stars ($12\,\Msun$,
$15\,\Msun$, $20\,\Msun$, $25\,\Msun$, $35\,\Msun$, $50\,\Msun$,
$75\,\Msun$ and $100\,\Msun$) and five choices of explosion energy
($0.6\,\B$, $1.2\,\B$, $2.4\,\B$, $5\,\B$, and $10\,\B$).  Standard
piston locations (the point where $SNAkB = 4.0$) and mixing
prescriptions (\Sect{mixing}) were employed.  It is not necessary to
tabulate yields above germanium because they were negligibly small in
all cases.  We do not include here, however, the nucleosynthesis from
the neutrino winds of all those stars that made neutron stars or from
any disks in those stars that made black holes.  To first order, the
\textsl{r}-process and perhaps part of the \textsl{p}-process would be
made in the neutrino winds of these first generation stars in the same
amounts that they are today.  That is, if the solar \textsl{r}-process
abundance pattern owes its origin to proto-neutron star winds, it
would not be surprising to see the same pattern in second generation
stars.  More extensive tables, available in the electronic version on
the paper, give the ejected masses of each isotope for all 1,440
explosion models with four choices of mixing (120 masses times 10
explosion energies with $SNAkB = 4.0$ plus two explosion energies with
pistons located at the edge of the iron core).  The principal mass
fractions in the ejecta of the mixed and unmixed Model z15D are given
in \Fig{mixed} which makes clear the sensitivity of the results to
fallback.

\subsection{Lithium and Deuterium Production by the Neutrino Process}

Deuterium, \I3{He} and \I7{Li} are known to be important products of
Big Bang nucleosynthesis, but interesting quantities can also be made
in stars.  \Fig{Li7} shows the distribution of these species and the
radioactive progenitor of \I7{Li}, \I7{Be}, in the ejecta of the
$1.2\,\B$ explosion of a $15\,\Msun$ star (Model z15D).

These species (as well as \I{11}B and \I{19}F) are made during the
explosion by neutrinos interacting with the supernova ejecta
\citep{Woo77,Dom78,Woo90}.  The deuterium production occurs in the
envelope due to the charged current reaction, $\nu_e(p,e^+)n$, with
neutrons immediately reacting with protons to make \I2H. A substantial
portion of this \I2H is turned into \I3{He}, and in any case the
abundance is not large enough compared with the Big Bang value, $\sim
4\E{-5}$, to be interesting.  In fact these stars are a net
\emph{sinks} for \I2H.  The same is true for \I3{He}.

The most interesting of the three is \I7{Li}, which is produced in
much greater quantities than is destroyed.  The reactions of interest
are the neutral current process
\I4{He}($\nu_{\mu,\tau},$\I{}n$)$\I3{He}($\alpha,\gamma)$\I7{Be} in
the helium and heavy element core, and the charged current sequence
$\anue($\I{}p$,$\I{}e$^+)$\I{}n$($\I{}p$,\gamma)$\I2H$($\I{}p$,\gamma)$\I3{He}$(\alpha,\gamma)$\I7{Be}
at the base of the hydrogen envelope (just outside $3.7\,\Msun$ in the
figure).  As we shall see later, the net yield averaged over an
initial mass function is an appreciable fraction of solar.

\subsection{Elemental yields as a function of mass}

\Figs{pf:A-F} show the production factors for all elements as a
function of stellar mass and explosion energy.  All calculations
assumed the standard prescription for mixing.  For all but the highest
energies, there is significant fallback above $30\,\Msun$ that
substantially decreases the production of heavy elements.

Taking Case D ($1.2\,\B$ explosions) as representative case, one sees
very limited production of elements above zinc and a strong odd-even
effect for elements lighter than silicon.  The lighter colored bands
for \El{Be} and \El{N} are indicative of substantial
underproduction. The trend continues to a lesser extent for \El{F},
\El{Na}, and \El{Al}.  Beryllium probably does not owe its origin to
massive stars and nitrogen may be underproduced because of the neglect
of rotation.  Fluorine is made in interesting quantities by the
neutrino process, but not quite in a solar ratio compared with other
major productions.  Sodium and aluminum are sensitive to the initial
metallicity because the neutron excess in carbon and neon burning is
smaller when the metallicity is low.

Some interesting odd $Z$ elements are made though.  The reddish band
for lithium shows that a lot is being mad in stars of many different
masses.  Boron is also produced in copious quantities by the neutrino
process.

Stars lighter than about $12\,\Msun$ produce large amounts of silicon
through zinc \emph{compared with magnesium}, though the total yield of
all heavy elements is low in such light stars.  Overall though, one is
stuck with the preponderance of blue and black in the figure, i.e.,
most elements above neon in most stars are made in nearly solar
proportions.

\subsection{Integrated Elemental Yields}

\subsubsection{Standard model and variations in IMF and mixing}

While the yields of individual stars can vary discontinuously due to
the location of various burning shells, the average over large numbers
of stars shows less fluctuation. \Fig{IMFyield} shows the integrated
yields of elements from hydrogen to selenium for various assumptions
regarding the initial mass function, explosion energy and mixing.

Panel A of this figure shows what may be regarded as the ``standard
solution''.  All explosion energies are taken to be $1.2\,\B$
regardless of mass and the standard mixing prescription is
employed.  The solid line is the result when all masses are
included.  One sees that the general trends discussed in the previous
section are reflected in the integral.  Odd-even effects tend to wash
out above silicon and nitrogen is underproduced compared with, e.g.,
oxygen in the sun by almost a factor of 10.  For these Population III
stars, the initial mass function is particularly uncertain, and might
have been ``top heavy'' compared with modern stars \cite{Tan04}, so
the dashed lines in Panel A explore the consequences of omitting
supernovae below a certain mass from the sample.  For the standard
choices of mixing and explosion energy, deleting stars on the light
end of the IMF tends to suppress the production of heavier
elements. These heavy elements are made, but a large fraction falls
back into the collapsed remnant. Primary nitrogen is augmented by as
much as a factor of three because it tends to be made chiefly in the
more massive stars and does not fallback.

Panel B shows that the effect of fallback on heavy element production
can be even more dramatic if mixing is small or absent and this is
certainly one solution to the existence of ultra-iron poor stars (see
\Sect{ultraFe}). Mixing is probably suppressed for explosions in
presupernova stars that are very compact compared with stars red
supergiants derived from the same main sequence mass
\citep{Her94,Chu08}.  Depending on the amount and timing of primary
nitrogen production,such compact progenitors may be common in
Population III stars (\Tab{pre-SN}).

\subsubsection{Effect of a mass-dependent explosion energy}

While we do not know the details of the explosion mechanism, it is
reasonable to expect that it will function differently in stars of
varying iron core mass and central density structure.  In particular,
the density gradients will be shallower in the silicon shells and
inner oxygen shells outside the iron cores of higher mass stars.
Because of the increased ram pressure from accretion of this material,
it will be harder to develop an explosion in stars of higher mass.  If
no outward shock is generated, the star just becomes a black hole and,
in the absence of rotation, simply disappears.  This is equivalent to
imposing an upper mass limit on the IMF Such stars would contribute to
re ionization, but nothing else.  On the other hand, if such stars do
explode, one expects their energy to be comparable to their binding
energy which increases with mass \cite{WHW02}.  The higher accretion
rate may also cause the buildup of a greater overpressure before the
shock moves out.  These considerations motivate exploring the effect
of an explosion energy, $\Ekin$, that increases with mass as some
power
\begin{equation}
\Ekin \ = \ F \left(\frac{M}{20 \ \Msun}\right)^n
\end{equation}
where $F$ is the kinetic energy at infinity for  $20\,\Msun$ supernova
and $n$ is some positive power.  Since we have calculated a large number
of stars and explosion energies, interpolation in energy is meaningful
and a continuous function can be employed.

Panel C of \Fig{IMFyield} explores the consequences of $F = 1.2\,\B$ and
$n = 1$ and Panel D shows what happens if $F$ is varied while $n = 0$.
In both cases raising the explosion energy for the heavier stars
decreases the amount of fallback, increases the iron-group yield, and
makes the results overall more robust to variations in the IMF and
fallback. The sensitivity of the integrated yields is much more
sensitive to mixing, fallback and explosion energy than they are to the 
placement of the piston (Panels E and F of \Fig{IMFyield}).

\section{An Automated Fitting Procedure to Abundance Patterns}
\lSect{robot}

Of course the solar abundance set was not exclusively produced by the
first generation of massive stars and it makes more sense to compare
the yields with observations of metal-poor stars.  Before doing so
however, we briefly discuss an algorithm that will facilitate the
objective comparison with both average date sets for metal poor stars
and individual objects.  The parameter space has indeed become rather
large - IMF, mixing, fallback, explosion energy, and piston location,
so an automated engine to search for the most appropriate models is
necessary. The parameters such an algorithm finds as optimal may have
interesting implications for the first stars as well as proving that a
good fit is in principle possible.

>From the compete isotopic yields from our nucleosynthesis calculations
we first follow all the decay chains including (ground state)
branchings to obtain yields of stable isotopes.  To compare with
observed abundance patterns we compute elemental yields by adding the
number fractions for each isotope.

To find the best fits to observed abundance patterns we compute a data
base of ten explosion energies (\Tab{explosions}) and $13$ different
mixtures ($0$ and $-0.6$, $-0.8$, $-1.0$ [default], ... \dex as
fraction of the \El{He} core size) for each of the $120$ stellar
models.  For the best star fits we search this data base.  For the IMF
fits we integrate yields for different Salpeter-like power-law IMFs
($\Gamma=-0.65$, $0.35$, $1.35$ [standard], $2.35$, $3.35$) and for
different lower and upper bounds of the IMF.  A second set assumes a
Gaussian IMF (in $\log M$) centered at different masses and with
widths of $(1\ldots20)\times 0.025\,\dex$.  For both IMFs we assume a
mass-dependent explosion energy such that
\begin{equation}
E=E_0 \times \left(\frac{M}{20\,\Msun}\right)^\EExp
\end{equation}
where $E_0$ can be any of our explosion energies and $\EExp$ is one
of $-1$, $-0.5$, $0$ [default], $0.5$, $1$.

For the $N$ elements present in each data set (HE0107, HE1327,
\citealt{cay04}) with $\log\epsilon$ values $D_i$ and uncertainties
$\sigma_i$ we then determine the quality of the fit from the standard
formula
\begin{equation}
  \chi^2=\left(\sum_{i=1}^N W_i\right)^{\!\!-1} \times 
    \sum_{i=1}^N
    \frac{\left(F_i+O-D_i\right)^2}{\sigma_i^2} W_i
\;.
\lEq{sigma}
\end{equation}
Extra weights $W_i$ can be employed; in this paper we always use
$W_i\equiv1$.  The fit value for the elements, $F_i$, are shifted up
or down by a constant offset $O$.  Note that the entire fits is done
in logarithmic abundance space.  The optimal value for $O$ is found by
differentiating \Eq{sigma} with respect to $O$ and setting it to zero
and solving for $O$ to find the minimum (there is no maximum).

Upper limits are only included in the determination of $\chi$ when
the observational data points lie below the theoretical abundance
pattern, otherwise they are ignored.  Algorithmically this is
implemented by first finding the offset $O$ when all upper limits are
ignored.  Then we successively add the data points of the lower limits
most below the theoretical data (this always will bring the fit down)
until only upper limits above the fit remain.  For the normalization
the upper limits always have to be considered.  If there are $N$
normal limits ($i=1\ldots N$) and $U$ upper limits ($i=N+1\ldots N+U$)
we hence compute $\chi$ from
\begin{eqnarray}
  \chi^2&=&\left(\sum_{i=1}^{N+U} W_i\right)^{\!\!-1} \!\!\!\times\; 
    \left(
\sum_{i=1}^N\frac{\left(F_i+O-D_i\right)^2}{\sigma_i^2} W_i
\; +\right.\\
\nonumber
& &\left.\sum_{i=N+1}^{N+U}\frac{\left(F_i+O-D_i\right)^2}{\sigma_i^2} 
  W_i\,\Theta\left(F_i+O-D_i\right)
\right)
\;,
\end{eqnarray}
where $\Theta(x)$ is the Heaviside function.  Lower limits for
theoretical data are mathematically equivalent and are treated in an
identical way.

For combining elements, e.g., carbon and nitrogen, ``C+N'', we add the
number fraction of the isotopes.  For production factors, we compare
the number fraction of isotopes with the total solar number fraction
for these elements.  The error bars, while given in dex, we combine in
linear space by multiplying them by the abundance, i.e., by the
product of production factor and solar abundance.  For a combined
species X which is the combination of $N$ species x$_i$,
X=''x$_1$+x$_2$+\ldots+x$_N$'' with production factors $P_i$, solar
abundances $S_i$, and error $\sigma_i$ in dex we compute the combined
solar number fraction $S_\mathrm{X}$, the combined production factor
$P_\mathrm{X}$, and the combined abundance error $\sigma_\mathrm{X}$ (in
dex) from
\begin{eqnarray}
S_\mathrm{X} &=& \sum_{i=1}^N S_i
\;,\quad\\
P_\mathrm{X} &=& S_\mathrm{X}^{-1}
  \;\times\; 
  \sum_{i=1}^N P_i\, S_i 
\;,\quad\\
\left(10^{\sigma_\mathrm{X}}-1\right)^2 &=& 
  P_\mathrm{X}^{-1} \, S_\mathrm{X}^{-1}
  \;\times\; 
  \sum_{i=1}^N \left(10^{\sigma_i}-1\right)^2\, P_i\, S_i 
\;.
\end{eqnarray}
Hence the absolute error of
combination of $N$ species of equal error $\sigma_i$ is $\sqrt{N}$
times bigger than that of each species, while the relative error
becomes smaller by a factor $1/\sqrt{N}$.  When combing species for
very different number fractions the smaller one does not contribute
much to the combined production factor $P_\mathrm{X}$ or the combined
relative error $\sigma_\mathrm{X}$ (in dex).  Note that a species with
a high production factor can still have a small number abundance due
to a small solar number fraction of this species.  For example, when
nitrogen has a production factor similar to that of oxygen, nitrogen
would contribute little to ``N+O'' and its error bar since it solar
number fraction is about 10$\times$ less than that of oxygen.

\section{Comparison With Observations}

\subsection{The \citeauthor{cay04} Data Set}
\lSect{Cayrel}

\cite{cay04} have summarized the abundances obtained from observations
of 35 giants, 30 of which have \SolRatEl{Fe}H in the range $-4.1$ to
$-2.7$.  In \Fig{Cayrel} we show the best matches between our model
database and these observations for various choices of physics.
\Tab{fit_table} gives further information on these and other similar
fits.  Note that we printed in bold font the parameters or limits of
each fit that were held fixed.  When carrying out these fits, we chose
to ignore the element chromium.  We simply could not fit the
\citeauthor{cay04} \El{Cr} abundance using any subset of our models
and did not want to contaminate the statistics for the other elements
by trying.  It seems likely to us that there are greater uncertainties
in the \citeauthor{cay04} abundance of Cr than indicated by the
published error bar \citep[see also][]{cay04,Sob07,Lai08}.  We also
treated our calculated abundances of scandium and copper as lower
bounds.  The zero metal stars studied here are singular in the absence
of any appreciable heavy element production by the \textsl{s}-process.
In a separate work in preparation, we have found that this is not the
case for stars with even a trace of initial metallicity (e.g., \Ep{-4}
solar).  Since the \citeauthor{cay04} set is for metallicity around
\Ep{-3} solar, it is reasonable that some contamination by second
generation stars may already have occurred.  Also, as previously
noted, we ignore nucleosynthesis in the neutrino-powered wind
\citep{Pru05}, GRB accretion disks \citep{Sur06} here, and any other
\textsl{r}-process site.

Panel E of \Fig{Cayrel}, by no means the best possible fit, shows the
``standard model'' - $1.2\,\B$ explosions for all masses, standard
mixing, and a Salpeter IMF with all stellar masses included.  Even
this ``first guess'' gives good overall agreement with the
observations.  Indeed, given all the uncertainties in both models and
observations, it may be that nothing beyond this set of assumptions is
really justified.  The elements \El{Cr} and \El{Na} are greatly
overproduced in the model.  We have discussed the uncertainty in
\El{Cr} and \El{Na} might have large non-LTE corrections or a
contribution from $Z > 0$ stars.  Or it may be that the odd-even
effect has been underestimated in our models.  As we shall see though,
it is difficult to simultaneously fit \El{Na}, \El{Al}, and \El{Mg}
for any selection of our model space.

Cobalt and zinc on the other hand are underproduced, though not by two
sigma.  Zinc has an uncertain contribution from the \textsl{s}-process
and the neutrino wind \citep{Hof96}.  Cobalt is chiefly a product of
the alpha-rich freeze out from nuclear statistical equilibrium and its
production is therefore particularly sensitive to how the inner zones
are treated in the explosion (i.e., using a piston or thermal energy
deposition to drive the expansion) and to fallback.  Note that simply
increasing the explosion energy is not a good solution for the cobalt
underproduction \citep{Kob06}.  Panel E shows the results when a
higher explosion energy is imposed on a standard IMF, $1.8\,\B$ at
$20\,\Msun$ and proportional to $M$ thereafter.  The production of
\El{Co} is not improved and the quality of the overall fit is reduced.
Since \El{Ni} and \El{Co} are both made by the alpha-rich freeze out
here, it is hard to raise \El{Co} without overproducing \El{Ni}.

Other possibilities are explored in Panels A - D.  The best overall
power law fit, not surprisingly, is achieved in Panel B when the code
is allowed to vary simultaneously all degrees of freedom - mixing,
explosion energy, slope of the IMF, and upper and lower mass limits.
Essentially everything is fit except \El{Na} and \El{Cr}.  The
parameters the code finds for this good fit are interesting.  It
prefers to have stars in the limited mass range $11\,\Msun$ to
$15\,\Msun$, a flatter IMF than Salpeter, very little mixing in the
explosion, and a \emph{low} explosion energy that decreases with
increasing mass.  The low energy and reduced mixing suppress the
contribution of the heavier stars because their synthesis mostly falls
back into the remnant.  Panel A shows again this preference for low
mass stars and low explosion energy when the IMF is allowed to be a
Gaussian rather than a power law.

It should be noted, however, that the goodness of the fit, both by eye
and formally, in the $\chi$, for Panels A - D and even Panel E, does
not differ greatly and no particular significance should be attached
to any one set of parameters.  Panel F on the other hand, where a
higher explosion energy is artificially imposed, is a decidedly
inferior fit.  Overall the \citeauthor{cay04} data set is consistent
with a normal IMF and explosion energy, but with a preference for
stars lighter than $15\,\Msun$ and reduced mixing in the explosion.
The reduced mixing may be a natural consequence of the compact nature
of the supernova progenitor stars in this mass range.

\subsection{Comparison with the Ultra-Iron-Poor Stars}
\lSect{ultraFe}

The stars HE0107-5240 \citep{Chr04,Bes04,Fre05,Col06,CBE07} and
HE1327-2326 \citep{Fre05,aok06,fre06,Col06,FL07} may be representative
of a larger class of stars that show both very low iron abundance and
extremely high oxygen to iron ratios.  A popular explanation is that
such stars represent the ejecta of a primitive stellar population in
which mixing was minimal and fallback was large.  The small iron
abundance thus reflects the fact that, in some supernova or set of
supernovae, iron was made, but failed to be mixed sufficiently far out
to be ejected \citep{Ume03,Lim03}.  A very similar scenario was put
forward in \citet{Dep02} based upon one of our unpublished models for
a $35\,\Msun$ zero-metallicity supernova, in order to explain
another low metal star, CS 22949-037, with a large oxygen to iron
ratio.  In fact, large oxygen to iron ratios due to fallback in low
metallicity supernovae were also published (Models Z35B and Z40B) by
\citet{Woo95}.

If the abundances in such stars do reflect the operation of an early
generation of supernovae, then the pattern might provide information
about those events.  Were they high mass stars or low, more energetic
or less, mixed or unmixed, and are the results consistent with a range
of supernova masses and metallicities or just a single event.

\subsection{Comparison with HE0107}

\Fig{HE0107} shows several model fits compared with data from
\citet{CBE07}.  The fitting program (\Sect{robot}) was again
instructed to find the best fit subject to various constraints.  As
discussed in \Sect{Cayrel}, the element \El{Cr} was ignored in the
fit, but the observational upper limit and was always above the model
value.  In general, the best fits were for supernovae in the mass
range $12\,\Msun$ to $30\,\Msun$ with the best single star fit
occurring for $20.5\,\Msun$ with an explosion energy of $0.6\,\B$ and
very little mixing (Panel A).  If the carbon and nitrogen abundances
are combined to reflect possible \El{C}\El{N} processing in the star,
the fit is modestly improved (Panel B) and the favored stellar mass
shifts to lower values.  Gaussian IMFs allow a spread in masses around
a preferred value and favor a relatively narrow range of masses around
$15\,\Msun$, again showing that low explosion energy and less mixing
are favored.  Panels E and F allow the code to seek a power law fit
that is either Salpeter in character (Panel F) or where the exponent
is allowed to float (Panel E).  In all cases, the best fit is for low
explosion energies, small fallback and stars in the $10\,\Msun$ to
$30\,\Msun$ range.

The low energy and preference for lighter supernova progenitors is
similar to what was inferred for the \citeauthor{cay04} data set, but
more extreme.  The \citeauthor{cay04} set definitely required more
mixing and was at least consistent with all masses from $10\,\Msun$ to
$100\,\Msun$ participating.  The small amount of mixing may relate to
the compact nature of the progenitor stars \citep{Chu08}, but could
also reflect envelope stripping in a binary system or simply low
explosion energy.  Unless the central engine produces a grossly
asymmetric explosion, one expects the least amount of mixing in Type
Ib and Ic supernovae derived from massive stars that have lost their
envelopes and this have no reverse shock in the explosion.

\subsubsection{Comparison with HE1327-2326}

Fits to observations of HE1327-2326 (\Fig{HE1327}) select similar
models.  Here data are taken from \citet{aok06,fre06} with new upper
limits for \El{Cr}, \El{Co}, and \El{Ni} from \citet{FL07}.  Again,
intermediate masses, low explosion energy, and a very limited amount
of fallback are preferred.  Markedly inferior fits results for both
HE0107 and HE1327 result if one insists on sampling heavier mass stars
up to $100\,\Msun$ using a Salpeter IMF (\Fig{HE1327norm}).  Carbon or
carbon plus nitrogen are grossly underproduced.  Magnesium is also
chronically overproduced.  A particularly bad fit ($\chi = 30.3$) is
obtained for the ``standard model'' (Panel F).  Clearly these stars
have a very different nucleosynthetic history from the
\citeauthor{cay04} set.  Supernova solutions to the problem are
feasible and require low energy, a relatively narrow range of masses,
and little mixing.  But just why these stars would sample such a
limited set possibilities is unclear.  Low energy, spherically
symmetric explosions in stars that have been stripped of their
envelopes in binaries or in compact blue stars like many of the models
computed here would be favored.  The numbers of such stars in a given
generation will be sensitive to the history of primary nitrogen
production since this affects whether the star dies as a compact star
or a red supergiant.  The frequency could thus be a function of both
metallicity and rotation rate.  It is by no means clear that the
ultra-iron poor stars are necessarily more primitive than the stars in
the \citeauthor{cay04} set.

\section{Conclusions}

We have systematically studied the complete evolution and
parameterized explosions of 120 stars of initially zero metallicity.
These stars have been exploded with a range of energies - ten choices
for kinetic energy at infinity between $0.3\,\B$ and $10\,\B$.  For
two energies, $1.2\,\B$ and $10\,\B$, the location of the piston has
been varied.  Nucleosynthesis has been accurately determined for all
isotopes with significant abundance in all these models, as have the
remnant masses that result after all fallback is over.

To first order, the nucleosynthesis of major elements has a solar
pattern.  That is to say, the helium core mass dominates in
determining the synthesis of abundant elements with even nuclear
charge and the helium core mass is not so different for the lighter,
most abundant stars when the metallicity is reduced.  There are four
important exceptions to this trend though.  First, as has long been
realized, reducing the metallicity reduces the synthesis of odd-$Z$
elements and neutron-rich isotopes because the neutron excess after
helium burning depends on the initial abundance of \El{C}\El{N}\El{O}.
This is particularly true for elements lighter than silicon.  Second,
since the mass loss is less, stars with larger helium cores can
survive to the presupernova stage and contribute to the
nucleosynthesis (if their explosion energy is large enough to unbind
them).  Since such large stars have shallow density gradients around
their iron cores, explosive nucleosynthesis contributes more to the
yield.  This can moderately increase the yield of iron group elements
and especially the $\alpha$-rich freeze out.  Third, because there in
no primordial iron in these stars, there is very little
\textsl{s}-process synthesis heavier than germanium.  Those few
neutrons that are produced during helium burning are absorbed on
carbon and oxygen and go to enhancing the production of intermediate
mass isotopes.  Finally, a significant number of stars, especially
those over $40\,\Msun$ (and presumably lower with rotation) produce
primary nitrogen.  This does not happen in non-rotating stars of solar
metallicity because the entropy at the base of the hydrogen shell is
much higher and the helium convective core does not extend as far.
Even here, in zero metal, non-rotating stars, the interpenetration of
shells is sensitive to zoning and an uncertain parameterization of
overshoot (and undershoot) mixing.

We do not include here any contribution from the neutrino wind or the
``hot convective bubble'' in those models that make neutron stars, and
most do.  Presumably the properties of those neutron stars and their
winds would be most sensitive to their mass and not to the star's
initial metallicity.  Therefore we expect the same nucleosynthesis in
the wind as in modern supernovae.  This probably means the production
of the \textsl{r}-process \citep{Woo94}, some light \textsl{p}-process
nuclei by the \textsl{rp}-process \citep{Pru06,Fro06}, and the
production of significant quantities of rare nuclei lighter than
molybdenum \citep{Hof96,Pru05}.  Most important in the last category
may be the elements scandium and zinc. These elements can also be
made, to some extent, by the \textsl{s}-process and $\alpha$-rich
freeze out, but one should therefore treat deficient production of
these elements in our models with caution.

It should also be noted that \emph{zero} metallicity is a singular
condition only realized for the very first generation of stars.  Stars
of even \Ep{-5} solar metallicity will have a different sort of
evolution, because they will not have to produce their own catalyst
for the \El{C}\El{N}\El{O} cycle and will have \emph{some} iron for
the \textsl{s}-process.  Our next survey, nearly completed, is for
\Ep{-4} solar metallicity.  Results will be reported elsewhere.

In addition to the above limitations and the omission of rotation, the
most fundamental limitations of the models is that they do not
incorporate a complete physical model for the explosion - because none
is known, and that they are one-dimensional.  Consequently, the highly
important parameter of ``mix'', how much the composition is stirred by
the explosion, pulsar remnant, and hydrodynamical instabilities,
remains artificial.  To compensate for this arbitrariness, we have
made a library of all our explosion models and developed an algorithm
that allows the user to specify any range of mixing and explosion
energy desired, or, conversely, to present a data set and find the
best fit.  The initial mass function (and whether it is Gaussian,
Salpeter, or single star), the power-law dependence of explosion
energy on mass, and the mixing can all be varied.  Upper and lower
mass limits (between $10\,\Msun$ and $100\,\Msun$ can be specified.
This powerful tool is discussed in \Sect{robot}.

To provide examples, as well as insight, we have applied these models
and this ``robotic'' search for best fits to three sets of data - the
``Cayrel'' set and two ultra-iron poor stars, HE0107 and HE1327.  For
the former we find (\Fig{Cayrel}) several good fits that are not
clearly distinguishable within the observational and theoretical error
bars.  What might be regarded as the ``standard model'' is not a bad
fit.  That is a constant kinetic energy at infinity of $1.2\,\B$ for
all stars, mixing using a moving box car average with width equal to
one-tenth the helium core mass, a Salpeter IMF and including all stars
from $10\,\Msun$ to $100\,\Msun$.  This model fit, and all others,
greatly overproduce chromium which may be more uncertain than reported
thus far by the observers.  The model also overproduces sodium and
underproduces cobalt and zinc by about two sigma.  The fit is
significantly improved by reducing the mixing and the explosion
energy, especially in the higher mass stars.  An explosion energy of
around $0.6\,\B$ to $0.9\,\B$ seems preferred.  That the mixing be less
than what seems to work well for making solar abundances in higher
metallicity stars \citep{Woo07} may not be surprising.  Many of these
zero metallicity stars are compact blue stars when they die and mixing
by the Rayleigh-Taylor instability is suppressed compared with red
supergiants \citep{Her94} where the instability has a longer time to
develop.  The automated fits find no need for a high energy
``hypernova'' component \citep{Tom07} despite the presence in our
library of stars with up to 10 B of explosion energy.

For the ultra-iron-poor stars we find, as did \citet{Ume03}, that the
low abundance of iron group elements is a consequence of fallback
coupled to small mixing.  This effect was also seen in the lower
energy calculations of metal free stars by \citet{Woo95} who had
\emph{no} mixing at early times, and a deficiency of iron in very low
metallicity stars was attributed to fallback in this models by
\citet{Nor01} and \citet{Dep02}.  Here we have considered a much
larger sample of models and confirm that low explosion energy (i.e.,
less than the canonical $1.2\,\B$) for stars above $20\,\Msun$
improves the fit and that the mixing should be smaller than for the
\citeauthor{cay04} set.

Taken together these calculations suggest several characteristics for
the ``first supernovae''.  First, no nucleosynthetic evidence is found
for either pair instability supernovae or hypernovae \citep[contrary
  to][]{Ume05}.  The same mass stars that make the solar abundances
today could have made the abundances in the most metal stars seen so
far quite well.  Mixing was less and fallback more important.  Both of
these effects are consistent with what is expected for more compact
supernova progenitors.  Fallback would also be greater from the very
massive stars that, here at least, had no mass loss and thus died with
much greater binding energy.  It will be interesting to see if the
abundance trends seen in HE0107 and HE1327 are characteristic of all
ultra-iron-poor stars when more are discovered.

The compact nature of some of the presupernova stars (\Tab{pre-SN})
also affects their light curves (\Fig{lc}).  Supernovae resembling
SN~1987A with its initially faint plateau and slow rise to a peak
powered by radioactivity might be much more common in the early
universe.  But this conclusion is sensitive to whether the stars make
primary nitrogen or not.  Here only the heaviest stars became red
supergiants because of nitrogen production, but the inclusion of
rotation may extend the critical mass for primary nitrogen production
downwards \citep{Heg00,Chi06,Mey06,Hir06}.  If so, the star in many,
and perhaps most cases will die as a red supergiant. This would make
the light curves more like ordinary Type IIp's.  We caution again,
though, that zero metallicity is a singularity that only characterized
the very first stars.  We find, in our survey of \Ep{-4} solar
metallicity stars that even a little metallicity can suppress primary
nitrogen production and that most of the supernova progenitors are
blue stars, though not so compact as for $Z = 0$.

For the upper end of the mass range studied, above about $90\,\Msun$,
the pulsational pair instability is encountered.  This leads to the
ejection of the envelope and part of the helium core in a series of
violent nuclear powered flashes \citep{Woo07b}.  Collisions between the
shells ejected in pairs of flashes can produce extraordinarily bright
supernovae.

Finally, the compact remnants of massive stellar evolution, especially
black holes, will be more massive for zero (and low) metallicity.
There are two reasons.  First, because the mass loss rate is low,
larger helium cores are possible at the time the star dies.  This both
increases the potential mass reservoir for making a black hole by
fallback (the hydrogen envelope is almost always ejected) and makes it
harder to explode the star.  Indeed, unless the explosion mechanism
can somehow provide a much greater energy than $1\,\B$, black holes
will be the typical result of stellar evolution for main sequence
masses above about $30\,\Msun$ \citep{Zha07}.  Black hole formation is
also favored in compact presupernova stars because the reverse shock
develops earlier and reaches the center when the density is still high
\citep{Che89,Zha07}.  The maximum black hole mass, though, will be about
$40\,\Msun$, even for very low explosion energies.  This is the largest
helium core that avoids the pulsational pair instability.

\acknowledgements

We are grateful to Thomas Janka for his physically motivated
suggestion to use the density jump at the base of the \El{O} shell as
most likely location for the initial piston that explodes the star.
We thank Rob Hoffman, Tommy Rauscher, Frank Timmes, Jason Pruet,
Karlheinz Langanke, and Gabriel Martinez-Pinedo for their
contributions to the reaction rate library, weak interaction rates,
and neutrino cross sections.  We are also grateful to Tim Beers, Mike
Bolte, and David Lai for discussions about the interpretation of UMP
star abundance and enrichment observation and theory.  We particularly
than Anna Frebel and Norbert Christlieb for providing us their most
recent abundance data for HE0107 and HE1327.  We than Dan Whalen for
helpful discussions about primordial gas chemistry and relevant
radiation bands, and we than Candace Church for providing us results
from her work in preparation on multi-dimensional simulations of
mixing in Pop III supernovae.

This work was supported by the NSF (AST 02-06111), and the DOE Program
for Scientific Discovery through Advanced Computing (SciDAC; grants
DOE-FC02-01ER41176 and DOE-FC02-06ER41438).  At LANL, Heger performed
this work under the auspices of the National Nuclear Security
Administration of the U.S.\ Department of Energy at Los Alamos
National Laboratory under Contract No.\ DE-AC52-06NA25396.

\clearpage

\onecolumn

\begin{table}
\renewcommand{\EE}{}
\centering
\caption{Ionizing Photons and Energies\lTab{ionize}}
\begin{tabular}{rrrrrrrrlrrlr}
\hline
\hline
\noalign{\smallskip}
mass & $E$(HII) & $N_\gamma$(HII) & $\bar{\lambda}$(HII) & $E$(HeII) & $N_\gamma$(HeII) & $\bar{\lambda}$(HeII) & $E$(HeIII) & $N_\gamma$(HeIII) & $\bar{\lambda}$(HeIII) & $E$(LW) & $N_\gamma$(LW) & $\bar{\lambda}$(LW) \\
\Msun & (erg) &  & ($\AA$) & (erg) &  & ($\AA$) & (erg) &  & ($\AA$) & (erg) &  & ($\AA$) \\
\noalign{\smallskip}
\hline
\noalign{\smallskip}
 10.0 & $1.23\E{ 52}$ & $3.96\E{ 62}$ &   637.94 & $2.82\E{ 51}$ & $5.92\E{ 61}$ &   417.56 & $1.18\E{ 49}$ & $1.25\E{ 59}$ &   210.23 & $3.33\E{ 51}$ & $1.69\E{ 62}$ &   1005.1 \\
 10.2 & $1.28\E{ 52}$ & $4.10\E{ 62}$ &   635.96 & $2.98\E{ 51}$ & $6.26\E{ 61}$ &   416.80 & $1.34\E{ 49}$ & $1.41\E{ 59}$ &   210.05 & $3.40\E{ 51}$ & $1.72\E{ 62}$ &   1005.0 \\
 10.4 & $1.36\E{ 52}$ & $4.36\E{ 62}$ &   636.00 & $3.18\E{ 51}$ & $6.67\E{ 61}$ &   416.50 & $1.49\E{ 49}$ & $1.57\E{ 59}$ &   209.87 & $3.64\E{ 51}$ & $1.84\E{ 62}$ &   1005.1 \\
 10.5 & $1.36\E{ 52}$ & $4.34\E{ 62}$ &   633.01 & $3.27\E{ 51}$ & $6.84\E{ 61}$ &   415.66 & $1.61\E{ 49}$ & $1.70\E{ 59}$ &   209.78 & $3.54\E{ 51}$ & $1.79\E{ 62}$ &   1004.9 \\
 10.6 & $1.37\E{ 52}$ & $4.37\E{ 62}$ &   631.94 & $3.33\E{ 51}$ & $6.96\E{ 61}$ &   415.27 & $1.70\E{ 49}$ & $1.79\E{ 59}$ &   209.70 & $3.53\E{ 51}$ & $1.79\E{ 62}$ &   1004.9 \\
 10.7 & $1.40\E{ 52}$ & $4.44\E{ 62}$ &   630.92 & $3.42\E{ 51}$ & $7.15\E{ 61}$ &   414.87 & $1.80\E{ 49}$ & $1.90\E{ 59}$ &   209.60 & $3.57\E{ 51}$ & $1.81\E{ 62}$ &   1004.9 \\
\hline
\end{tabular}
\end{table}
\addtocounter{table}{-1}
\begin{table}
\renewcommand{\EE}{}
\centering
\caption{Ionizing Photons and Energies (continued)}
\begin{tabular}{rrrrrrrrlrrlr}
\hline
\hline
\noalign{\smallskip}
mass & $E$(HII) & $N_\gamma$(HII) & $\bar{\lambda}$(HII) & $E$(HeII) & $N_\gamma$(HeII) & $\bar{\lambda}$(HeII) & $E$(HeIII) & $N_\gamma$(HeIII) & $\bar{\lambda}$(HeIII) & $E$(LW) & $N_\gamma$(LW) & $\bar{\lambda}$(LW) \\
\Msun & (erg) &  & ($\AA$) & (erg) &  & ($\AA$) & (erg) &  & ($\AA$) & (erg) &  & ($\AA$) \\
\noalign{\smallskip}
\hline
\noalign{\smallskip}
 19.2 & $4.02\E{ 52}$ & $1.18\E{ 63}$ &   581.71 & $1.45\E{ 52}$ & $2.89\E{ 62}$ &   395.16 & $2.83\E{ 50}$ & $2.92\E{ 60}$ &   204.87 & $6.91\E{ 51}$ & $3.49\E{ 62}$ &   1003.5 \\
 19.4 & $4.09\E{ 52}$ & $1.19\E{ 63}$ &   580.38 & $1.49\E{ 52}$ & $2.96\E{ 62}$ &   394.67 & $2.97\E{ 50}$ & $3.07\E{ 60}$ &   204.78 & $6.95\E{ 51}$ & $3.51\E{ 62}$ &   1003.4 \\
 19.6 & $4.16\E{ 52}$ & $1.21\E{ 63}$ &   579.56 & $1.52\E{ 52}$ & $3.02\E{ 62}$ &   394.33 & $3.10\E{ 50}$ & $3.19\E{ 60}$ &   204.70 & $7.03\E{ 51}$ & $3.55\E{ 62}$ &   1003.4 \\
 19.8 & $4.28\E{ 52}$ & $1.25\E{ 63}$ &   579.80 & $1.57\E{ 52}$ & $3.11\E{ 62}$ &   394.33 & $3.20\E{ 50}$ & $3.30\E{ 60}$ &   204.65 & $7.26\E{ 51}$ & $3.67\E{ 62}$ &   1003.4 \\
 20.0 & $4.46\E{ 52}$ & $1.31\E{ 63}$ &   581.83 & $1.61\E{ 52}$ & $3.20\E{ 62}$ &   394.51 & $3.31\E{ 50}$ & $3.41\E{ 60}$ &   204.56 & $7.76\E{ 51}$ & $3.92\E{ 62}$ &   1003.6 \\
 20.5 & $4.64\E{ 52}$ & $1.35\E{ 63}$ &   579.33 & $1.70\E{ 52}$ & $3.38\E{ 62}$ &   393.66 & $3.65\E{ 50}$ & $3.75\E{ 60}$ &   204.38 & $7.90\E{ 51}$ & $3.99\E{ 62}$ &   1003.5 \\
\hline
\end{tabular}
\end{table}

\begin{table}
\renewcommand{\EE}{}
\centering
\caption{Binnerd Photon Numbers\lTab{photon}}
\begin{tabular}{rrrrrrrrr}
\hline
\hline
\noalign{\smallskip}
mass & 0.755\eV $\ldots$ & 2.650\eV $\ldots$ & 11.18\eV $\ldots$ & 13.60\eV $\ldots$ & 15.42\eV $\ldots$ & 24.59\eV $\ldots$ & 30.00\eV $\ldots$ & $>$54.42 \eV \\
\Msun & $\ldots$2.650 \eV & $\ldots$11.18 \eV & $\ldots$13.60 \eV & $\ldots$15.42 \eV & $\ldots$24.59 \eV & $\ldots$30.00 \eV & $\ldots$54.42 \eV &  \\
\noalign{\smallskip}
\hline
\noalign{\smallskip}
 10.0 & $1.12\E{ 62}$ & $8.30\E{ 62}$ & $1.69\E{ 62}$ & $9.84\E{ 61}$ & $2.38\E{ 62}$ & $3.84\E{ 61}$ & $2.07\E{ 61}$ & $1.25\E{ 59}$ \\
 10.2 & $1.12\E{ 62}$ & $8.40\E{ 62}$ & $1.72\E{ 62}$ & $1.01\E{ 62}$ & $2.46\E{ 62}$ & $4.03\E{ 61}$ & $2.21\E{ 61}$ & $1.41\E{ 59}$ \\
 10.4 & $1.21\E{ 62}$ & $9.04\E{ 62}$ & $1.84\E{ 62}$ & $1.08\E{ 62}$ & $2.62\E{ 62}$ & $4.29\E{ 61}$ & $2.37\E{ 61}$ & $1.57\E{ 59}$ \\
 10.5 & $1.14\E{ 62}$ & $8.62\E{ 62}$ & $1.79\E{ 62}$ & $1.05\E{ 62}$ & $2.60\E{ 62}$ & $4.37\E{ 61}$ & $2.46\E{ 61}$ & $1.70\E{ 59}$ \\
 10.6 & $1.12\E{ 62}$ & $8.56\E{ 62}$ & $1.79\E{ 62}$ & $1.06\E{ 62}$ & $2.62\E{ 62}$ & $4.43\E{ 61}$ & $2.51\E{ 61}$ & $1.79\E{ 59}$ \\
 10.7 & $1.12\E{ 62}$ & $8.61\E{ 62}$ & $1.81\E{ 62}$ & $1.07\E{ 62}$ & $2.66\E{ 62}$ & $4.54\E{ 61}$ & $2.60\E{ 61}$ & $1.90\E{ 59}$ \\
\hline
\end{tabular}
\end{table}
\addtocounter{table}{-1}
\begin{table}
\renewcommand{\EE}{}
\centering
\caption{Binnerd Photon Numbers (continued)}
\begin{tabular}{rrrrrrrrr}
\hline
\hline
\noalign{\smallskip}
mass & 0.755\eV $\ldots$ & 2.650\eV $\ldots$ & 11.18\eV $\ldots$ & 13.60\eV $\ldots$ & 15.42\eV $\ldots$ & 24.59\eV $\ldots$ & 30.00\eV $\ldots$ & $>$54.42 \eV \\
\Msun & $\ldots$2.650 \eV & $\ldots$11.18 \eV & $\ldots$13.60 \eV & $\ldots$15.42 \eV & $\ldots$24.59 \eV & $\ldots$30.00 \eV & $\ldots$54.42 \eV &  \\
\noalign{\smallskip}
\hline
\noalign{\smallskip}
 19.2 & $1.57\E{ 62}$ & $1.39\E{ 63}$ & $3.49\E{ 62}$ & $2.23\E{ 62}$ & $6.64\E{ 62}$ & $1.56\E{ 62}$ & $1.30\E{ 62}$ & $2.92\E{ 60}$ \\
 19.4 & $1.56\E{ 62}$ & $1.40\E{ 63}$ & $3.51\E{ 62}$ & $2.25\E{ 62}$ & $6.73\E{ 62}$ & $1.59\E{ 62}$ & $1.34\E{ 62}$ & $3.07\E{ 60}$ \\
 19.6 & $1.57\E{ 62}$ & $1.41\E{ 63}$ & $3.55\E{ 62}$ & $2.28\E{ 62}$ & $6.83\E{ 62}$ & $1.62\E{ 62}$ & $1.37\E{ 62}$ & $3.19\E{ 60}$ \\
 19.8 & $1.63\E{ 62}$ & $1.46\E{ 63}$ & $3.67\E{ 62}$ & $2.35\E{ 62}$ & $7.04\E{ 62}$ & $1.67\E{ 62}$ & $1.41\E{ 62}$ & $3.30\E{ 60}$ \\
 20.0 & $1.81\E{ 62}$ & $1.59\E{ 63}$ & $3.92\E{ 62}$ & $2.50\E{ 62}$ & $7.38\E{ 62}$ & $1.72\E{ 62}$ & $1.44\E{ 62}$ & $3.41\E{ 60}$ \\
 20.5 & $1.80\E{ 62}$ & $1.60\E{ 63}$ & $3.99\E{ 62}$ & $2.55\E{ 62}$ & $7.62\E{ 62}$ & $1.81\E{ 62}$ & $1.53\E{ 62}$ & $3.75\E{ 60}$ \\
\hline
\end{tabular}
\end{table}

\clearpage

\begin{table}
\caption{Summary of Presupernova Model Data\lTab{pre-SN}}
\begin{tabular}{rrrrrrrrrrrrrrr}
\hline
\hline
\noalign{\smallskip}
mass & \Tc & \rhoc & \Sc & \Yec & He core & C/O core & O shell & \Ye core & \BEYecore & \MS & \BES & \Reff & \Teff & \Leff \\
(\Msun) & (\GK) & (\kt/\ccm) & (\kB/\amu) &  & (\Msun) & (\Msun) & (\Msun) & (\Msun) & (\B) & (\Msun) & (\B) & (\Rsun) & (\kK) & (\kLsun) \\
\noalign{\smallskip}
\hline
\noalign{\smallskip}
 10.0 & 6.023 & 16.52 & 0.493 & 0.426 &  1.90 &  1.43 &  1.27 &  1.27 &  0.09 &  1.28 &  0.09 &      62 &  8.02 &         14 \\
 10.2 & 6.642 & 18.43 & 0.515 & 0.424 &  1.97 &  1.45 &  1.29 &  1.18 &  0.27 &  1.38 &  0.04 &      38 & 10.51 &         15 \\
 10.4 & 6.731 & 17.01 & 0.531 & 0.422 &  2.29 &  1.59 &  1.32 &  1.18 &  0.34 &  1.32 &  0.11 &      34 & 11.58 &         18 \\
 10.5 & 6.488 & 14.79 & 0.528 & 0.424 &  2.15 &  1.54 &  1.33 &  1.20 &  0.34 &  1.41 &  0.07 &      27 & 12.77 &         17 \\
 10.6 & 6.468 & 18.41 & 0.505 & 0.425 &  2.09 &  1.52 &  1.31 &  1.20 &  0.30 &  1.40 &  0.06 &      21 & 14.86 &         18 \\
 10.7 & 6.499 & 14.85 & 0.528 & 0.425 &  2.17 &  1.54 &  1.32 &  1.19 &  0.35 &  1.41 &  0.08 &      20 & 15.12 &         18 \\
\hline
\end{tabular}
\end{table}
\addtocounter{table}{-1}
\begin{table}
\caption{Summary of Presupernova Model Data (continued)}
\begin{tabular}{rrrrrrrrrrrrrrr}
\hline
\hline
\noalign{\smallskip}
mass & \Tc & \rhoc & \Sc & \Yec & He core & C/O core & O shell & \Ye core & \BEYecore & \MS & \BES & \Reff & \Teff & \Leff \\
(\Msun) & (\GK) & (\kt/\ccm) & (\kB/\amu) &  & (\Msun) & (\Msun) & (\Msun) & (\Msun) & (\B) & (\Msun) & (\B) & (\Rsun) & (\kK) & (\kLsun) \\
\noalign{\smallskip}
\hline
\noalign{\smallskip}
 18.6 & 7.099 &  7.32 & 0.657 & 0.437 &  4.92 &  3.63 &  1.51 &  1.41 &  0.57 &  1.51 &  0.42 &      22 & 21.56 &         95 \\
 18.7 & 7.316 &  6.86 & 0.681 & 0.436 &  4.99 &  3.69 &  1.55 &  1.41 &  0.66 &  1.55 &  0.48 &      10 & 32.31 &         98 \\
 18.8 & 7.400 &  6.42 & 0.699 & 0.438 &  5.00 &  3.68 &  1.57 &  1.42 &  0.69 &  1.57 &  0.50 &      11 & 31.38 &        105 \\
 18.9 & 7.498 &  5.33 & 0.740 & 0.439 &  5.26 &  4.02 &  1.63 &  1.47 &  0.76 &  1.63 &  0.56 &      11 & 31.76 &        101 \\
 19.0 & 7.575 &  5.68 & 0.736 & 0.439 &  5.30 &  4.05 &  1.63 &  1.44 &  0.80 &  1.63 &  0.57 &      11 & 31.28 &        105 \\
 19.2 & 7.427 &  5.94 & 0.714 & 0.438 &  5.19 &  3.88 &  1.59 &  1.44 &  0.72 &  1.59 &  0.53 &      10 & 32.29 &        106 \\
\hline
\end{tabular}
\end{table}

\clearpage

\begin{table}
\caption{Summary of Explosion Models \lTab{explosions}}
\centering
\begin{tabular}{rr@{$\qquad$}l}
\hline\noalign{\smallskip}
explosion &
energy &
piston location\\
model &
(\B) &
(initial mass cut)\\
\noalign{\smallskip}\hline\noalign{\smallskip}
A &  0.3 &  $S=4$ (base of O shell) \\
B &  0.6 &  $S=4$ (base of O shell) \\
C &  0.9 &  $S=4$ (base of O shell) \\
D &  1.2 &  $S=4$ (base of O shell) \\
E &  1.5 &  $S=4$ (base of O shell) \\
F &  1.8 &  $S=4$ (base of O shell) \\
G &  2.4 &  $S=4$ (base of O shell) \\
H &  3.0 &  $S=4$ (base of O shell) \\
I &  5.0 &  $S=4$ (base of O shell) \\
J & 10.0 &  $S=4$ (base of O shell) \\
\noalign{\smallskip}
P &  1.2 &  iron core (drop in $\Ye$) \\
V & 10.0 &  iron core (drop in $\Ye$) \\
\noalign{\smallskip}\hline
\end{tabular}
\end{table}

\begin{table}
\centering
\caption{Adopted Solar Abundaces}
\begin{tabular}{lr|lr|lr|lr|lr|lr}
\hline
\hline
\noalign{\smallskip}
ion & mass frac. & ion & mass frac. & ion & mass frac. & ion & mass frac. & ion & mass frac. & ion & mass frac. \\
\noalign{\smallskip}
\hline
\noalign{\smallskip}
\I{  1}{ H} &  7.11\E{- 1} & \I{ 17}{ O} &  2.62\E{- 6} & \I{ 31}{ P} &  7.59\E{- 6} & \I{ 42}{Ca} &  5.00\E{- 7} & \I{ 52}{Cr} &  1.64\E{- 5} & \I{ 63}{Cu} &  6.72\E{- 7} \\
\I{  2}{ H} &  2.76\E{- 5} & \I{ 18}{ O} &  1.49\E{- 5} & \I{ 32}{ S} &  3.96\E{- 4} & \I{ 43}{Ca} &  1.07\E{- 7} & \I{ 53}{Cr} &  1.89\E{- 6} & \I{ 65}{Cu} &  3.09\E{- 7} \\
\I{  3}{He} &  3.41\E{- 5} & \I{ 19}{ F} &  4.67\E{- 7} & \I{ 33}{ S} &  3.22\E{- 6} & \I{ 44}{Ca} &  1.69\E{- 6} & \I{ 54}{Cr} &  4.80\E{- 7} & \I{ 64}{Zn} &  1.12\E{- 6} \\
\I{  4}{He} &  2.74\E{- 1} & \I{ 20}{Ne} &  1.17\E{- 3} & \I{ 34}{ S} &  1.86\E{- 5} & \I{ 46}{Ca} &  3.36\E{- 9} & \I{ 55}{Mn} &  1.47\E{- 5} & \I{ 66}{Zn} &  6.60\E{- 7} \\
\I{  6}{Li} &  7.39\E{-10} & \I{ 21}{Ne} &  2.94\E{- 6} & \I{ 36}{ S} &  8.00\E{- 8} & \I{ 48}{Ca} &  1.66\E{- 7} & \I{ 54}{Fe} &  7.73\E{- 5} & \I{ 67}{Zn} &  9.86\E{- 8} \\
\I{  7}{Li} &  1.05\E{- 8} & \I{ 22}{Ne} &  9.44\E{- 5} & \I{ 35}{Cl} &  4.06\E{- 6} & \I{ 45}{Sc} &  4.50\E{- 8} & \I{ 56}{Fe} &  1.26\E{- 3} & \I{ 68}{Zn} &  4.57\E{- 7} \\
\I{  9}{Be} &  1.94\E{-10} & \I{ 23}{Na} &  3.87\E{- 5} & \I{ 37}{Cl} &  1.37\E{- 6} & \I{ 46}{Ti} &  2.69\E{- 7} & \I{ 57}{Fe} &  2.96\E{- 5} & \I{ 70}{Zn} &  1.56\E{- 8} \\
\hline
\end{tabular}
\end{table}

\clearpage

\begin{table}
\centering
\caption{$^{14}$N Yields, 1.2\,B, Standard Mixing\lTab{n14}}
\begin{tabular}{rr|rr|rr|rr|rr|rr|rr}
\hline
\hline
\noalign{\smallskip}
mass & $^{14}$N & mass & $^{14}$N & mass & $^{14}$N & mass & $^{14}$N & mass & $^{14}$N & mass & $^{14}$N & mass & $^{14}$N \\
(\Msun) & (\Msun) & (\Msun) & (\Msun) & (\Msun) & (\Msun) & (\Msun) & (\Msun) & (\Msun) & (\Msun) & (\Msun) & (\Msun) & (\Msun) & (\Msun) \\
\noalign{\smallskip}
\hline
\noalign{\smallskip}
 10.0 & $9.39\E{- 7}$ &  12.0 & $1.66\E{- 6}$ &  15.6 & $4.46\E{- 6}$ &  18.1 & $2.18\E{- 6}$ &  22.0 & $3.19\E{- 5}$ &  31.0 & $8.89\E{- 7}$ &  45.0 & $3.09\E{- 2}$ \\
 10.2 & $4.06\E{- 7}$ &  12.2 & $2.94\E{- 6}$ &  15.8 & $2.21\E{- 6}$ &  18.2 & $2.53\E{- 6}$ &  22.5 & $1.46\E{- 5}$ &  31.5 & $1.42\E{- 6}$ &  50.0 & $1.05\E{- 1}$ \\
 10.4 & $2.07\E{- 6}$ &  12.4 & $3.39\E{- 6}$ &  16.0 & $2.04\E{- 6}$ &  18.3 & $2.35\E{- 6}$ &  23.0 & $1.34\E{- 4}$ &  32.0 & $1.06\E{- 6}$ &  55.0 & $1.21\E{- 1}$ \\
 10.5 & $1.17\E{- 6}$ &  12.6 & $2.79\E{- 6}$ &  16.2 & $2.01\E{- 6}$ &  18.4 & $2.54\E{- 6}$ &  23.5 & $1.09\E{- 5}$ &  32.5 & $1.23\E{- 6}$ &  60.0 & $3.74\E{- 5}$ \\
 10.6 & $1.29\E{- 6}$ &  12.8 & $2.57\E{- 6}$ &  16.4 & $1.97\E{- 6}$ &  18.5 & $2.21\E{- 6}$ &  24.0 & $8.00\E{- 6}$ &  33.0 & $1.05\E{- 6}$ &  65.0 & $9.27\E{- 2}$ \\
 10.7 & $1.39\E{- 6}$ &  13.0 & $2.58\E{- 6}$ &  16.6 & $2.30\E{- 6}$ &  18.6 & $2.16\E{- 6}$ &  24.5 & $8.02\E{- 6}$ &  33.5 & $1.66\E{- 6}$ &  70.0 & $4.38\E{- 5}$ \\
\hline
\end{tabular}
\end{table}

\begin{table}
\centering
\renewcommand{\EE}{}
\renewcommand{\I}[2]{$^{#1}$&#2}
\caption{Postsupernova Yields in Solar Masses, 0.6\,B, Standard Mixing, $S=4$ Piston\lTab{0.6-yield-mix}}
\begin{tabular}{r@{}llllllllll}
\hline
\hline
\noalign{\smallskip}
\multicolumn{2}{c}{isotope} & $10\,\Msun$ & $12\,\Msun$ & $15\,\Msun$ & $20\,\Msun$ & $25\,\Msun$ & $35\,\Msun$ & $50\,\Msun$ & $75\,\Msun$ & $100\,\Msun$ \\
\noalign{\smallskip}
\hline
\noalign{\smallskip}
\I{  1}{ H} & $5.29\EE    $ & $6.23\EE    $ & $7.43\EE    $ & $9.03\EE    $ & $1.03\E{  1}$ & $1.13\E{  1}$ & $1.63\E{  1}$ & $2.10\E{  1}$ & $2.74\E{  1}$ \\
\I{  2}{ H} & $2.22\E{- 7}$ & $1.25\E{- 6}$ & $1.04\E{- 6}$ & $9.26\E{- 7}$ & $8.71\E{- 7}$ & $7.60\E{- 7}$ & $3.10\E{- 8}$ & $3.40\E{- 8}$ & $4.91\E{- 9}$ \\
\I{  3}{He} & $5.83\E{- 5}$ & $6.18\E{- 5}$ & $6.58\E{- 5}$ & $7.22\E{- 5}$ & $6.98\E{- 5}$ & $8.07\E{- 5}$ & $9.13\E{- 5}$ & $1.15\E{- 4}$ & $1.53\E{- 4}$ \\
\I{  4}{He} & $3.25\EE    $ & $3.82\EE    $ & $4.34\EE    $ & $5.44\EE    $ & $6.50\EE    $ & $5.65\EE    $ & $1.52\E{  1}$ & $2.35\E{  1}$ & $2.81\E{  1}$ \\
\I{  6}{Li} & $1.63\E{-12}$ & $6.26\E{-12}$ & $1.72\E{-12}$ & $4.17\E{-13}$ & $1.71\E{-12}$ & $5.98\E{-14}$ & $7.49\E{-12}$ & $6.82\E{-13}$ & $1.36\E{-12}$ \\
\I{  7}{Li} & $8.24\E{- 8}$ & $1.66\E{- 7}$ & $4.14\E{- 7}$ & $5.29\E{- 7}$ & $1.93\E{- 7}$ & $2.93\E{- 7}$ & $4.94\E{- 8}$ & $5.22\E{- 8}$ & $2.17\E{- 9}$ \\
\I{  9}{Be} & $7.87\E{-12}$ & $1.05\E{-11}$ & $1.32\E{-12}$ & $9.56\E{-13}$ & $7.03\E{-12}$ & $1.49\E{-14}$ & $2.05\E{-11}$ & $3.22\E{-12}$ & $7.65\E{-12}$ \\
\hline
\end{tabular}
\end{table}
\addtocounter{table}{-1}
\begin{table}
\centering
\renewcommand{\EE}{}
\renewcommand{\I}[2]{$^{#1}$&#2}
\caption{Postsupernova Yields in Solar Masses, 0.6\,B, Standard Mixing, $S=4$ Piston (continued)}
\begin{tabular}{r@{}llllllllll}
\hline
\hline
\noalign{\smallskip}
\multicolumn{2}{c}{isotope} & $10\,\Msun$ & $12\,\Msun$ & $15\,\Msun$ & $20\,\Msun$ & $25\,\Msun$ & $35\,\Msun$ & $50\,\Msun$ & $75\,\Msun$ & $100\,\Msun$ \\
\noalign{\smallskip}
\hline
\noalign{\smallskip}
\I{ 59}{Co} & $1.17\E{- 4}$ & $4.94\E{- 5}$ & $1.17\E{- 5}$ & $4.82\E{- 6}$ & $1.80\E{- 6}$ & $3.54\E{- 8}$ & $8.01\E{- 6}$ & $2.16\E{- 6}$ & $2.82\E{- 6}$ \\
\I{ 58}{Ni} & $7.97\E{- 4}$ & $2.51\E{- 4}$ & $5.52\E{- 5}$ & $5.69\E{- 5}$ & $1.09\E{- 5}$ & $1.16\E{- 7}$ & $1.08\E{- 4}$ & $1.15\E{- 4}$ & $6.65\E{- 5}$ \\
\I{ 60}{Ni} & $1.08\E{- 3}$ & $4.68\E{- 4}$ & $1.36\E{- 4}$ & $4.21\E{- 5}$ & $3.99\E{- 5}$ & $9.68\E{- 7}$ & $1.42\E{- 4}$ & $8.23\E{- 5}$ & $2.24\E{- 5}$ \\
\I{ 61}{Ni} & $5.37\E{- 5}$ & $2.18\E{- 5}$ & $6.41\E{- 6}$ & $2.42\E{- 6}$ & $1.87\E{- 6}$ & $3.81\E{- 8}$ & $7.41\E{- 6}$ & $5.95\E{- 6}$ & $1.50\E{- 6}$ \\
\I{ 62}{Ni} & $2.58\E{- 4}$ & $9.42\E{- 5}$ & $2.28\E{- 5}$ & $1.23\E{- 5}$ & $4.90\E{- 6}$ & $8.59\E{- 8}$ & $2.69\E{- 5}$ & $2.63\E{- 5}$ & $1.08\E{- 5}$ \\
\I{ 64}{Ni} & $1.20\E{-12}$ & $1.67\E{-12}$ & $5.90\E{-13}$ & $2.19\E{-13}$ & $7.55\E{-14}$ & $2.50\E{-15}$ & $1.98\E{-11}$ & $1.78\E{-12}$ & $8.20\E{-14}$ \\
\I{ 63}{Cu} & $6.50\E{- 6}$ & $4.36\E{- 6}$ & $9.78\E{- 7}$ & $1.45\E{- 7}$ & $1.03\E{- 7}$ & $3.59\E{- 9}$ & $2.50\E{- 7}$ & $5.83\E{- 8}$ & $8.01\E{- 8}$ \\
\hline
\end{tabular}
\end{table}

\begin{table}
\centering
\renewcommand{\EE}{}
\renewcommand{\I}[2]{$^{#1}$&#2}
\caption{Postsupernova Yields in Solar Masses, 1.2\,B, Standard Mixing, $S=4$ Piston\lTab{1.2-yield-mix}}
\begin{tabular}{r@{}llllllllll}
\hline
\hline
\noalign{\smallskip}
\multicolumn{2}{c}{isotope} & $10\,\Msun$ & $12\,\Msun$ & $15\,\Msun$ & $20\,\Msun$ & $25\,\Msun$ & $35\,\Msun$ & $50\,\Msun$ & $75\,\Msun$ & $100\,\Msun$ \\
\noalign{\smallskip}
\hline
\noalign{\smallskip}
\I{  1}{ H} & $5.29\EE    $ & $6.23\EE    $ & $7.44\EE    $ & $9.11\EE    $ & $1.08\E{  1}$ & $1.30\E{  1}$ & $1.63\E{  1}$ & $2.13\E{  1}$ & $2.74\E{  1}$ \\
\I{  2}{ H} & $1.85\E{- 7}$ & $1.21\E{- 6}$ & $1.05\E{- 6}$ & $9.24\E{- 7}$ & $8.69\E{- 7}$ & $7.60\E{- 7}$ & $2.81\E{- 8}$ & $3.75\E{- 8}$ & $4.91\E{- 9}$ \\
\I{  3}{He} & $5.84\E{- 5}$ & $6.30\E{- 5}$ & $6.56\E{- 5}$ & $7.28\E{- 5}$ & $7.33\E{- 5}$ & $8.59\E{- 5}$ & $9.16\E{- 5}$ & $1.17\E{- 4}$ & $1.53\E{- 4}$ \\
\I{  4}{He} & $3.27\EE    $ & $3.91\EE    $ & $4.74\EE    $ & $6.49\EE    $ & $7.54\EE    $ & $8.48\EE    $ & $1.53\E{  1}$ & $2.41\E{  1}$ & $2.81\E{  1}$ \\
\I{  6}{Li} & $1.59\E{-12}$ & $7.40\E{-12}$ & $3.70\E{-12}$ & $3.78\E{-12}$ & $1.43\E{-11}$ & $5.29\E{-13}$ & $8.95\E{-12}$ & $6.92\E{-13}$ & $2.23\E{-12}$ \\
\I{  7}{Li} & $7.42\E{- 8}$ & $3.34\E{- 7}$ & $1.00\E{- 6}$ & $1.47\E{- 6}$ & $7.95\E{- 7}$ & $3.54\E{- 6}$ & $6.18\E{- 8}$ & $5.53\E{- 8}$ & $3.43\E{- 9}$ \\
\I{  9}{Be} & $6.30\E{-12}$ & $1.29\E{-11}$ & $4.18\E{-12}$ & $8.66\E{-12}$ & $3.54\E{-11}$ & $2.81\E{-13}$ & $2.29\E{-11}$ & $3.46\E{-12}$ & $1.20\E{-11}$ \\
\hline
\end{tabular}
\end{table}
\addtocounter{table}{-1}
\begin{table}
\centering
\renewcommand{\EE}{}
\renewcommand{\I}[2]{$^{#1}$&#2}
\caption{Postsupernova Yields in Solar Masses, 1.2\,B, Standard Mixing, $S=4$ Piston (continued)}
\begin{tabular}{r@{}llllllllll}
\hline
\hline
\noalign{\smallskip}
\multicolumn{2}{c}{isotope} & $10\,\Msun$ & $12\,\Msun$ & $15\,\Msun$ & $20\,\Msun$ & $25\,\Msun$ & $35\,\Msun$ & $50\,\Msun$ & $75\,\Msun$ & $100\,\Msun$ \\
\noalign{\smallskip}
\hline
\noalign{\smallskip}
\I{ 59}{Co} & $2.15\E{- 4}$ & $1.65\E{- 4}$ & $1.66\E{- 4}$ & $1.74\E{- 4}$ & $6.27\E{- 5}$ & $1.01\E{- 6}$ & $1.58\E{- 5}$ & $4.40\E{- 6}$ & $6.12\E{- 6}$ \\
\I{ 58}{Ni} & $7.58\E{- 4}$ & $4.87\E{- 4}$ & $3.87\E{- 4}$ & $1.95\E{- 3}$ & $2.10\E{- 4}$ & $1.93\E{- 6}$ & $1.29\E{- 4}$ & $1.65\E{- 4}$ & $1.28\E{- 4}$ \\
\I{ 60}{Ni} & $1.52\E{- 3}$ & $1.32\E{- 3}$ & $1.64\E{- 3}$ & $1.83\E{- 3}$ & $1.48\E{- 3}$ & $2.61\E{- 5}$ & $4.20\E{- 4}$ & $1.87\E{- 4}$ & $4.21\E{- 5}$ \\
\I{ 61}{Ni} & $6.07\E{- 5}$ & $5.21\E{- 5}$ & $6.30\E{- 5}$ & $9.64\E{- 5}$ & $5.98\E{- 5}$ & $8.48\E{- 7}$ & $1.88\E{- 5}$ & $1.27\E{- 5}$ & $2.52\E{- 6}$ \\
\I{ 62}{Ni} & $3.17\E{- 4}$ & $2.43\E{- 4}$ & $2.38\E{- 4}$ & $4.57\E{- 4}$ & $1.41\E{- 4}$ & $1.63\E{- 6}$ & $5.35\E{- 5}$ & $3.87\E{- 5}$ & $1.91\E{- 5}$ \\
\I{ 64}{Ni} & $5.97\E{-13}$ & $2.40\E{-12}$ & $4.18\E{-12}$ & $5.60\E{-12}$ & $3.34\E{-12}$ & $7.80\E{-14}$ & $2.75\E{-11}$ & $2.56\E{-12}$ & $1.75\E{-13}$ \\
\I{ 63}{Cu} & $2.05\E{- 5}$ & $1.47\E{- 5}$ & $1.52\E{- 5}$ & $6.21\E{- 6}$ & $5.65\E{- 6}$ & $1.67\E{- 7}$ & $1.23\E{- 6}$ & $9.95\E{- 8}$ & $2.14\E{- 7}$ \\
\hline
\end{tabular}
\end{table}

\begin{table}
\centering
\renewcommand{\EE}{}
\renewcommand{\I}[2]{$^{#1}$&#2}
\caption{Postsupernova Yields in Solar Masses, 2.4\,B, Standard Mixing, $S=4$ Piston\lTab{2.4-yield-mix}}
\begin{tabular}{r@{}llllllllll}
\hline
\hline
\noalign{\smallskip}
\multicolumn{2}{c}{isotope} & $10\,\Msun$ & $12\,\Msun$ & $15\,\Msun$ & $20\,\Msun$ & $25\,\Msun$ & $35\,\Msun$ & $50\,\Msun$ & $75\,\Msun$ & $100\,\Msun$ \\
\noalign{\smallskip}
\hline
\noalign{\smallskip}
\I{  1}{ H} & $5.29\EE    $ & $6.23\EE    $ & $7.44\EE    $ & $9.11\EE    $ & $1.08\E{  1}$ & $1.34\E{  1}$ & $1.63\E{  1}$ & $2.15\E{  1}$ & $2.74\E{  1}$ \\
\I{  2}{ H} & $1.53\E{- 7}$ & $1.05\E{- 6}$ & $1.04\E{- 6}$ & $9.22\E{- 7}$ & $8.72\E{- 7}$ & $7.61\E{- 7}$ & $2.14\E{- 8}$ & $3.96\E{- 8}$ & $4.91\E{- 9}$ \\
\I{  3}{He} & $5.84\E{- 5}$ & $6.44\E{- 5}$ & $6.51\E{- 5}$ & $7.24\E{- 5}$ & $7.28\E{- 5}$ & $8.74\E{- 5}$ & $9.16\E{- 5}$ & $1.18\E{- 4}$ & $1.53\E{- 4}$ \\
\I{  4}{He} & $3.28\EE    $ & $3.92\EE    $ & $4.76\EE    $ & $6.52\EE    $ & $7.57\EE    $ & $1.04\E{  1}$ & $1.54\E{  1}$ & $2.46\E{  1}$ & $2.81\E{  1}$ \\
\I{  6}{Li} & $1.32\E{-12}$ & $3.57\E{-12}$ & $3.49\E{-12}$ & $4.74\E{-12}$ & $1.85\E{-11}$ & $2.53\E{-12}$ & $9.67\E{-12}$ & $7.33\E{-13}$ & $6.95\E{-12}$ \\
\I{  7}{Li} & $6.57\E{- 8}$ & $6.76\E{- 7}$ & $2.10\E{- 6}$ & $2.55\E{- 6}$ & $1.37\E{- 6}$ & $9.49\E{- 6}$ & $6.67\E{- 8}$ & $6.10\E{- 8}$ & $8.89\E{- 9}$ \\
\I{  9}{Be} & $5.37\E{-12}$ & $5.39\E{-12}$ & $4.28\E{-12}$ & $1.01\E{-11}$ & $4.27\E{-11}$ & $7.43\E{-12}$ & $3.43\E{-11}$ & $4.61\E{-12}$ & $2.74\E{-11}$ \\
\hline
\end{tabular}
\end{table}
\addtocounter{table}{-1}
\begin{table}
\centering
\renewcommand{\EE}{}
\renewcommand{\I}[2]{$^{#1}$&#2}
\caption{Postsupernova Yields in Solar Masses, 2.4\,B, Standard Mixing, $S=4$ Piston (continued)}
\begin{tabular}{r@{}llllllllll}
\hline
\hline
\noalign{\smallskip}
\multicolumn{2}{c}{isotope} & $10\,\Msun$ & $12\,\Msun$ & $15\,\Msun$ & $20\,\Msun$ & $25\,\Msun$ & $35\,\Msun$ & $50\,\Msun$ & $75\,\Msun$ & $100\,\Msun$ \\
\noalign{\smallskip}
\hline
\noalign{\smallskip}
\I{ 59}{Co} & $3.83\E{- 4}$ & $2.73\E{- 4}$ & $3.55\E{- 4}$ & $5.19\E{- 4}$ & $3.34\E{- 4}$ & $3.82\E{- 4}$ & $1.35\E{- 4}$ & $2.49\E{- 5}$ & $5.07\E{- 5}$ \\
\I{ 58}{Ni} & $7.57\E{- 4}$ & $5.08\E{- 4}$ & $4.58\E{- 4}$ & $4.46\E{- 3}$ & $5.96\E{- 4}$ & $7.89\E{- 4}$ & $4.68\E{- 4}$ & $7.02\E{- 4}$ & $8.44\E{- 4}$ \\
\I{ 60}{Ni} & $1.99\E{- 3}$ & $1.90\E{- 3}$ & $3.10\E{- 3}$ & $5.89\E{- 3}$ & $7.05\E{- 3}$ & $7.68\E{- 3}$ & $4.14\E{- 3}$ & $9.50\E{- 4}$ & $3.02\E{- 4}$ \\
\I{ 61}{Ni} & $5.51\E{- 5}$ & $6.05\E{- 5}$ & $1.08\E{- 4}$ & $2.88\E{- 4}$ & $2.37\E{- 4}$ & $2.06\E{- 4}$ & $1.44\E{- 4}$ & $5.86\E{- 5}$ & $1.71\E{- 5}$ \\
\I{ 62}{Ni} & $3.42\E{- 4}$ & $3.17\E{- 4}$ & $3.86\E{- 4}$ & $1.28\E{- 3}$ & $5.04\E{- 4}$ & $3.12\E{- 4}$ & $3.01\E{- 4}$ & $1.58\E{- 4}$ & $1.32\E{- 4}$ \\
\I{ 64}{Ni} & $2.78\E{-13}$ & $1.57\E{-12}$ & $4.42\E{-12}$ & $1.10\E{-11}$ & $1.28\E{-11}$ & $2.11\E{-11}$ & $4.41\E{-11}$ & $5.53\E{-12}$ & $7.53\E{-13}$ \\
\I{ 63}{Cu} & $4.70\E{- 5}$ & $2.82\E{- 5}$ & $4.78\E{- 5}$ & $2.65\E{- 5}$ & $5.66\E{- 5}$ & $8.93\E{- 5}$ & $2.95\E{- 5}$ & $4.31\E{- 7}$ & $1.12\E{- 6}$ \\
\hline
\end{tabular}
\end{table}

\begin{table}
\centering
\renewcommand{\EE}{}
\renewcommand{\I}[2]{$^{#1}$&#2}
\caption{Postsupernova Yields in Solar Masses, 5.0\,B, Standard Mixing, $S=4$ Piston\lTab{5.0-yield-mix}}
\begin{tabular}{r@{}lllll|r@{}lllll}
\hline
\hline
\noalign{\smallskip}
\multicolumn{2}{c}{isotope} & $35\,\Msun$ & $50\,\Msun$ & $75\,\Msun$ & $100\,\Msun$ & \multicolumn{2}{c}{isotope} & $35\,\Msun$ & $50\,\Msun$ & $75\,\Msun$ & $100\,\Msun$ \\
\noalign{\smallskip}
\hline
\noalign{\smallskip}
\I{  1}{ H} & $1.34\E{  1}$ & $1.63\E{  1}$ & $2.15\E{  1}$ & $2.74\E{  1}$ & \I{ 42}{Ca} & $3.03\E{- 6}$ & $2.41\E{- 6}$ & $7.67\E{- 6}$ & $1.42\E{- 4}$ \\
\I{  2}{ H} & $7.62\E{- 7}$ & $1.72\E{- 8}$ & $3.39\E{- 8}$ & $4.91\E{- 9}$ & \I{ 43}{Ca} & $2.85\E{- 6}$ & $3.82\E{- 7}$ & $1.14\E{- 6}$ & $5.24\E{- 6}$ \\
\I{  3}{He} & $8.65\E{- 5}$ & $9.16\E{- 5}$ & $1.18\E{- 4}$ & $1.53\E{- 4}$ & \I{ 44}{Ca} & $1.24\E{- 4}$ & $1.53\E{- 4}$ & $1.85\E{- 4}$ & $2.09\E{- 4}$ \\
\I{  4}{He} & $1.04\E{  1}$ & $1.55\E{  1}$ & $2.46\E{  1}$ & $2.82\E{  1}$ & \I{ 46}{Ca} & $5.62\E{-13}$ & $5.33\E{-11}$ & $1.80\E{-11}$ & $1.87\E{-11}$ \\
\I{  6}{Li} & $1.88\E{-12}$ & $4.46\E{-12}$ & $6.26\E{-13}$ & $1.28\E{-11}$ & \I{ 48}{Ca} & $3.53\E{-15}$ & $6.57\E{-12}$ & $1.22\E{-13}$ & $4.68\E{-13}$ \\
\I{  7}{Li} & $1.18\E{- 5}$ & $5.06\E{- 8}$ & $5.99\E{- 8}$ & $1.41\E{- 8}$ & \I{ 45}{Sc} & $1.21\E{- 6}$ & $5.29\E{- 7}$ & $3.33\E{- 7}$ & $1.19\E{- 5}$ \\
\I{  9}{Be} & $7.52\E{-12}$ & $3.62\E{-11}$ & $4.90\E{-12}$ & $3.46\E{-11}$ & \I{ 46}{Ti} & $1.62\E{- 6}$ & $1.63\E{- 6}$ & $3.98\E{- 6}$ & $9.58\E{- 5}$ \\
\hline
\end{tabular}
\end{table}

\begin{table}
\centering
\renewcommand{\EE}{}
\renewcommand{\I}[2]{$^{#1}$&#2}
\caption{Postsupernova Yields in Solar Masses, 10.0\,B, Standard Mixing, $S=4$ Piston\lTab{10.0-yield-mix}}
\begin{tabular}{r@{}lllll|r@{}lllll}
\hline
\hline
\noalign{\smallskip}
\multicolumn{2}{c}{isotope} & $35\,\Msun$ & $50\,\Msun$ & $75\,\Msun$ & $100\,\Msun$ & \multicolumn{2}{c}{isotope} & $35\,\Msun$ & $50\,\Msun$ & $75\,\Msun$ & $100\,\Msun$ \\
\noalign{\smallskip}
\hline
\noalign{\smallskip}
\I{  1}{ H} & $1.34\E{  1}$ & $1.63\E{  1}$ & $2.15\E{  1}$ & $2.74\E{  1}$ & \I{ 42}{Ca} & $4.91\E{- 6}$ & $4.52\E{- 6}$ & $2.29\E{- 5}$ & $1.43\E{- 4}$ \\
\I{  2}{ H} & $7.62\E{- 7}$ & $1.57\E{- 8}$ & $2.88\E{- 8}$ & $4.91\E{- 9}$ & \I{ 43}{Ca} & $6.84\E{- 6}$ & $2.96\E{- 6}$ & $4.42\E{- 6}$ & $9.70\E{- 6}$ \\
\I{  3}{He} & $8.58\E{- 5}$ & $9.16\E{- 5}$ & $1.18\E{- 4}$ & $1.53\E{- 4}$ & \I{ 44}{Ca} & $1.86\E{- 4}$ & $2.09\E{- 4}$ & $4.68\E{- 4}$ & $2.53\E{- 4}$ \\
\I{  4}{He} & $1.05\E{  1}$ & $1.55\E{  1}$ & $2.47\E{  1}$ & $2.83\E{  1}$ & \I{ 46}{Ca} & $5.53\E{-13}$ & $5.33\E{-11}$ & $1.89\E{-11}$ & $1.46\E{-11}$ \\
\I{  6}{Li} & $1.57\E{-12}$ & $3.55\E{-12}$ & $5.27\E{-13}$ & $9.78\E{-12}$ & \I{ 48}{Ca} & $7.06\E{-15}$ & $6.57\E{-12}$ & $1.32\E{-13}$ & $3.85\E{-13}$ \\
\I{  7}{Li} & $1.34\E{- 5}$ & $3.64\E{- 8}$ & $6.23\E{- 8}$ & $1.29\E{- 8}$ & \I{ 45}{Sc} & $3.05\E{- 6}$ & $2.28\E{- 6}$ & $1.30\E{- 6}$ & $1.23\E{- 5}$ \\
\I{  9}{Be} & $7.10\E{-12}$ & $3.14\E{-11}$ & $6.12\E{-12}$ & $3.48\E{-11}$ & \I{ 46}{Ti} & $2.49\E{- 6}$ & $2.64\E{- 6}$ & $1.10\E{- 5}$ & $9.99\E{- 5}$ \\
\hline
\end{tabular}
\end{table}

\clearpage

\begin{table}
\centering
\renewcommand{\EE}{}
\renewcommand{\I}[2]{$^{#1}$&#2}
\caption{Postsupernova Yields in Solar Masses, 1.2\,B, No Mixing, $S=4$ Piston\lTab{1.2-yield-nomix}}
\begin{tabular}{r@{}llllllllll}
\hline
\hline
\noalign{\smallskip}
\multicolumn{2}{c}{isotope} & $10\,\Msun$ & $12\,\Msun$ & $15\,\Msun$ & $20\,\Msun$ & $25\,\Msun$ & $35\,\Msun$ & $50\,\Msun$ & $75\,\Msun$ & $100\,\Msun$ \\
\noalign{\smallskip}
\hline
\noalign{\smallskip}
\I{  1}{ H} & $5.29\EE    $ & $6.23\EE    $ & $7.44\EE    $ & $9.11\EE    $ & $1.08\E{  1}$ & $1.33\E{  1}$ & $1.63\E{  1}$ & $2.15\E{  1}$ & $2.74\E{  1}$ \\
\I{  2}{ H} & $1.85\E{- 7}$ & $1.21\E{- 6}$ & $1.05\E{- 6}$ & $9.24\E{- 7}$ & $8.69\E{- 7}$ & $7.60\E{- 7}$ & $2.81\E{- 8}$ & $4.55\E{- 8}$ & $4.91\E{- 9}$ \\
\I{  3}{He} & $5.84\E{- 5}$ & $6.30\E{- 5}$ & $6.56\E{- 5}$ & $7.28\E{- 5}$ & $7.33\E{- 5}$ & $8.80\E{- 5}$ & $9.16\E{- 5}$ & $1.18\E{- 4}$ & $1.53\E{- 4}$ \\
\I{  4}{He} & $3.27\EE    $ & $3.91\EE    $ & $4.73\EE    $ & $6.48\EE    $ & $7.54\EE    $ & $8.64\EE    $ & $1.54\E{  1}$ & $2.46\E{  1}$ & $2.81\E{  1}$ \\
\I{  6}{Li} & $1.59\E{-12}$ & $7.43\E{-12}$ & $4.01\E{-12}$ & $4.10\E{-12}$ & $1.57\E{-11}$ & $1.93\E{-15}$ & $1.71\E{-12}$ & $8.82\E{-13}$ & $8.20\E{-13}$ \\
\I{  7}{Li} & $7.32\E{- 8}$ & $3.26\E{- 7}$ & $9.92\E{- 7}$ & $1.47\E{- 6}$ & $7.96\E{- 7}$ & $3.36\E{- 6}$ & $6.48\E{- 8}$ & $6.40\E{- 8}$ & $2.32\E{- 9}$ \\
\I{  9}{Be} & $6.31\E{-12}$ & $1.30\E{-11}$ & $4.64\E{-12}$ & $9.29\E{-12}$ & $3.88\E{-11}$ & $6.73\E{-18}$ & $1.84\E{-11}$ & $4.72\E{-12}$ & $1.14\E{-11}$ \\
\hline
\end{tabular}
\end{table}
\addtocounter{table}{-1}
\begin{table}
\centering
\renewcommand{\EE}{}
\renewcommand{\I}[2]{$^{#1}$&#2}
\caption{Postsupernova Yields in Solar Masses, 1.2\,B, No Mixing, $S=4$ Piston (continued)}
\begin{tabular}{r@{}llllllllll}
\hline
\hline
\noalign{\smallskip}
\multicolumn{2}{c}{isotope} & $10\,\Msun$ & $12\,\Msun$ & $15\,\Msun$ & $20\,\Msun$ & $25\,\Msun$ & $35\,\Msun$ & $50\,\Msun$ & $75\,\Msun$ & $100\,\Msun$ \\
\noalign{\smallskip}
\hline
\noalign{\smallskip}
\I{ 59}{Co} & $2.09\E{- 4}$ & $1.37\E{- 4}$ & $9.92\E{-16}$ & $1.10\E{-15}$ & $1.51\E{-14}$ & $1.84\E{-35}$ & $1.25\E{-11}$ & $1.98\E{-12}$ & $3.93\E{-14}$ \\
\I{ 58}{Ni} & $7.55\E{- 4}$ & $4.71\E{- 4}$ & $4.55\E{-17}$ & $1.36\E{-19}$ & $1.10\E{-18}$ & $1.06\E{-43}$ & $3.18\E{-15}$ & $9.79\E{-15}$ & $2.54\E{-20}$ \\
\I{ 60}{Ni} & $1.50\E{- 3}$ & $1.23\E{- 3}$ & $1.58\E{-15}$ & $1.28\E{-15}$ & $1.53\E{-14}$ & $1.04\E{-38}$ & $1.62\E{-11}$ & $2.04\E{-12}$ & $3.20\E{-14}$ \\
\I{ 61}{Ni} & $6.04\E{- 5}$ & $5.10\E{- 5}$ & $5.19\E{-16}$ & $3.58\E{-16}$ & $2.78\E{-15}$ & $1.63\E{-41}$ & $6.16\E{-12}$ & $7.03\E{-13}$ & $9.28\E{-15}$ \\
\I{ 62}{Ni} & $3.17\E{- 4}$ & $2.43\E{- 4}$ & $5.84\E{-16}$ & $1.63\E{-16}$ & $4.16\E{-15}$ & $6.33\E{-45}$ & $1.25\E{-11}$ & $1.27\E{-12}$ & $1.41\E{-14}$ \\
\I{ 64}{Ni} & $5.55\E{-13}$ & $1.27\E{-12}$ & $2.30\E{-16}$ & $2.27\E{-17}$ & $9.40\E{-15}$ & $8.19\E{-54}$ & $3.19\E{-11}$ & $2.11\E{-12}$ & $9.93\E{-15}$ \\
\I{ 63}{Cu} & $1.87\E{- 5}$ & $8.62\E{- 6}$ & $2.56\E{-16}$ & $4.87\E{-17}$ & $4.48\E{-15}$ & $4.42\E{-50}$ & $1.19\E{-11}$ & $6.70\E{-13}$ & $4.38\E{-15}$ \\
\hline
\end{tabular}
\end{table}

\begin{table}
\centering
\renewcommand{\EE}{}
\renewcommand{\I}[2]{$^{#1}$&#2}
\caption{Postsupernova Yields in Solar Masses, 1.2\,B, Standard Mixing, \Ye Piston\lTab{1.2-yield-mix-Ye}}
\begin{tabular}{r@{}llllllllll}
\hline
\hline
\noalign{\smallskip}
\multicolumn{2}{c}{isotope} & $10\,\Msun$ & $12\,\Msun$ & $15\,\Msun$ & $20\,\Msun$ & $25\,\Msun$ & $35\,\Msun$ & $50\,\Msun$ & $75\,\Msun$ & $100\,\Msun$ \\
\noalign{\smallskip}
\hline
\noalign{\smallskip}
\I{  1}{ H} & $5.29\EE    $ & $6.23\EE    $ & $7.44\EE    $ & $9.11\EE    $ & $1.07\E{  1}$ & $1.29\E{  1}$ & $1.63\E{  1}$ & $2.11\E{  1}$ & $2.74\E{  1}$ \\
\I{  2}{ H} & $1.85\E{- 7}$ & $1.21\E{- 6}$ & $1.04\E{- 6}$ & $9.25\E{- 7}$ & $8.71\E{- 7}$ & $7.63\E{- 7}$ & $2.72\E{- 8}$ & $2.86\E{- 8}$ & $4.91\E{- 9}$ \\
\I{  3}{He} & $5.84\E{- 5}$ & $6.27\E{- 5}$ & $6.56\E{- 5}$ & $7.31\E{- 5}$ & $7.37\E{- 5}$ & $8.55\E{- 5}$ & $9.16\E{- 5}$ & $1.15\E{- 4}$ & $1.53\E{- 4}$ \\
\I{  4}{He} & $3.27\EE    $ & $3.90\EE    $ & $4.71\EE    $ & $6.49\EE    $ & $7.30\EE    $ & $8.31\EE    $ & $1.53\E{  1}$ & $2.36\E{  1}$ & $2.81\E{  1}$ \\
\I{  6}{Li} & $1.59\E{-12}$ & $4.62\E{-12}$ & $2.56\E{-12}$ & $1.92\E{-12}$ & $4.19\E{-12}$ & $4.22\E{-13}$ & $8.21\E{-12}$ & $5.64\E{-13}$ & $2.18\E{-12}$ \\
\I{  7}{Li} & $7.40\E{- 8}$ & $3.20\E{- 7}$ & $1.01\E{- 6}$ & $1.49\E{- 6}$ & $7.53\E{- 7}$ & $3.23\E{- 6}$ & $5.93\E{- 8}$ & $4.94\E{- 8}$ & $3.37\E{- 9}$ \\
\I{  9}{Be} & $6.30\E{-12}$ & $5.93\E{-12}$ & $1.68\E{-12}$ & $4.13\E{-12}$ & $1.09\E{-11}$ & $1.78\E{-13}$ & $2.28\E{-11}$ & $2.75\E{-12}$ & $1.17\E{-11}$ \\
\hline
\end{tabular}
\end{table}
\addtocounter{table}{-1}
\begin{table}
\centering
\renewcommand{\EE}{}
\renewcommand{\I}[2]{$^{#1}$&#2}
\caption{Postsupernova Yields in Solar Masses, 1.2\,B, Standard Mixing, \Ye Piston (continued)}
\begin{tabular}{r@{}llllllllll}
\hline
\hline
\noalign{\smallskip}
\multicolumn{2}{c}{isotope} & $10\,\Msun$ & $12\,\Msun$ & $15\,\Msun$ & $20\,\Msun$ & $25\,\Msun$ & $35\,\Msun$ & $50\,\Msun$ & $75\,\Msun$ & $100\,\Msun$ \\
\noalign{\smallskip}
\hline
\noalign{\smallskip}
\I{ 59}{Co} & $2.14\E{- 4}$ & $1.21\E{- 4}$ & $2.57\E{- 4}$ & $1.64\E{- 4}$ & $5.44\E{- 5}$ & $3.90\E{- 6}$ & $5.09\E{- 5}$ & $2.55\E{- 6}$ & $8.00\E{- 6}$ \\
\I{ 58}{Ni} & $7.63\E{- 4}$ & $6.18\E{- 4}$ & $7.60\E{- 3}$ & $1.92\E{- 3}$ & $1.27\E{- 3}$ & $7.88\E{- 5}$ & $8.17\E{- 4}$ & $1.05\E{- 4}$ & $2.50\E{- 4}$ \\
\I{ 60}{Ni} & $1.52\E{- 3}$ & $1.26\E{- 3}$ & $1.49\E{- 3}$ & $1.72\E{- 3}$ & $4.19\E{- 4}$ & $2.93\E{- 5}$ & $4.78\E{- 4}$ & $1.04\E{- 4}$ & $5.97\E{- 5}$ \\
\I{ 61}{Ni} & $6.04\E{- 5}$ & $6.09\E{- 5}$ & $1.31\E{- 4}$ & $9.21\E{- 5}$ & $2.97\E{- 5}$ & $1.88\E{- 6}$ & $2.73\E{- 5}$ & $6.99\E{- 6}$ & $4.06\E{- 6}$ \\
\I{ 62}{Ni} & $3.16\E{- 4}$ & $2.79\E{- 4}$ & $1.43\E{- 3}$ & $4.45\E{- 4}$ & $2.52\E{- 4}$ & $1.54\E{- 5}$ & $1.68\E{- 4}$ & $2.38\E{- 5}$ & $3.73\E{- 5}$ \\
\I{ 64}{Ni} & $5.88\E{-13}$ & $1.43\E{-12}$ & $3.96\E{-11}$ & $4.78\E{-12}$ & $5.36\E{-12}$ & $2.92\E{-13}$ & $2.84\E{-11}$ & $1.86\E{-12}$ & $1.99\E{-13}$ \\
\I{ 63}{Cu} & $2.04\E{- 5}$ & $7.18\E{- 6}$ & $8.16\E{- 6}$ & $4.33\E{- 6}$ & $1.73\E{- 6}$ & $1.22\E{- 7}$ & $1.38\E{- 6}$ & $6.04\E{- 8}$ & $1.50\E{- 7}$ \\
\hline
\end{tabular}
\end{table}

\begin{table}
\centering
\renewcommand{\EE}{}
\renewcommand{\I}[2]{$^{#1}$&#2}
\caption{Postsupernova Yields in Solar Masses, 1.2\,B, No Mixing, \Ye Piston\lTab{1.2-yield-nomix-Ye}}
\begin{tabular}{r@{}llllllllll}
\hline
\hline
\noalign{\smallskip}
\multicolumn{2}{c}{isotope} & $10\,\Msun$ & $12\,\Msun$ & $15\,\Msun$ & $20\,\Msun$ & $25\,\Msun$ & $35\,\Msun$ & $50\,\Msun$ & $75\,\Msun$ & $100\,\Msun$ \\
\noalign{\smallskip}
\hline
\noalign{\smallskip}
\I{  1}{ H} & $5.29\EE    $ & $6.23\EE    $ & $7.44\EE    $ & $9.11\EE    $ & $1.08\E{  1}$ & $1.32\E{  1}$ & $1.63\E{  1}$ & $2.15\E{  1}$ & $2.74\E{  1}$ \\
\I{  2}{ H} & $1.85\E{- 7}$ & $1.21\E{- 6}$ & $1.04\E{- 6}$ & $9.25\E{- 7}$ & $8.71\E{- 7}$ & $7.63\E{- 7}$ & $2.73\E{- 8}$ & $4.31\E{- 8}$ & $4.91\E{- 9}$ \\
\I{  3}{He} & $5.84\E{- 5}$ & $6.27\E{- 5}$ & $6.56\E{- 5}$ & $7.31\E{- 5}$ & $7.41\E{- 5}$ & $8.74\E{- 5}$ & $9.16\E{- 5}$ & $1.18\E{- 4}$ & $1.53\E{- 4}$ \\
\I{  4}{He} & $3.27\EE    $ & $3.90\EE    $ & $4.73\EE    $ & $6.48\EE    $ & $7.54\EE    $ & $8.43\EE    $ & $1.54\E{  1}$ & $2.46\E{  1}$ & $2.81\E{  1}$ \\
\I{  6}{Li} & $1.59\E{-12}$ & $4.77\E{-12}$ & $2.68\E{-12}$ & $2.07\E{-12}$ & $4.05\E{-12}$ & $1.94\E{-15}$ & $1.71\E{-12}$ & $8.64\E{-13}$ & $8.19\E{-13}$ \\
\I{  7}{Li} & $7.30\E{- 8}$ & $3.14\E{- 7}$ & $1.01\E{- 6}$ & $1.49\E{- 6}$ & $8.31\E{- 7}$ & $1.44\E{- 6}$ & $6.37\E{- 8}$ & $6.33\E{- 8}$ & $2.27\E{- 9}$ \\
\I{  9}{Be} & $6.31\E{-12}$ & $6.25\E{-12}$ & $1.91\E{-12}$ & $4.43\E{-12}$ & $1.10\E{-11}$ & $6.75\E{-18}$ & $1.82\E{-11}$ & $3.59\E{-12}$ & $1.12\E{-11}$ \\
\hline
\end{tabular}
\end{table}
\addtocounter{table}{-1}
\begin{table}
\centering
\renewcommand{\EE}{}
\renewcommand{\I}[2]{$^{#1}$&#2}
\caption{Postsupernova Yields in Solar Masses, 1.2\,B, No Mixing, \Ye Piston (continued)}
\begin{tabular}{r@{}llllllllll}
\hline
\hline
\noalign{\smallskip}
\multicolumn{2}{c}{isotope} & $10\,\Msun$ & $12\,\Msun$ & $15\,\Msun$ & $20\,\Msun$ & $25\,\Msun$ & $35\,\Msun$ & $50\,\Msun$ & $75\,\Msun$ & $100\,\Msun$ \\
\noalign{\smallskip}
\hline
\noalign{\smallskip}
\I{ 59}{Co} & $2.08\E{- 4}$ & $8.22\E{- 5}$ & $9.81\E{-16}$ & $1.06\E{-15}$ & $1.52\E{-14}$ & $9.15\E{-40}$ & $1.19\E{-11}$ & $9.57\E{-13}$ & $3.89\E{-14}$ \\
\I{ 58}{Ni} & $7.59\E{- 4}$ & $4.10\E{- 4}$ & $3.27\E{-24}$ & $1.27\E{-19}$ & $2.28\E{-20}$ & $5.17\E{-46}$ & $5.08\E{-17}$ & $3.28\E{-18}$ & $2.46\E{-20}$ \\
\I{ 60}{Ni} & $1.50\E{- 3}$ & $9.34\E{- 4}$ & $1.58\E{-15}$ & $1.24\E{-15}$ & $1.53\E{-14}$ & $7.53\E{-44}$ & $1.58\E{-11}$ & $9.50\E{-13}$ & $3.16\E{-14}$ \\
\I{ 61}{Ni} & $6.01\E{- 5}$ & $4.47\E{- 5}$ & $5.20\E{-16}$ & $3.45\E{-16}$ & $2.79\E{-15}$ & $3.50\E{-48}$ & $5.95\E{-12}$ & $2.67\E{-13}$ & $9.19\E{-15}$ \\
\I{ 62}{Ni} & $3.16\E{- 4}$ & $1.80\E{- 4}$ & $5.84\E{-16}$ & $1.57\E{-16}$ & $4.15\E{-15}$ & $1.54\E{-52}$ & $1.21\E{-11}$ & $5.20\E{-13}$ & $1.39\E{-14}$ \\
\I{ 64}{Ni} & $5.48\E{-13}$ & $6.75\E{-13}$ & $2.31\E{-16}$ & $2.19\E{-17}$ & $9.41\E{-15}$ & $4.60\E{-61}$ & $3.08\E{-11}$ & $9.76\E{-13}$ & $9.83\E{-15}$ \\
\I{ 63}{Cu} & $1.86\E{- 5}$ & $2.66\E{- 6}$ & $2.58\E{-16}$ & $4.69\E{-17}$ & $4.48\E{-15}$ & $7.89\E{-57}$ & $1.15\E{-11}$ & $3.70\E{-13}$ & $4.34\E{-15}$ \\
\hline
\end{tabular}
\end{table}

\begin{table}
\renewcommand{\EE}{}
\centering
\caption{$^{56}$Ni Yields (Solar Masses), $S=4$ Piston\lTab{ni56-S4}}
\begin{tabular}{rllllllllll}
\hline
\hline
\noalign{\smallskip}
mass & 0.3\,\B & 0.3\,\B & 0.6\,\B & 0.6\,\B & 1.2\,\B & 1.2\,\B & 2.4\,\B & 2.4\,\B & 5\,\B & 10\,\B \\
(\Msun) & (no mix) & (mixed) & (no mix) & (mixed) & (no mix) & (mixed) & (no mix) & (mixed) & (mixed) & (mixed) \\
\noalign{\smallskip}
\hline
\noalign{\smallskip}
 10.0 & $1.98\E{-10}$ & $2.29\E{- 2}$ & $3.04\E{- 2}$ & $3.39\E{- 2}$ & $3.69\E{- 2}$ & $3.72\E{- 2}$ & $3.47\E{- 2}$ & $3.49\E{- 2}$ & $2.98\E{- 2}$ & $1.09\E{- 2}$ \\
 10.2 & $1.76\E{-18}$ & $5.08\E{- 3}$ & $1.46\E{- 3}$ & $1.71\E{- 3}$ & $1.04\E{- 3}$ & $1.09\E{- 3}$ & $1.80\E{- 3}$ & $1.80\E{- 3}$ & $4.02\E{- 3}$ & $3.68\E{- 3}$ \\
 10.4 & $1.18\E{-20}$ & $1.58\E{- 2}$ & $3.04\E{- 2}$ & $3.45\E{- 2}$ & $4.30\E{- 2}$ & $4.36\E{- 2}$ & $5.00\E{- 2}$ & $5.02\E{- 2}$ & $5.15\E{- 2}$ & $4.41\E{- 2}$ \\
 10.5 & $9.08\E{-29}$ & $8.77\E{- 3}$ & $2.27\E{- 3}$ & $1.18\E{- 2}$ & $8.23\E{- 3}$ & $8.40\E{- 3}$ & $1.13\E{- 2}$ & $1.13\E{- 2}$ & $1.06\E{- 2}$ & $1.04\E{- 2}$ \\
 10.6 & $8.54\E{-24}$ & $4.86\E{- 3}$ & $3.30\E{- 3}$ & $6.01\E{- 3}$ & $4.10\E{- 3}$ & $4.24\E{- 3}$ & $4.26\E{- 3}$ & $4.26\E{- 3}$ & $7.39\E{- 3}$ & $7.96\E{- 3}$ \\
 10.7 & $4.58\E{-29}$ & $8.61\E{- 3}$ & $1.66\E{- 3}$ & $1.18\E{- 2}$ & $7.90\E{- 3}$ & $8.42\E{- 3}$ & $1.11\E{- 2}$ & $1.12\E{- 2}$ & $1.08\E{- 2}$ & $1.09\E{- 2}$ \\
\hline
\end{tabular}
\end{table}
\addtocounter{table}{-1}
\begin{table}
\renewcommand{\EE}{}
\centering
\caption{$^{56}$N Yields (Solar Masses), $S=4$ Piston (continued)}
\begin{tabular}{rllllllllll}
\hline
\hline
\noalign{\smallskip}
mass & 0.3\,\B & 0.3\,\B & 0.6\,\B & 0.6\,\B & 1.2\,\B & 1.2\,\B & 2.4\,\B & 2.4\,\B & 5\,\B & 10\,\B \\
(\Msun) & (no mix) & (mixed) & (no mix) & (mixed) & (no mix) & (mixed) & (no mix) & (mixed) & (mixed) & (mixed) \\
\noalign{\smallskip}
\hline
\noalign{\smallskip}
 19.2 & $6.44\E{-44}$ & $5.78\E{- 5}$ & $1.14\E{-29}$ & $1.43\E{- 3}$ & $1.36\E{-26}$ & $9.93\E{- 3}$ & $1.43\E{- 1}$ & $1.44\E{- 1}$ & $1.94\E{- 1}$ & $2.48\E{- 1}$ \\
 19.4 & $7.89\E{-44}$ & $5.14\E{- 5}$ & $2.58\E{-30}$ & $1.46\E{- 3}$ & $1.19\E{-26}$ & $1.05\E{- 2}$ & $1.52\E{- 1}$ & $1.55\E{- 1}$ & $2.05\E{- 1}$ & $2.61\E{- 1}$ \\
 19.6 & $7.05\E{-44}$ & $5.38\E{- 5}$ & $2.47\E{-37}$ & $1.41\E{- 3}$ & $7.70\E{-27}$ & $9.10\E{- 3}$ & $1.54\E{- 1}$ & $1.56\E{- 1}$ & $2.12\E{- 1}$ & $2.69\E{- 1}$ \\
 19.8 & $3.23\E{-44}$ & $5.28\E{- 5}$ & $2.79\E{-38}$ & $1.27\E{- 3}$ & $4.27\E{-23}$ & $1.34\E{- 2}$ & $1.57\E{- 1}$ & $1.59\E{- 1}$ & $2.10\E{- 1}$ & $2.71\E{- 1}$ \\
 20.0 & $3.06\E{-45}$ & $5.65\E{- 5}$ & $2.17\E{-29}$ & $1.85\E{- 3}$ & $2.00\E{-21}$ & $7.56\E{- 2}$ & $2.15\E{- 1}$ & $2.19\E{- 1}$ & $2.85\E{- 1}$ & $3.06\E{- 1}$ \\
 20.5 & $1.21\E{-44}$ & $5.34\E{- 5}$ & $2.57\E{-28}$ & $1.40\E{- 3}$ & $2.41\E{-22}$ & $1.83\E{- 2}$ & $1.52\E{- 1}$ & $1.55\E{- 1}$ & $2.06\E{- 1}$ & $2.57\E{- 1}$ \\
\hline
\end{tabular}
\end{table}

\begin{table}
\renewcommand{\EE}{}
\centering
\caption{$^{56}$Ni Yields (Solar Masses), Piston at $\Ye$ core\lTab{ni56-Ye}}
\begin{tabular}{rll|rll|rll|rll}
\hline
\hline
\noalign{\smallskip}
mass & 1.2\,\B & 1.2\,\B & mass & 1.2\,\B & 1.2\,\B & mass & 1.2\,\B & 1.2\,\B & mass & 1.2\,\B & 1.2\,\B \\
(\Msun) & (no mix) & (mixed) & (\Msun) & (no mix) & (mixed) & (\Msun) & (no mix) & (mixed) & (\Msun) & (no mix) & (mixed) \\
\noalign{\smallskip}
\hline
\noalign{\smallskip}
 10.0 & $3.70\E{- 2}$ & $3.73\E{- 2}$ &  14.4 & $8.45\E{-27}$ & $5.05\E{- 2}$ &  18.7 & $9.71\E{-27}$ & $1.41\E{- 2}$ &  31.0 & $6.14\E{-28}$ & $2.04\E{- 3}$ \\
 10.2 & $1.16\E{- 1}$ & $1.16\E{- 1}$ &  14.6 & $1.61\E{-27}$ & $4.15\E{- 2}$ &  18.8 & $1.58\E{-26}$ & $2.02\E{- 2}$ &  31.5 & $1.14\E{-27}$ & $2.23\E{- 3}$ \\
 10.4 & $4.41\E{- 2}$ & $4.24\E{- 2}$ &  14.8 & $1.47\E{-27}$ & $4.08\E{- 2}$ &  18.9 & $2.30\E{-26}$ & $1.81\E{- 2}$ &  32.0 & $1.36\E{-39}$ & $1.72\E{- 3}$ \\
 10.5 & $2.16\E{- 2}$ & $2.15\E{- 2}$ &  15.0 & $1.51\E{-25}$ & $5.71\E{- 2}$ &  19.0 & $1.03\E{-21}$ & $2.64\E{- 2}$ &  32.5 & $7.43\E{-31}$ & $1.79\E{- 3}$ \\
 10.6 & $2.10\E{- 3}$ & $2.09\E{- 3}$ &  15.2 & $6.96\E{-26}$ & $6.19\E{- 2}$ &  19.2 & $1.11\E{-26}$ & $1.54\E{- 2}$ &  33.0 & $1.61\E{-40}$ & $1.56\E{- 3}$ \\
 10.7 & $1.90\E{- 2}$ & $1.89\E{- 2}$ &  15.4 & $3.86\E{-26}$ & $5.07\E{- 2}$ &  19.4 & $9.22\E{-27}$ & $1.39\E{- 2}$ &  33.5 & $2.74\E{-28}$ & $2.09\E{- 3}$ \\
\hline
\end{tabular}
\end{table}

\clearpage

\begin{table*}
\newcommand{\T}[1]{{\textbf{#1}}}
\caption{Summary of fits/figures \lTab{fit_table}}
\centering
\fontfamily{ptm}\selectfont
\begin{tabular}{lccrrrrrll}
\hline\noalign{\smallskip}
star or &
mass &
IMF &
$E$ &
\EExp &
mixing &
$\chi$ &
rank &
comment &
figure \\
compilation&
(\Msun) &
&
(B) &
&
&
&
(\#) &
&
\\
\noalign{\smallskip}
\hline
\noalign{\smallskip}
HE0107 &       20.5 &              -- & 0.6 &   -- & 0.0631 &  3.845 &  1 &  & \FIG{A}{HE0107} \\
HE0107 &         27 &              -- & 0.6 &   -- & 0.1000 &  4.297 &  2 &  & \\
HE0107 &       10.5 &              -- & 0.3 &   -- & 0.0063 &  1.856 &  1 & C+N & \FIG{B}{HE0107} \\
HE0107 &       12.8 &              -- & 0.6 &   -- & 0.0251 &  2.183 &  2 & C+N & \\
HE0107 &       18.1 &              -- & 0.3 &   -- & 0.0631 &  2.786 & 10 & C+N & \\
HE0107 &       10.5 &              -- & 0.3 &   -- & 0.0063 &  1.958 &  1 & C+N+O & \\
\end{tabular}
\end{table*}

\clearpage 
\begin{figure} 
\centering 
\includegraphics[angle=0,width=\columnwidth]{\figurepath 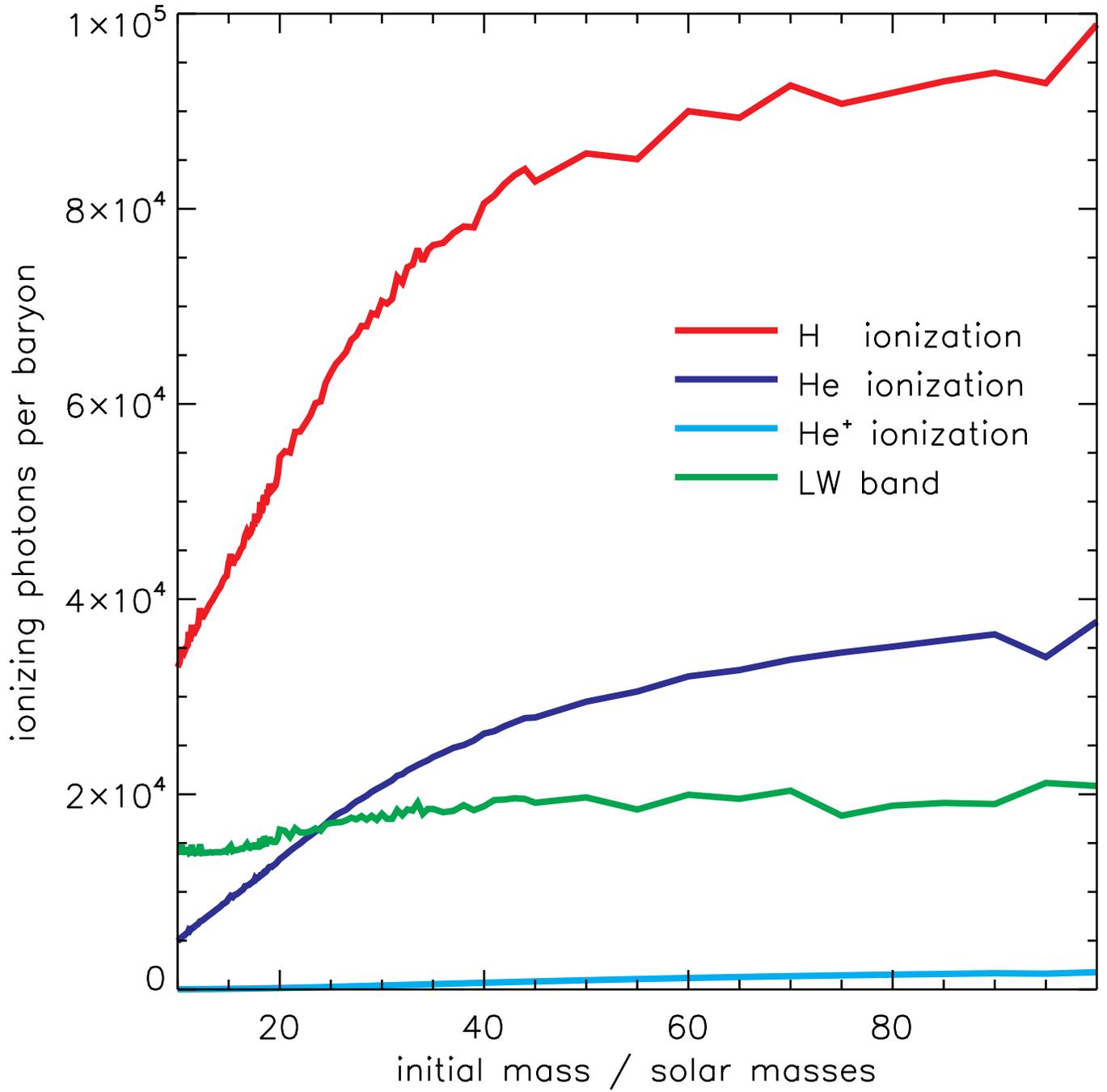}
\caption{Ionizing radiation as a function of mass.  Shown is the
number of ionizing photons per baryon as a function on initial mass,
for atomic hydrogen (\textsl{red}) and helium ionization (HeI to HeII,
\textsl{blue}; HeII to HeIII, \textsl{cyan}), and for H$_2$ dissociation
(Lyman-Werner band, \textsl{green}).  \lFig{ionize}}
\end{figure}

\clearpage 
\begin{figure*} 
\centering 
\includegraphics[angle=0,width=\textwidth]{\figurepath 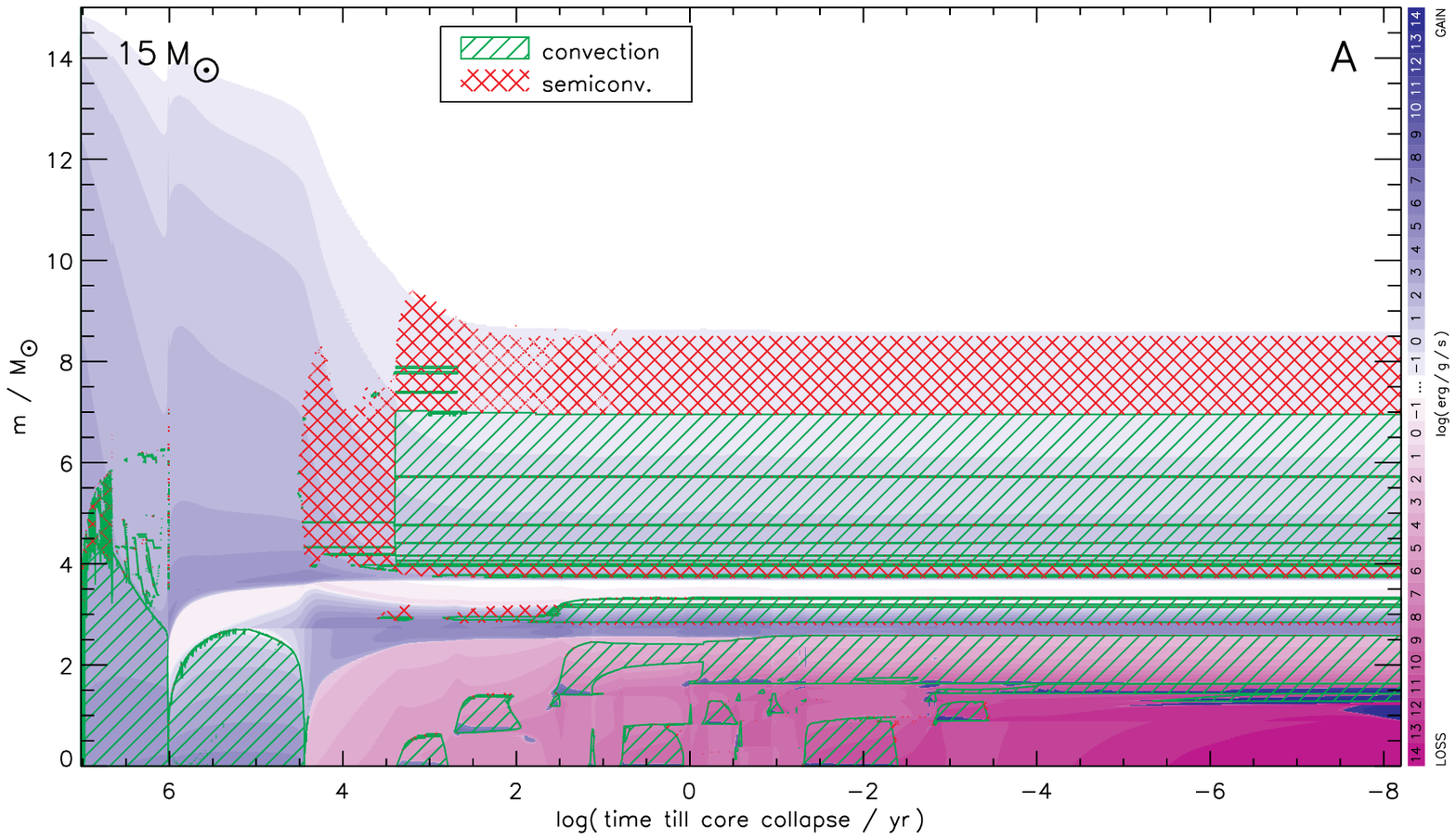}\\
\bigskip
\includegraphics[angle=0,width=\textwidth]{\figurepath 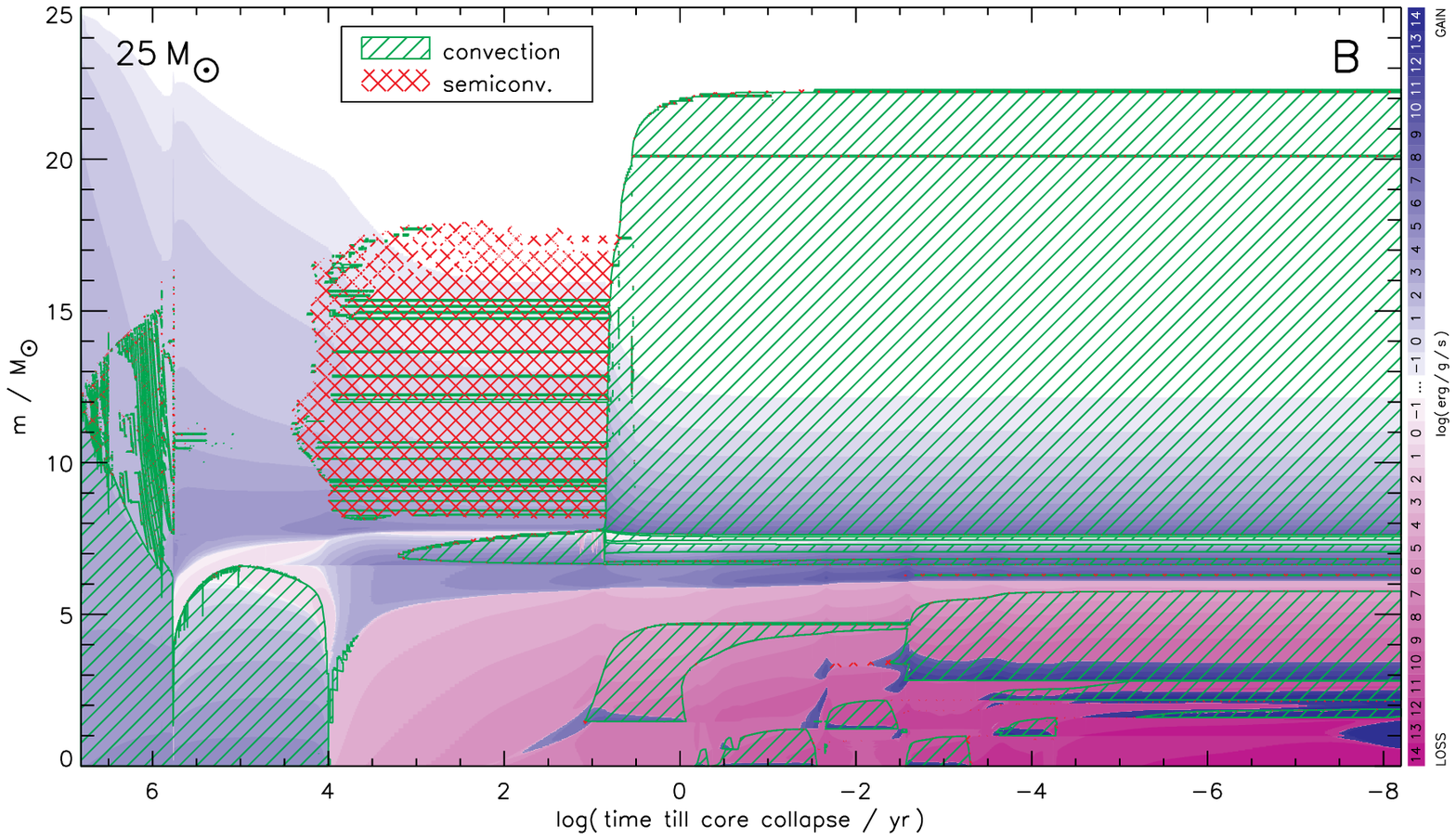}
\caption{Kippenhahn diagram of $15\,\Msun$ and $25\,\Msun$ stars.  As
  a function of time till core collapse (\textsl{X-Axis}, logarithmic
  scale) and mass coordinate of the star (\textsl{Y-Axis}) we show net
  specific energy generation (\textsl{blue shades}), net energy loss
  (\textsl{purple shades}), convective regions (\textsl{green
    hatching}) and semiconvective regions (\textsl{red cross
    hatching}).  Energy generations and loss increase by one order of
  magnitude for each level of darker shading (scale on right hand
  side), ranging from $10^{-1}\,\erggs$ (lightest colors) to
  $10^{14}\,\erggs$ (darkest colors).  \lFig{conv15+25}}
\end{figure*}

\clearpage 
\begin{figure*} 
\centering 
\includegraphics[angle=0,width=\textwidth]{\figurepath 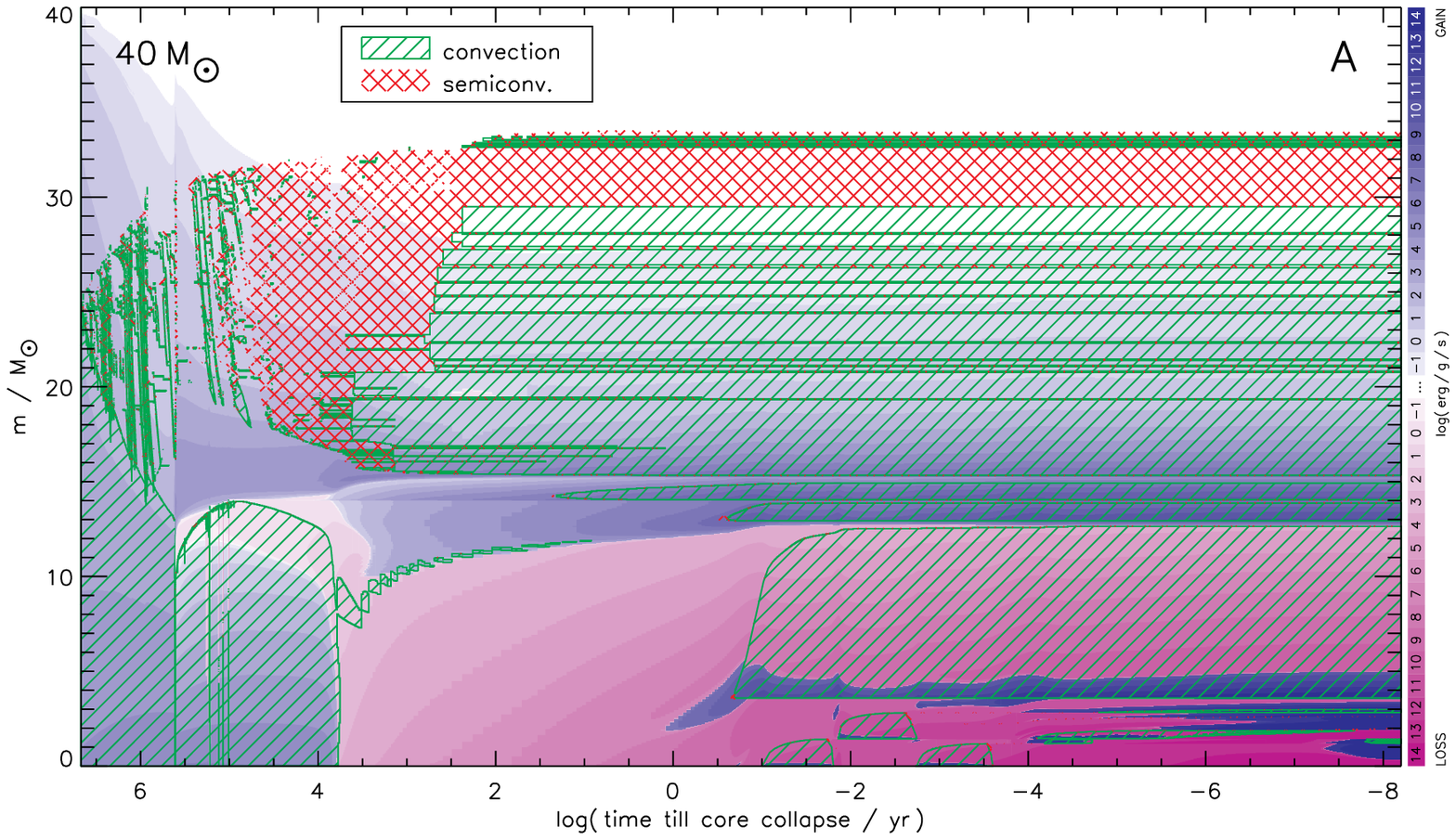}\\
\bigskip
\includegraphics[angle=0,width=\textwidth]{\figurepath 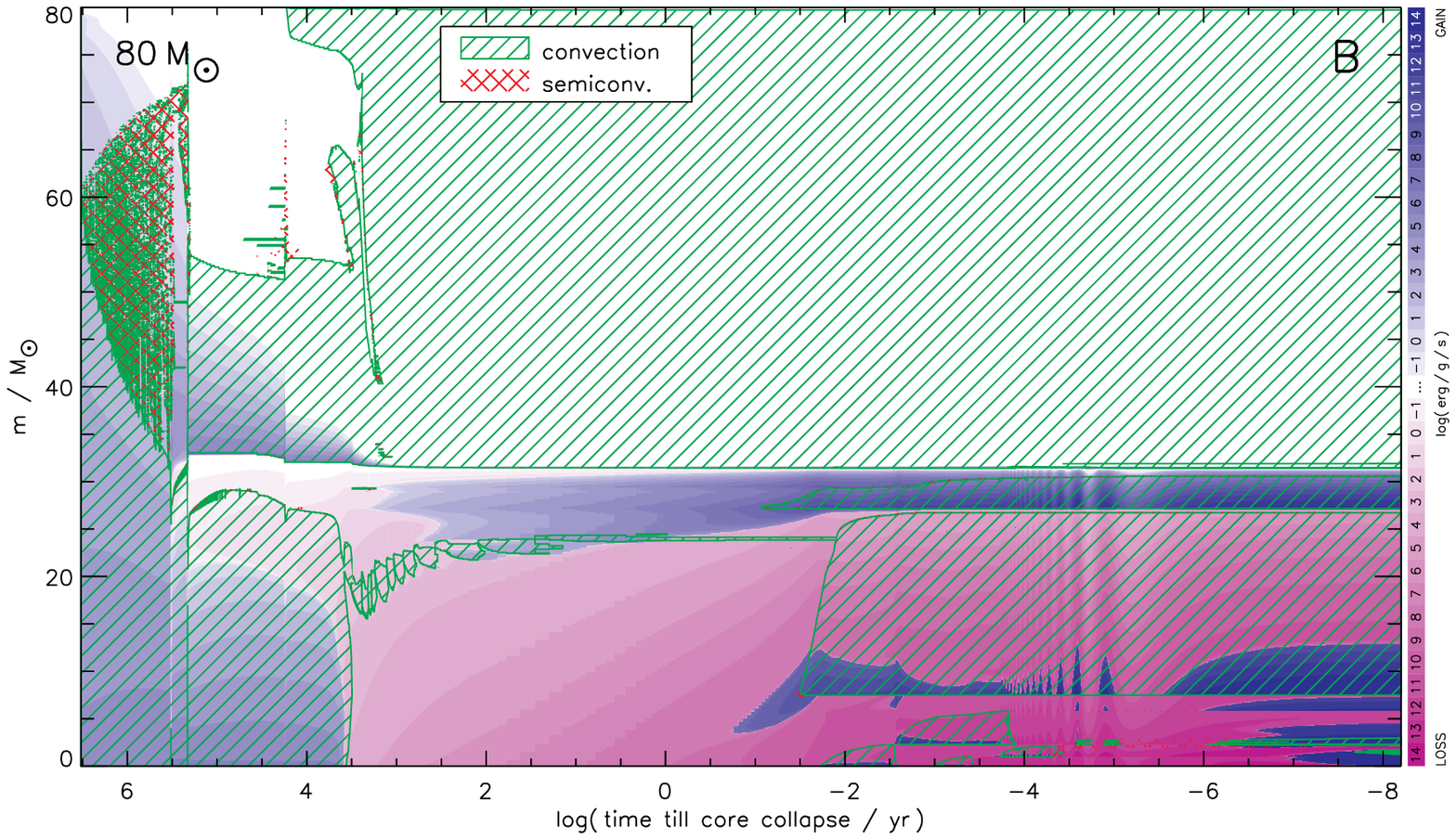}
\caption{Kippenhahn diagram of $40\,\Msun$ and $80\,\Msun$ stars.  See
  \Fig{conv15+25} for a detailed explanation.  \lFig{conv40+80}}
\end{figure*}

\clearpage 
\begin{figure*} 
\centering 
\includegraphics[angle=0,width=\textwidth]{\figurepath 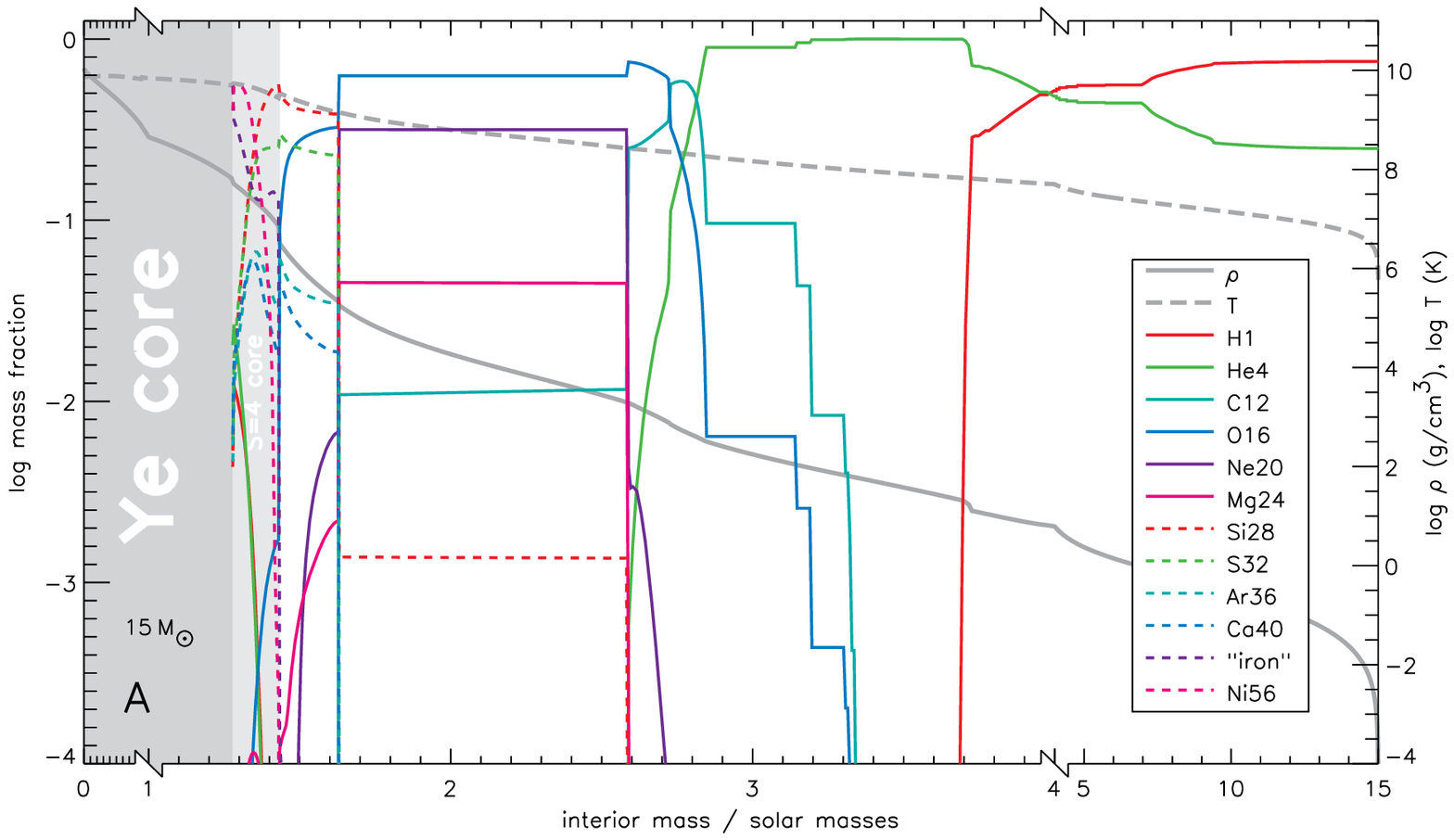}\\
\bigskip
\includegraphics[angle=0,width=\textwidth]{\figurepath 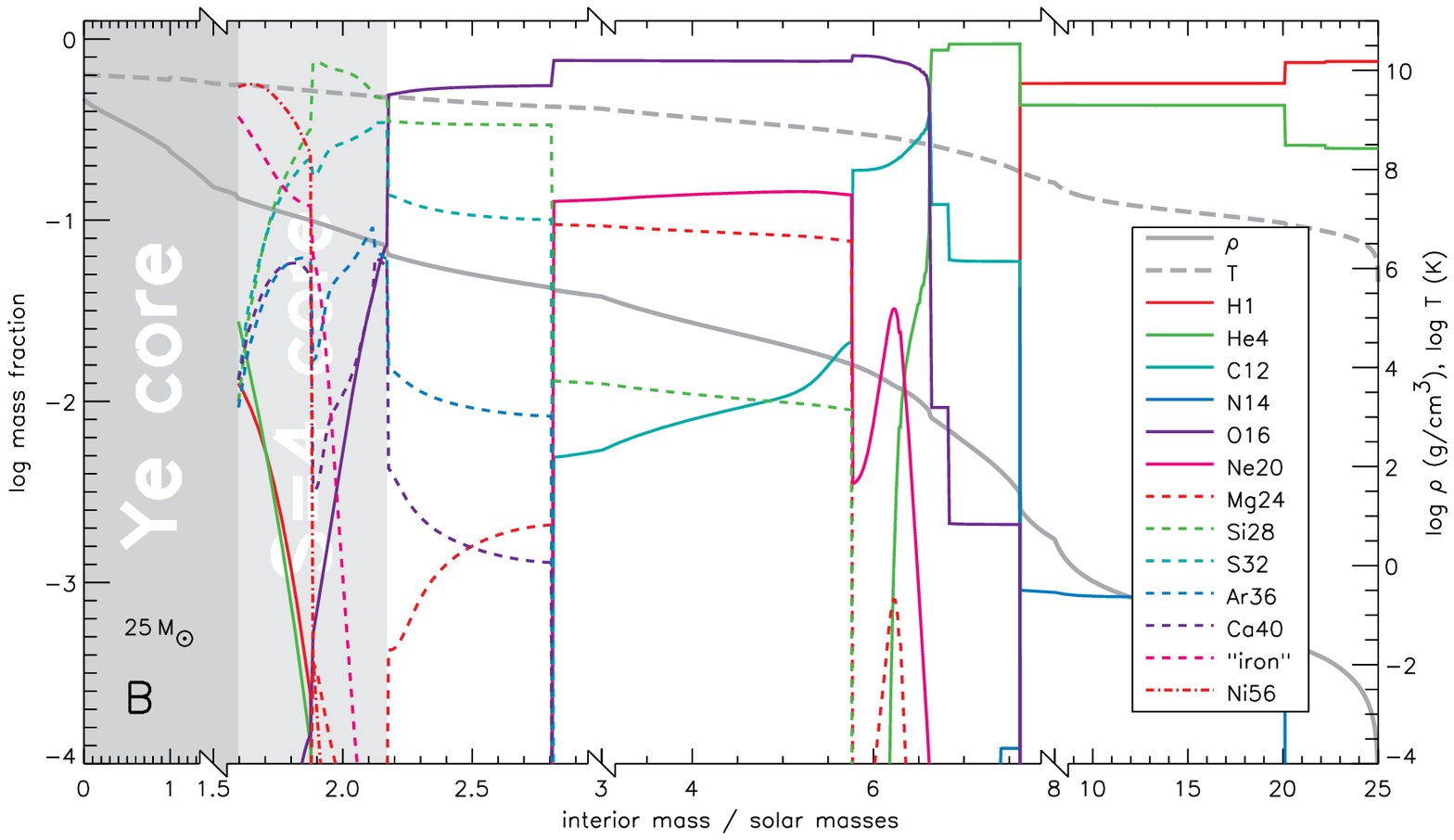}
\caption{Composition and structure of the $15\,\Msun$ (\textbf{Panel~A})
  and $25\,\Msun$ (\textbf{Panel~B}) presupernova stars.  The left
  darker background shading indicates the ``$\Ye$'' core, the lighter
  shading right of it the $S=4$ core.  \lFig{presn}}
\end{figure*}

\clearpage 
\begin{figure} 
\centering 
\includegraphics[angle=0,width=\columnwidth]{\figurepath 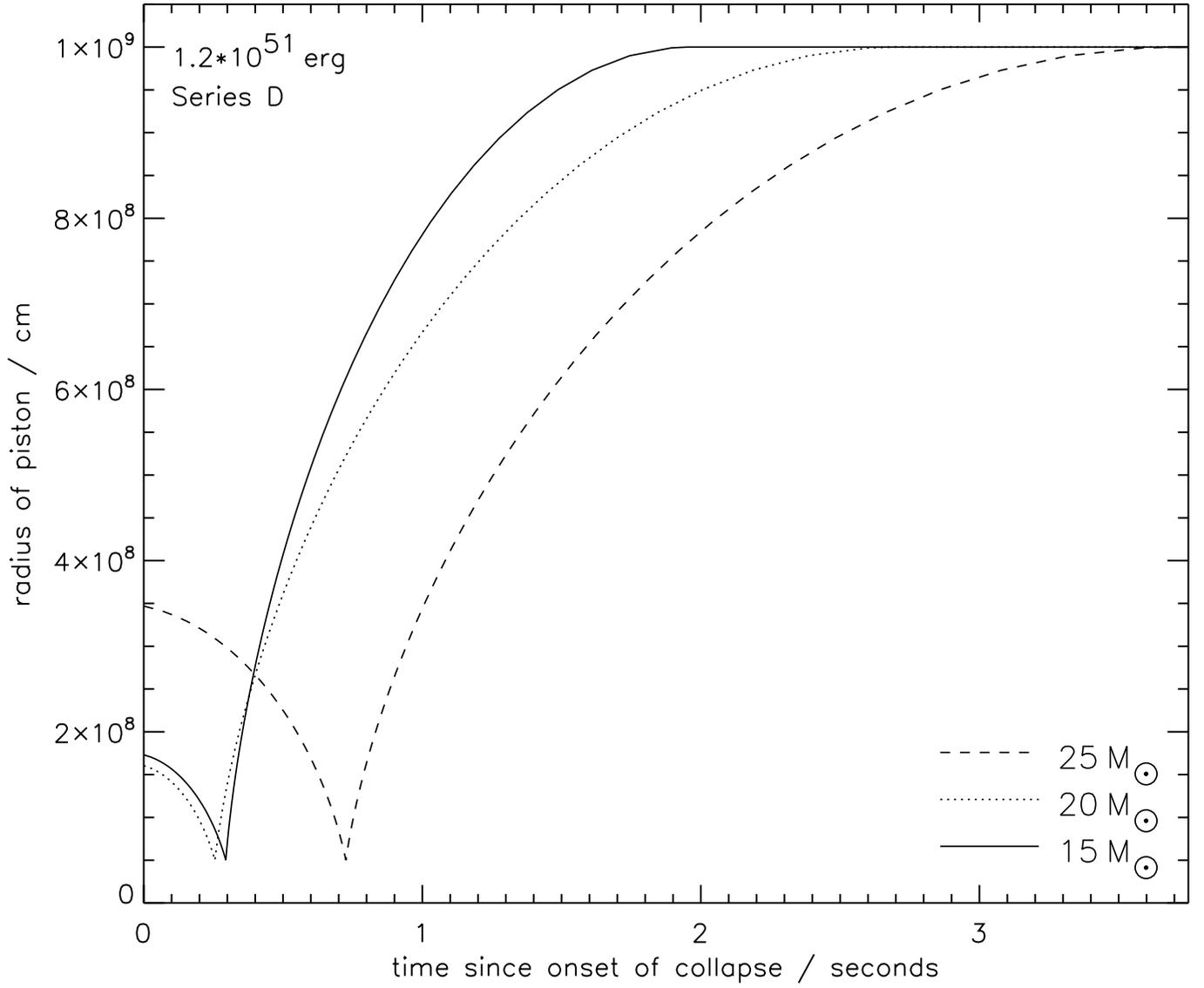}
\caption{Piston location as a function of time for the 15, 20, and
25\,\Msun stars.  When a piston location of \Ep9\,\cm is reached the
piston is stopped. ``$\!$\emph{Onset of collapse}'' is defined as the time
when a peak infall velocity of 900\,\kms is reached. \lFig{piston}}
\end{figure}

\clearpage 
\begin{figure*} 
\centering
\includegraphics[width=\textwidth,clip=true]{\figurepath 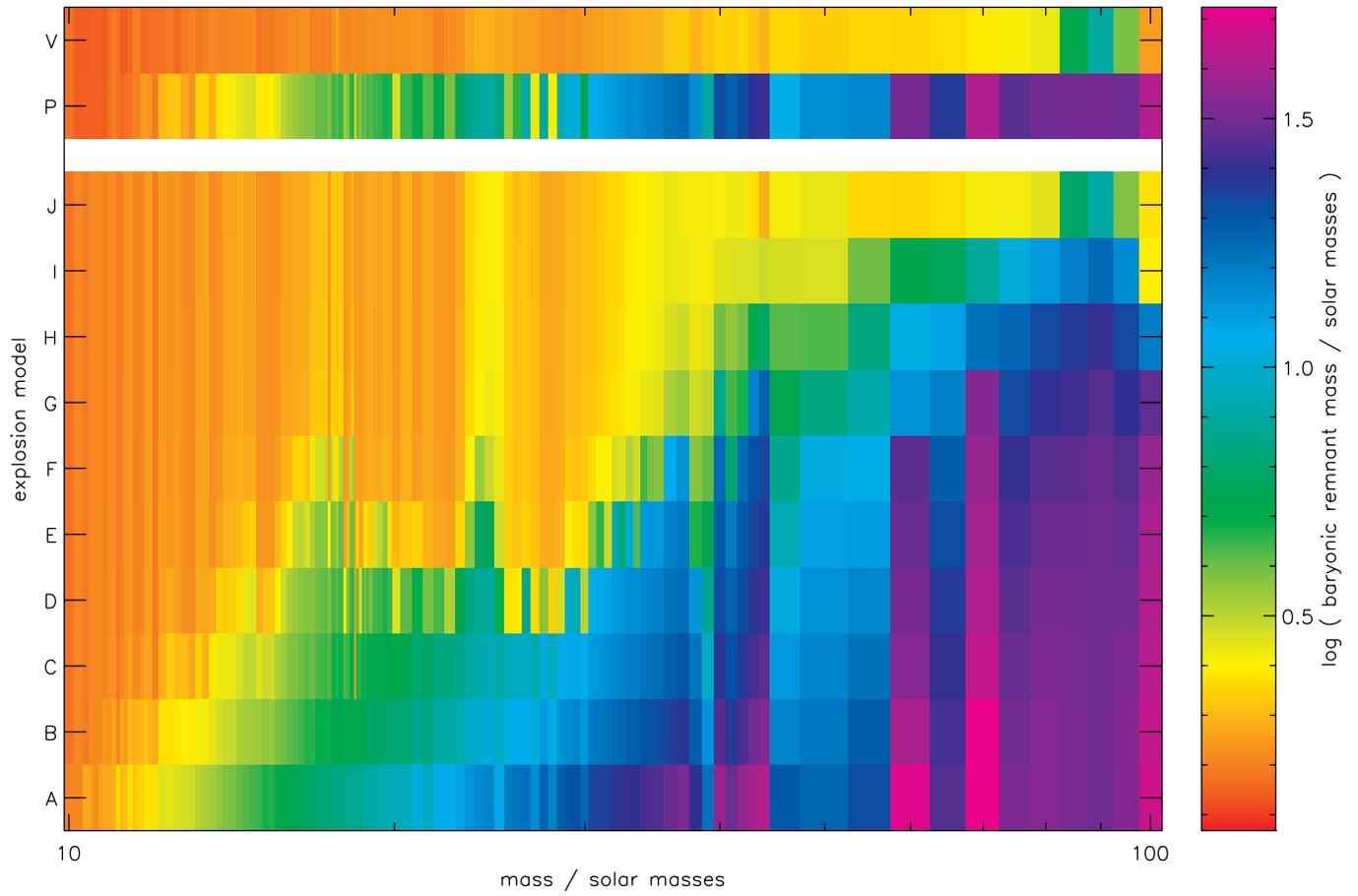}
\caption{Remnant mass as a function of stellar mass (\textsl{x-axis})
  and explosion model (\textsl{y-axis}).  Remnant masses are
  color-coded.  As a result of essentially negligible fallback, at low
  mass and high explosion energy the color remains constant for a
  given stellar mass, however, there are still star-to-star
  variations.  \lFig{remnant}}
\end{figure*}

\clearpage 
\begin{figure*} 
\centering 
\includegraphics[angle=0,width=\textwidth]{\figurepath 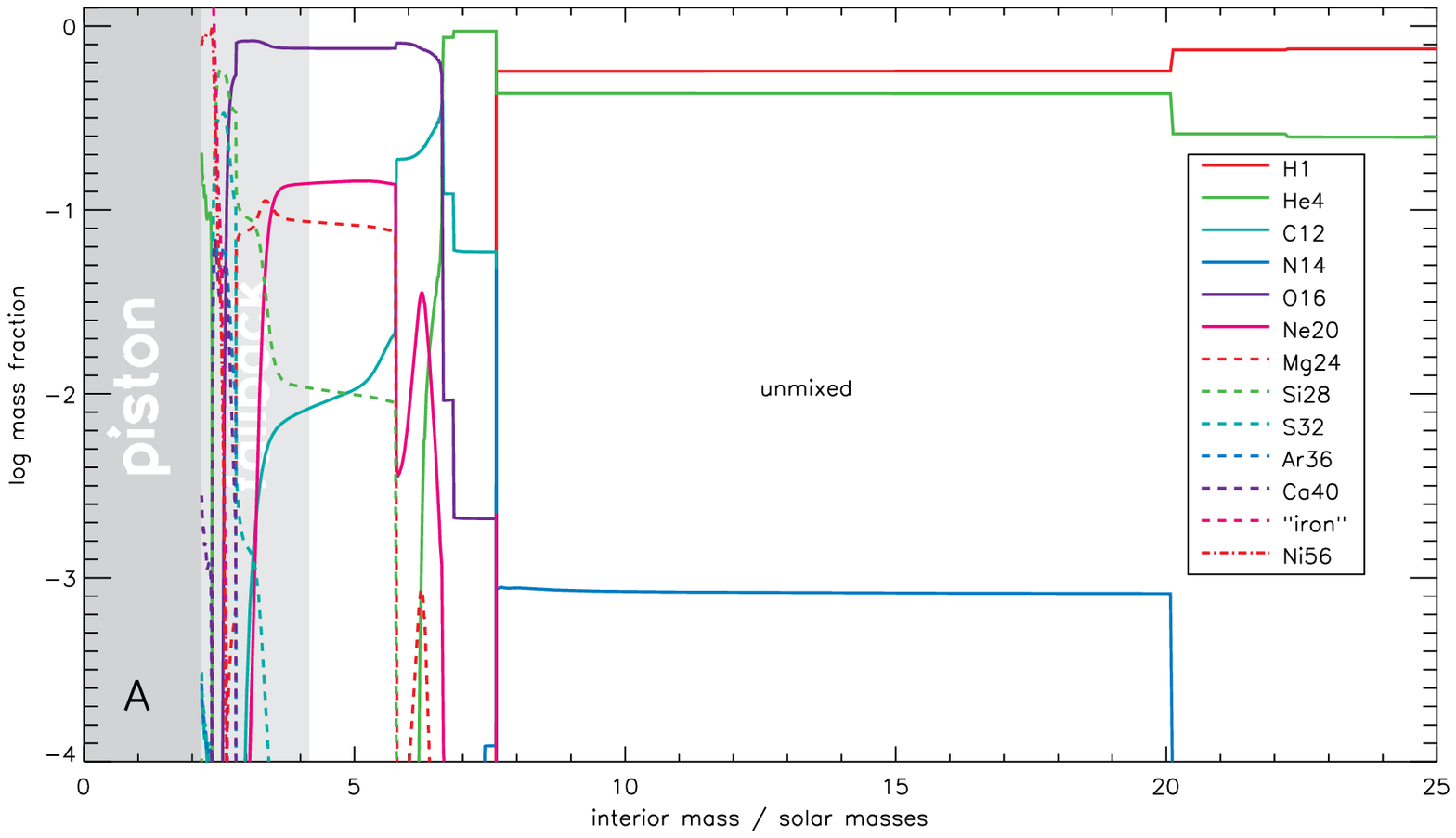}\\
\bigskip
\includegraphics[angle=0,width=\textwidth]{\figurepath 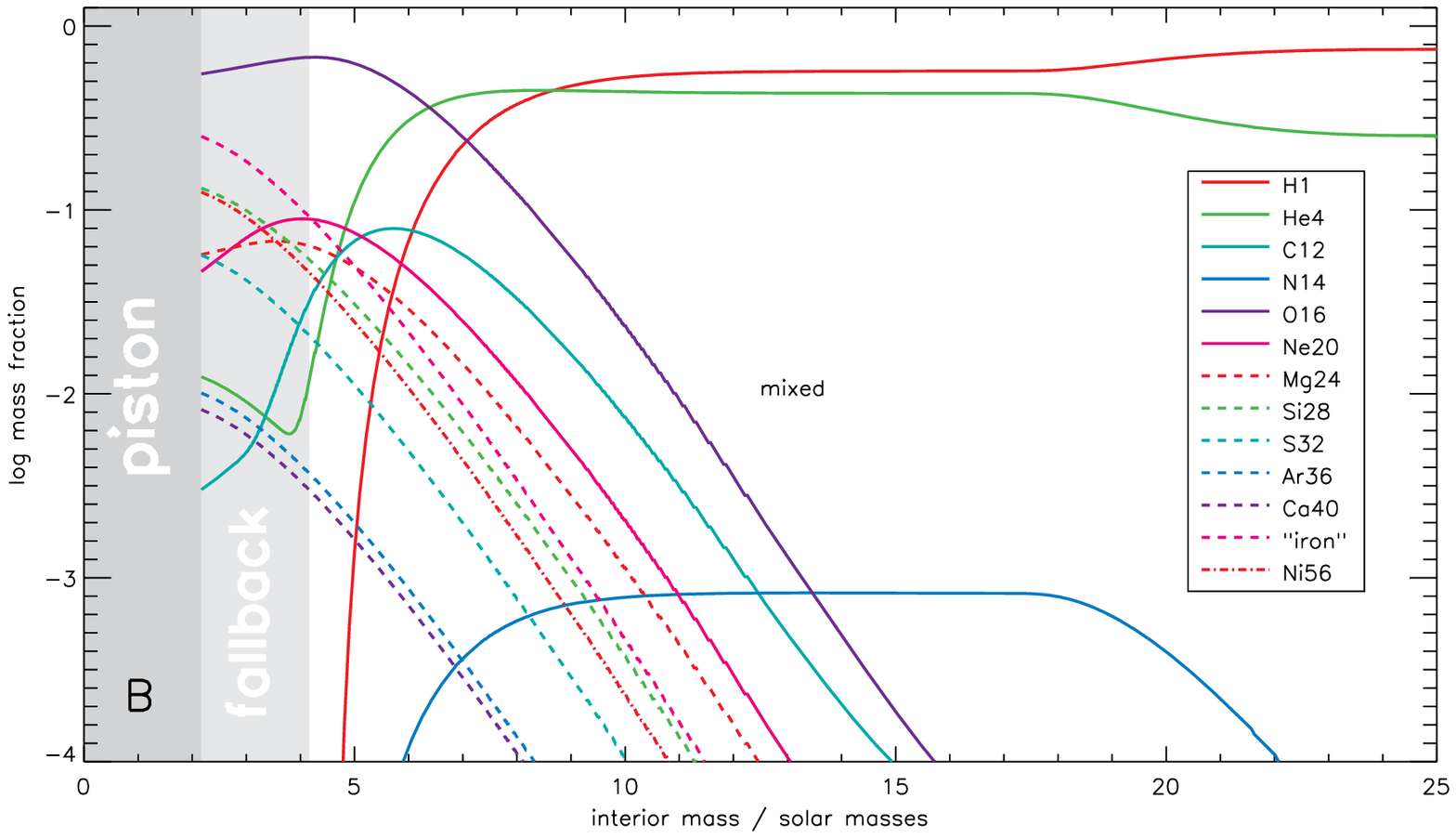}
\caption{Composition of Model z25D ($25\,\Msun$, $1.2\,\B$, piston at
  $S/\NA\kB = 4$) before (\textbf{Panel~A}) and after
  (\textbf{Panel~B}) the standard mixing operation. All explosions
  were routinely mixed in this fashion unless otherwise specified in
  the text. The mixing operation is carried out 100 s into the
  explosion before any significant fallback occurs.  The leftmost
  dark shading indicates the mass of the piston (remnant before
  fallback); the lighter gray region to its right indicates the mass
  that falls back.  None of the $^{56}$Ni comes out without the
  mixing. The ``iron'' composition compromises everything heavier than
  calcium except for $^{56}$Ni.  \lFig{mixed}}
\end{figure*}

\clearpage 
\begin{figure*} 
\centering 
\includegraphics[angle=0,width=\textwidth]{\figurepath 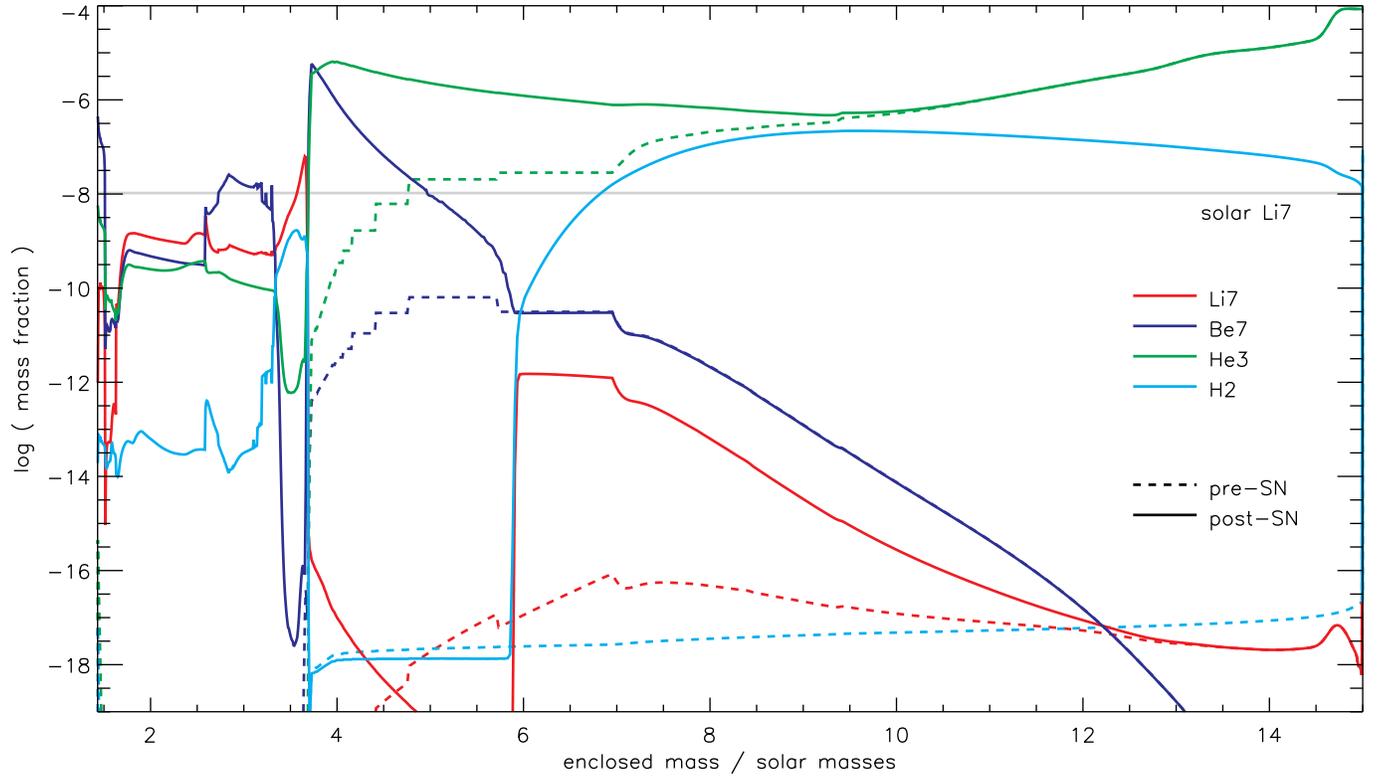}\\
\caption{Supernova $^7$Li production in the $15\,\Msun$ star for an
  explosion energy of $1.2\,\B$.  The \textsl{dashed curves} shows the
  mass fractions of $^2$H, (\textsl{magenta line}), $^3$He
  (\textsl{green}), $^7$Li (\textsl{red}) and $^7$Be (\textsl{blue})
  as a function of the enclosed mass at onset of core collapse (before
  the supernova).  The \textsl{sold lines} show these mass fractions
  $100\,$s after core collapse and without mixing.  At this time the
  supernova shock has propagated to a mass coordinate of about
  $5.9\,\Msun$; the edge of the helium core is located at about
  $3.7\,\Msun$.  The post-SN $^7$Be, which later decays to $^7$Li, is
  well above the solar value for $^7$Li (\textsl{gray line}) at the
  inner edge of the hydrogen envelope and causes a significant
  overproduction of $^7$Li (observe logarithmic scale of
  \textsl{y-axis}).  The $^7$Be is made by the reaction
  $^1$H($\bar{\nu_\mathrm{e}}$,e$^+$)n(p,$\gamma$)$^2$H(p,$\gamma$)$
  ^3$He($\alpha$,$\gamma$)$^7$Be initiated by supernova neutrinos.
  This reaction is only efficient here due to the compactness of the
  hydrogen envelope of the primordial stars.  The neutrons made from
  neutrino interactions quickly from $^2$H.  Below $\sim8\,\Msun$ most
  of the $^2$H has quickly reacted to make $^3$He.  Much of the $^7$Be
  is then synthesized in the supernova shock.  \lFig{Li7}}
\end{figure*}

\clearpage 
\begin{figure*} 
\centering
\includegraphics[width=0.475\textwidth,clip=true]{\figurepath 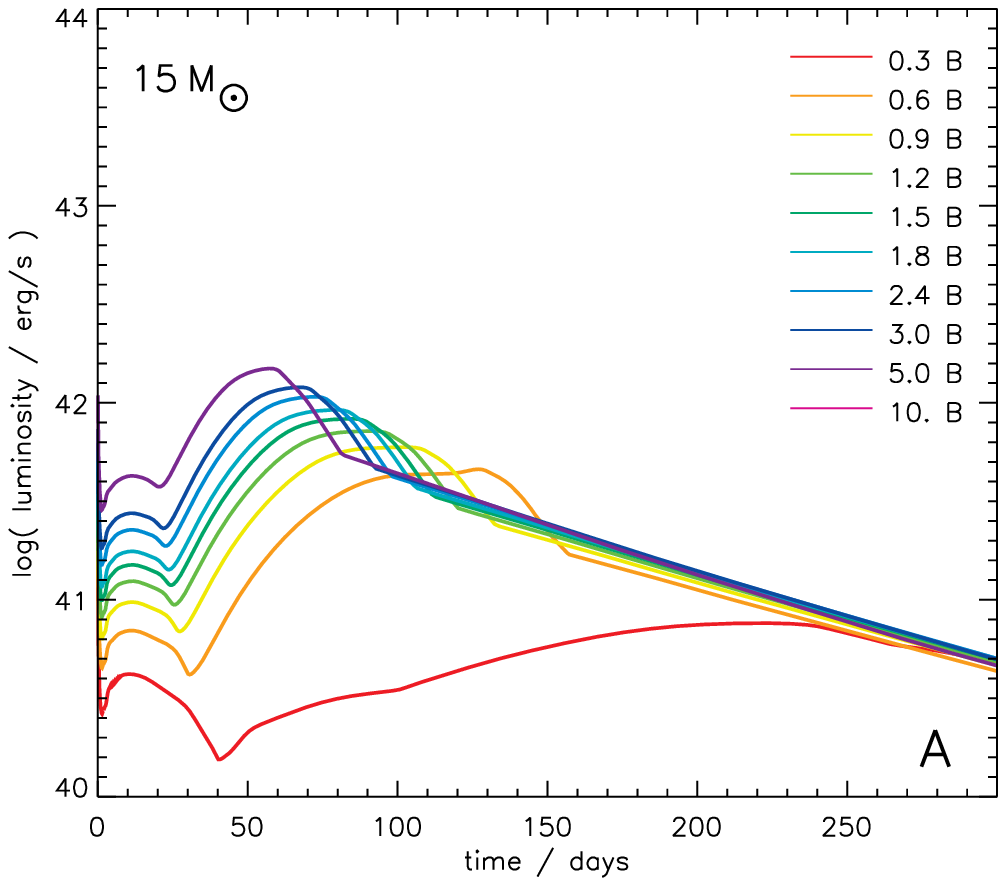}
\hfill
\includegraphics[width=0.475\textwidth,clip=true]{\figurepath 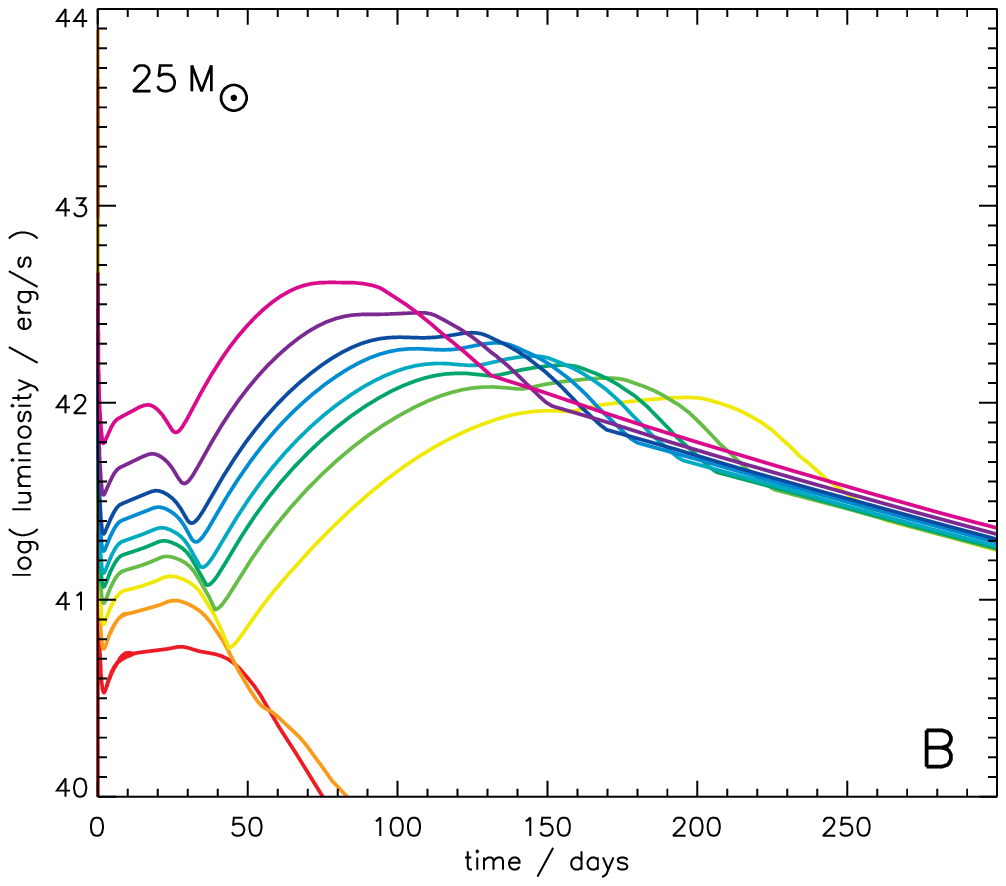}
\vskip 24pt
\includegraphics[width=0.475\textwidth,clip=true]{\figurepath 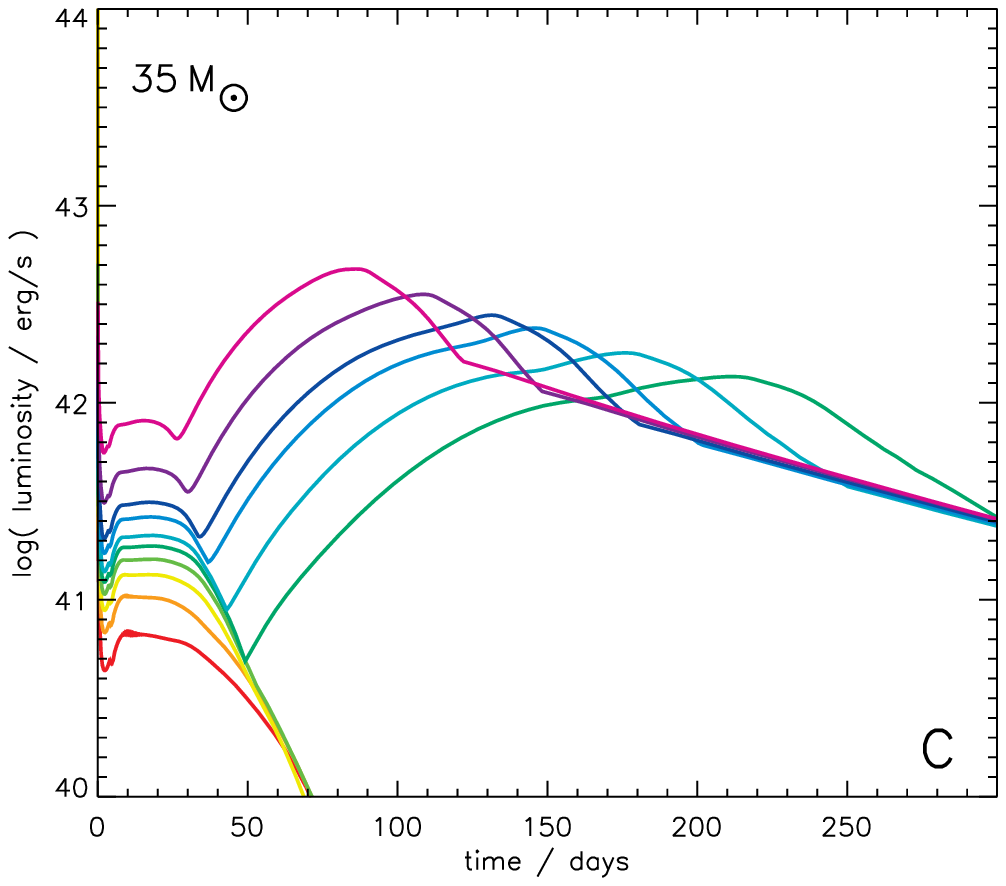}
\hfill
\includegraphics[width=0.475\textwidth,clip=true]{\figurepath 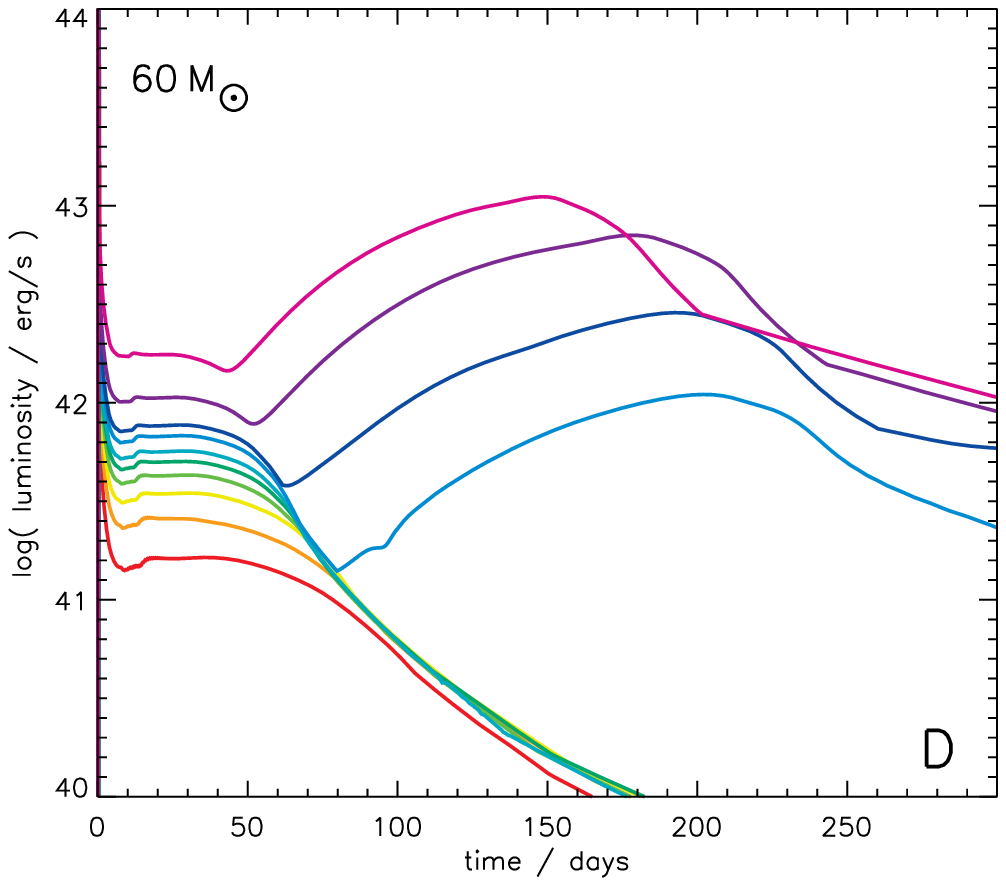}
\vskip 24pt
\caption{Bolometric light curves for the $15\,\Msun$ (\textsl{Panel
    A}), $25\,\Msun$ (\textsl{Panel B}), $35\,\Msun$ (\textsl{Panel
    C}), and $60\,\Msun$ (\textsl{Panel D}) stars calculated for a
  variety of explosion energies. \lFig{lc}}
\end{figure*}

\clearpage

\newcounter{fignum}

\begin{figure*} 
\setcounter{fignum}{\thefigure}
\centering
\includegraphics[width=0.475\textwidth,clip=true]{\figurepath 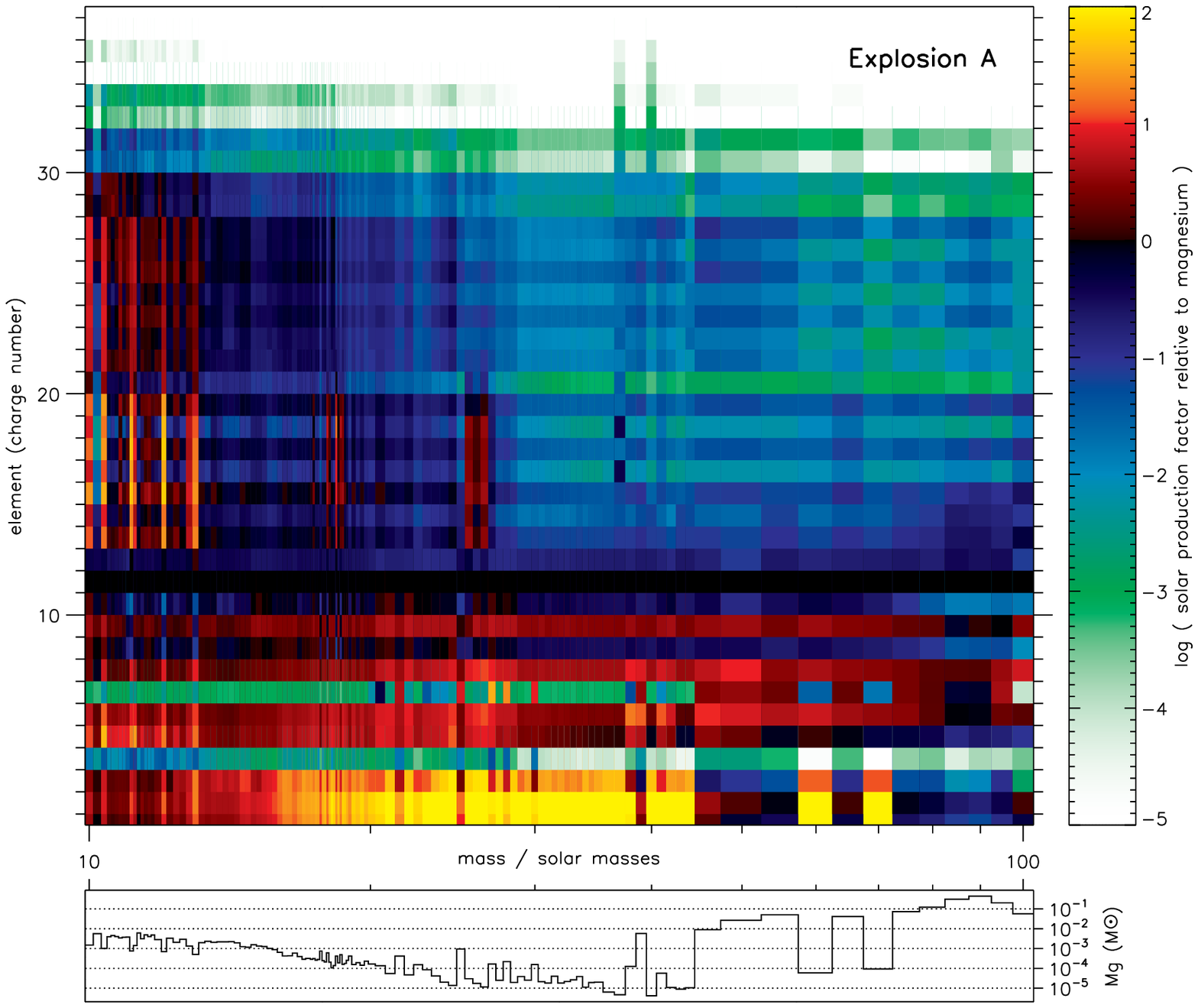}
\hfill
\includegraphics[width=0.475\textwidth,clip=true]{\figurepath 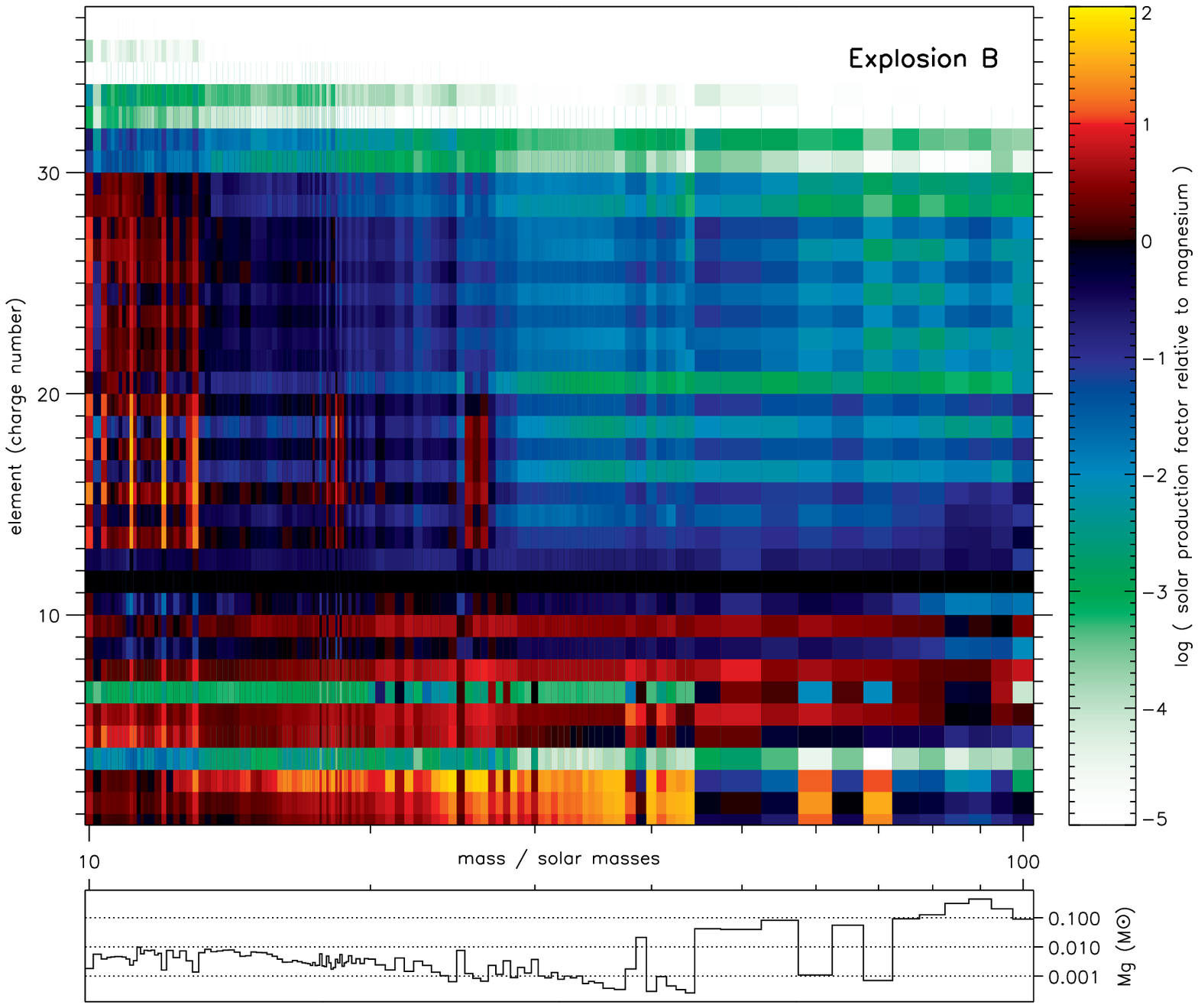}
\vskip 24pt
\includegraphics[width=0.475\textwidth,clip=true]{\figurepath 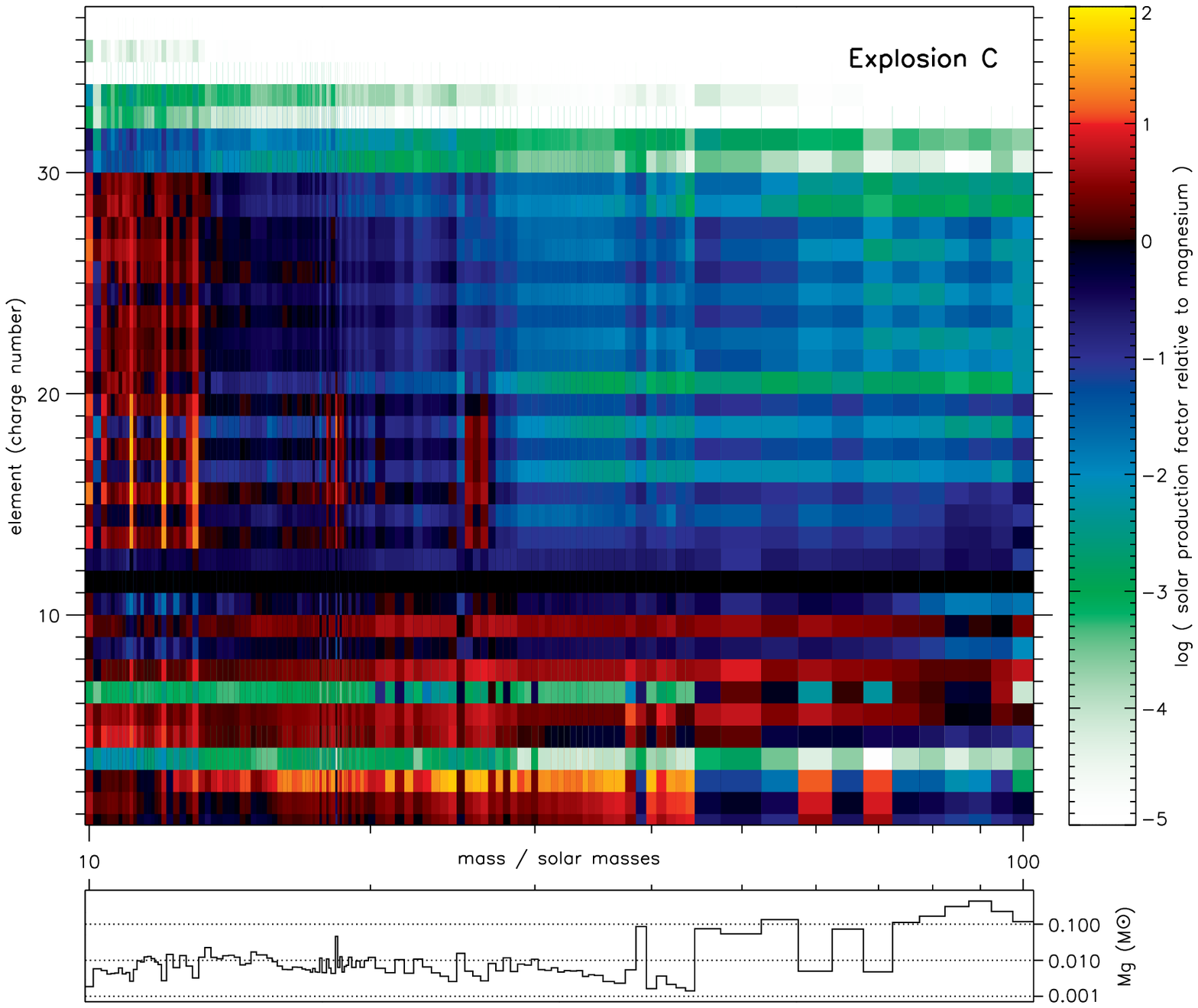}
\hfill
\includegraphics[width=0.475\textwidth,clip=true]{\figurepath 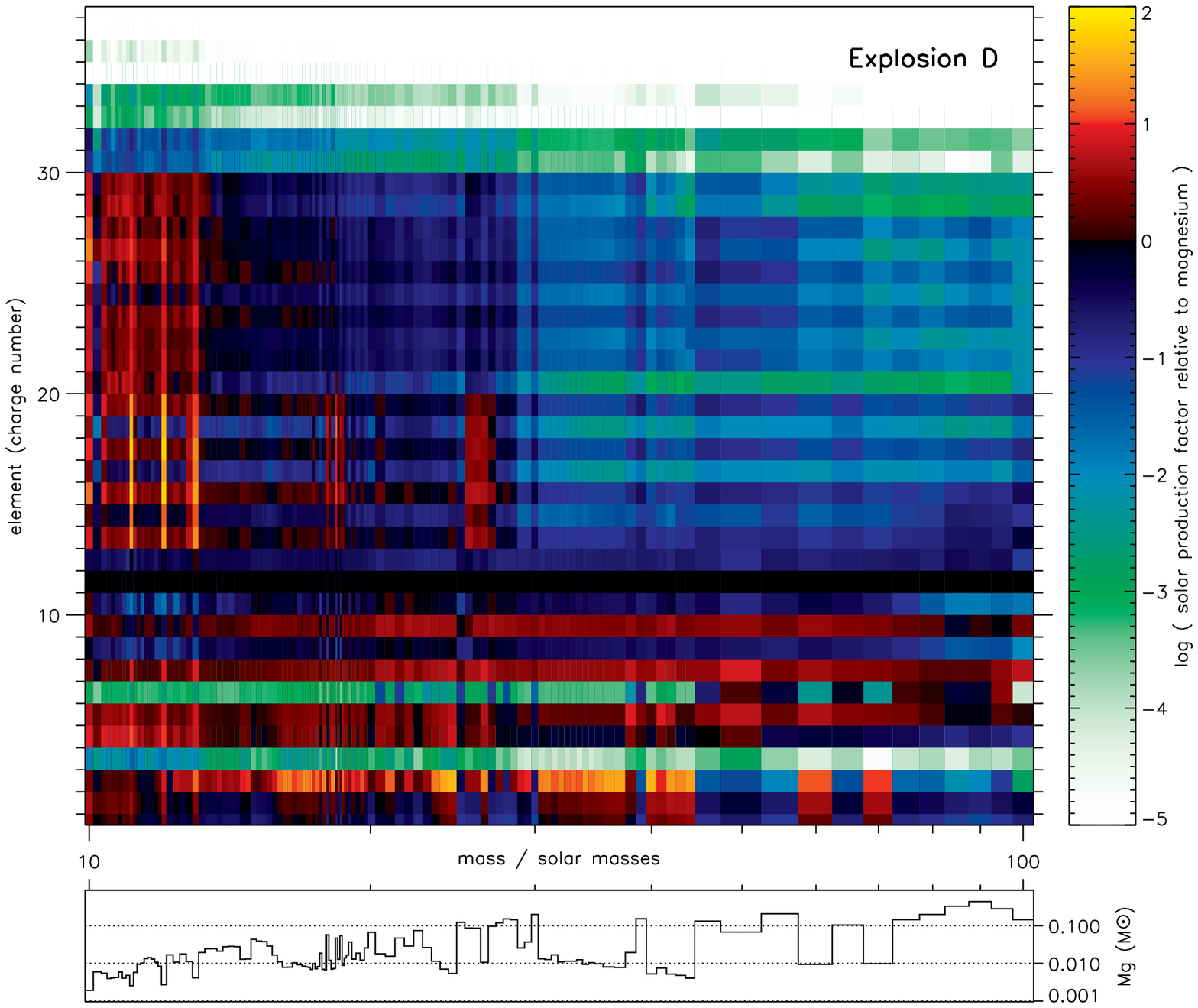}
\vskip 24pt
\includegraphics[width=0.475\textwidth,clip=true]{\figurepath 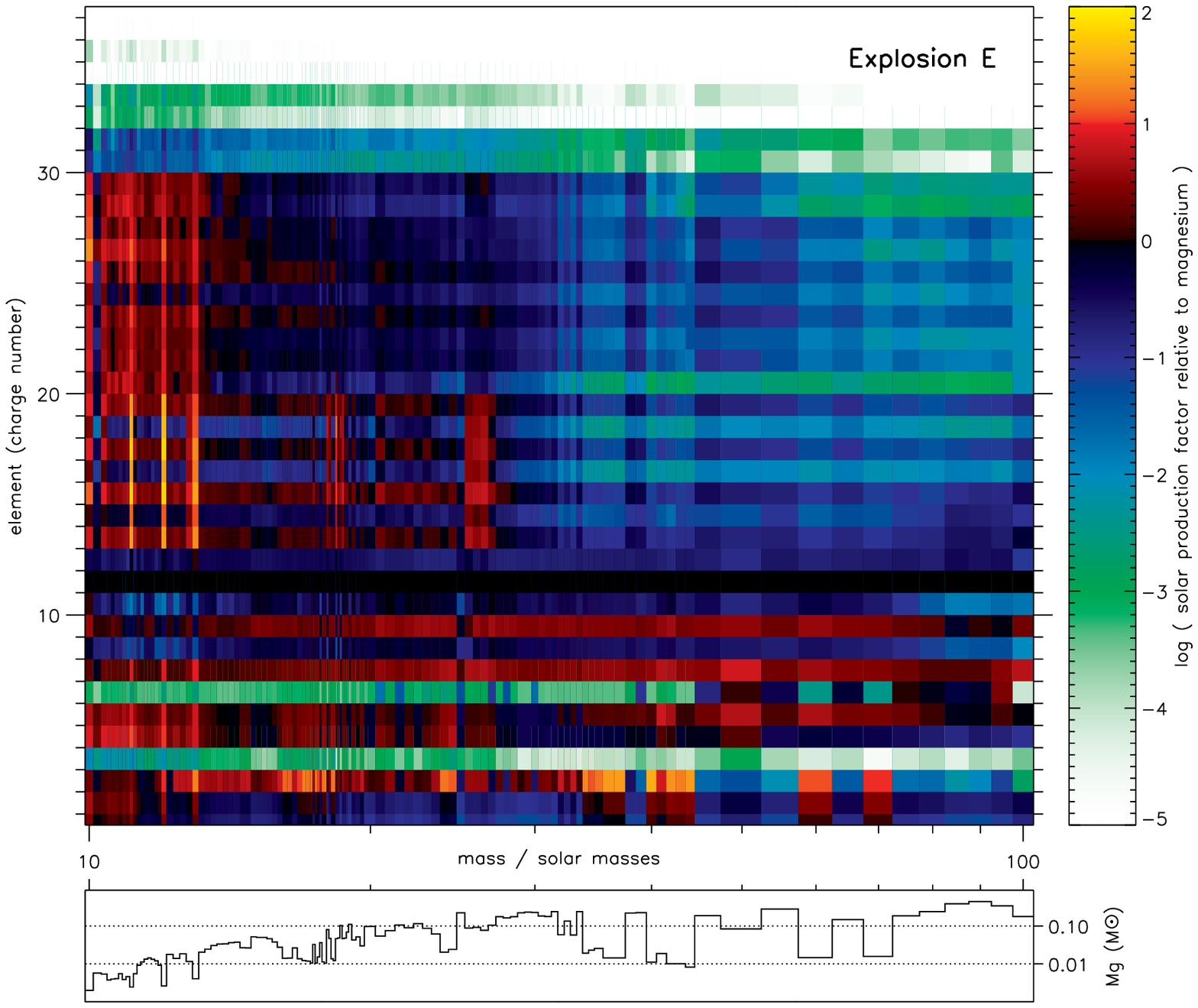}
\hfill
\includegraphics[width=0.475\textwidth,clip=true]{\figurepath 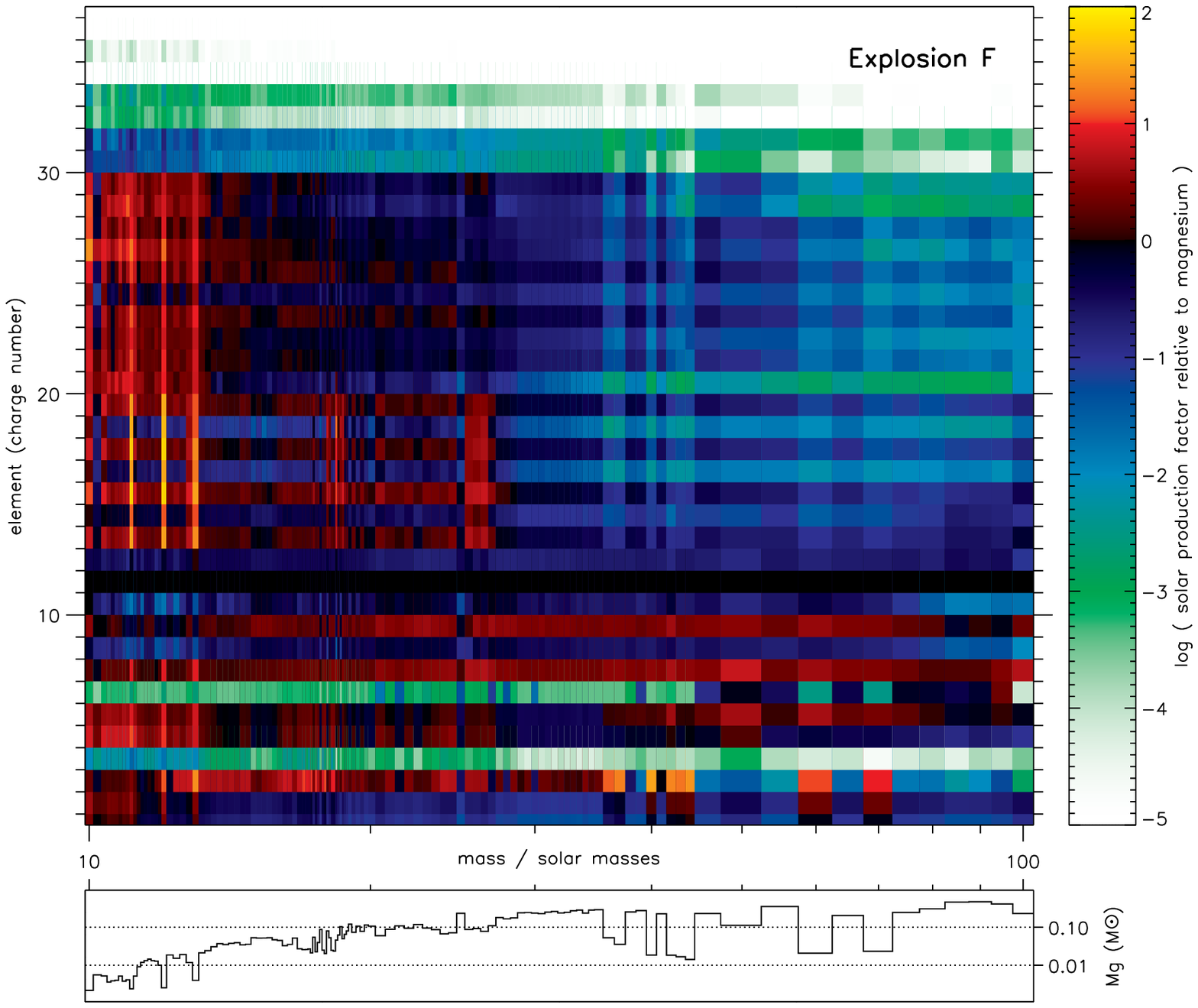}
\caption{Elemental production factors relative to solar as a function
  of mass for the $0.3$--$1.8\,\B$ explosions (Explosion Models A--F).
  The production factors are normalized to a solar yield of magnesium
  and color coded according to the scale on the right hand side.  The
  absolute yield of magnesium as a function of mass is given below
  each map.  Darker colors correspond to about solar co-production
  with magnesium; red through yellow, to overproduction, and blue
  through green and white, to underproduction. \textsc{(Continued on
    next page.)}  \lFig{pf:A-F}}
\end{figure*}

\begin{figure*} 
\setcounter{figure}{\thefignum}
\centering
\includegraphics[width=0.475\textwidth,clip=true]{\figurepath 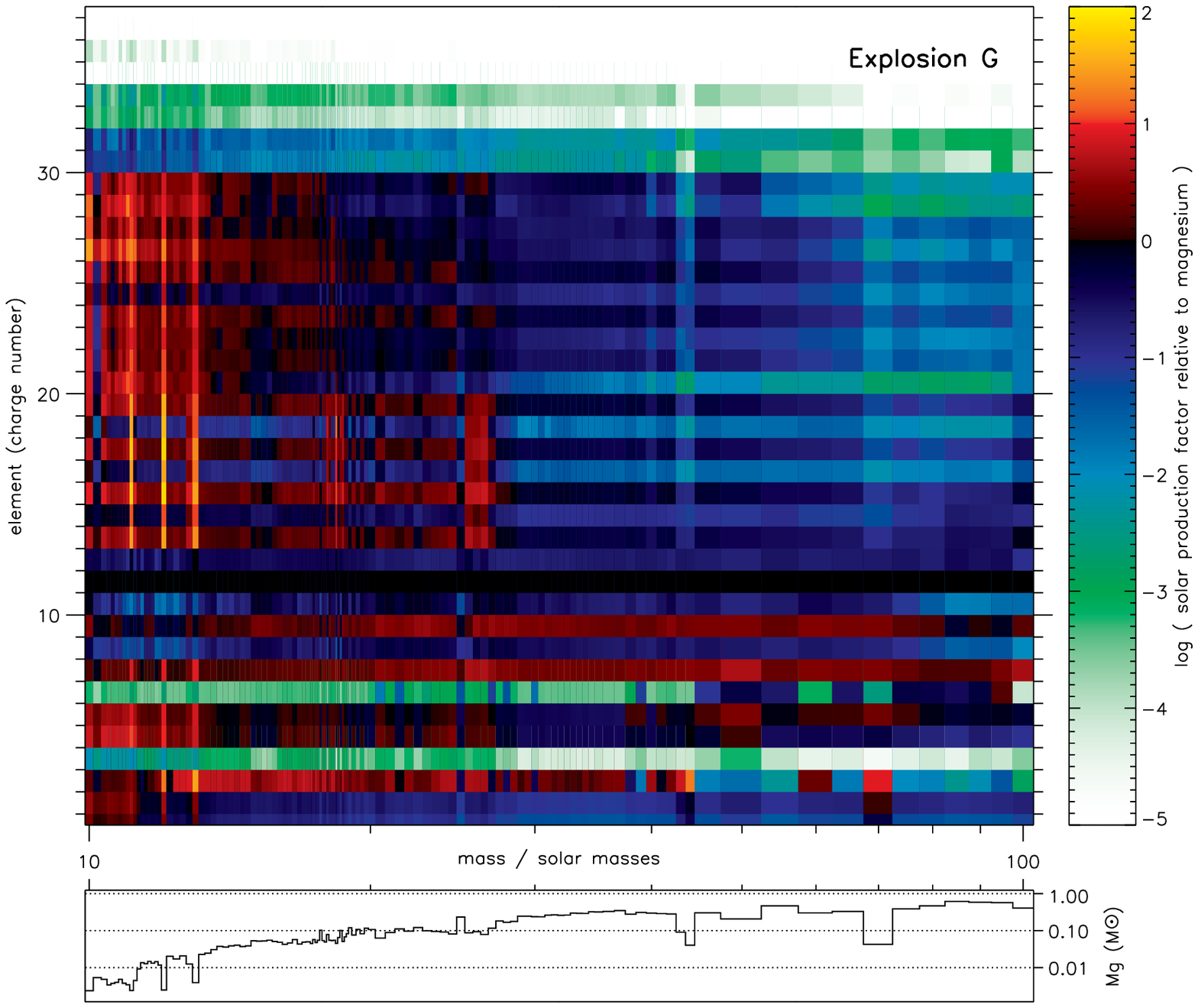}
\hfill
\includegraphics[width=0.475\textwidth,clip=true]{\figurepath 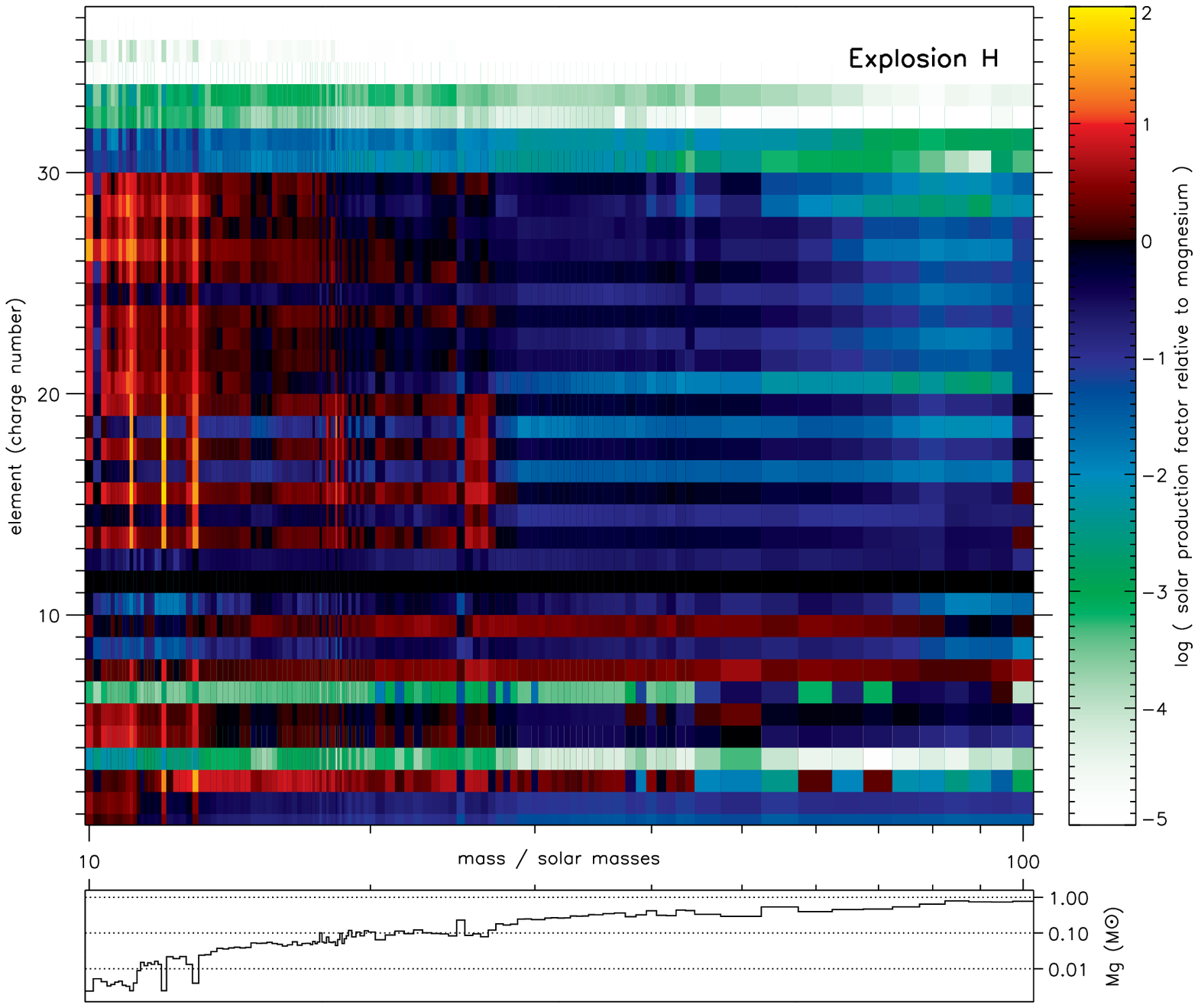}
\vskip 24pt
\includegraphics[width=0.475\textwidth,clip=true]{\figurepath 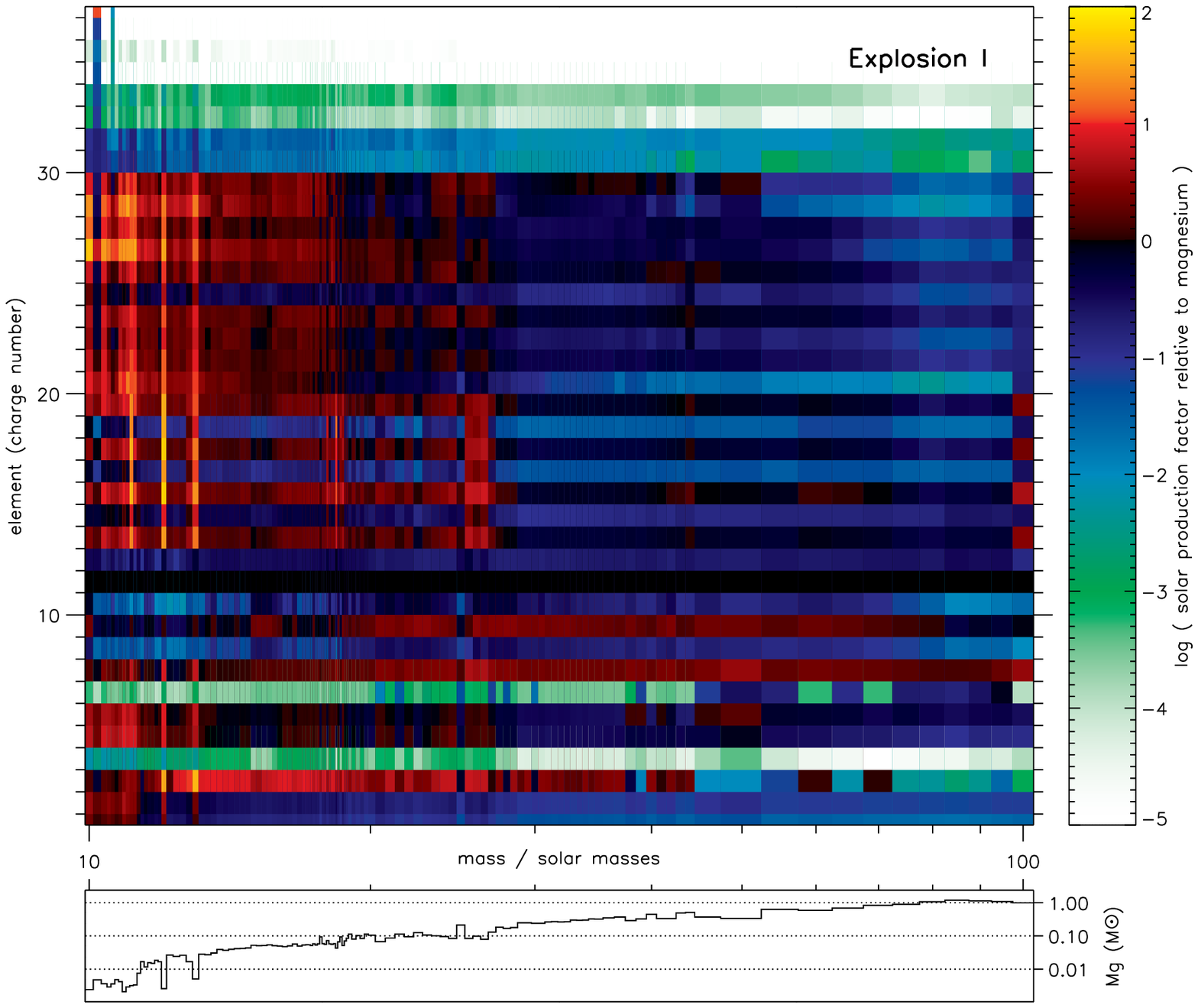}
\hfill
\includegraphics[width=0.475\textwidth,clip=true]{\figurepath 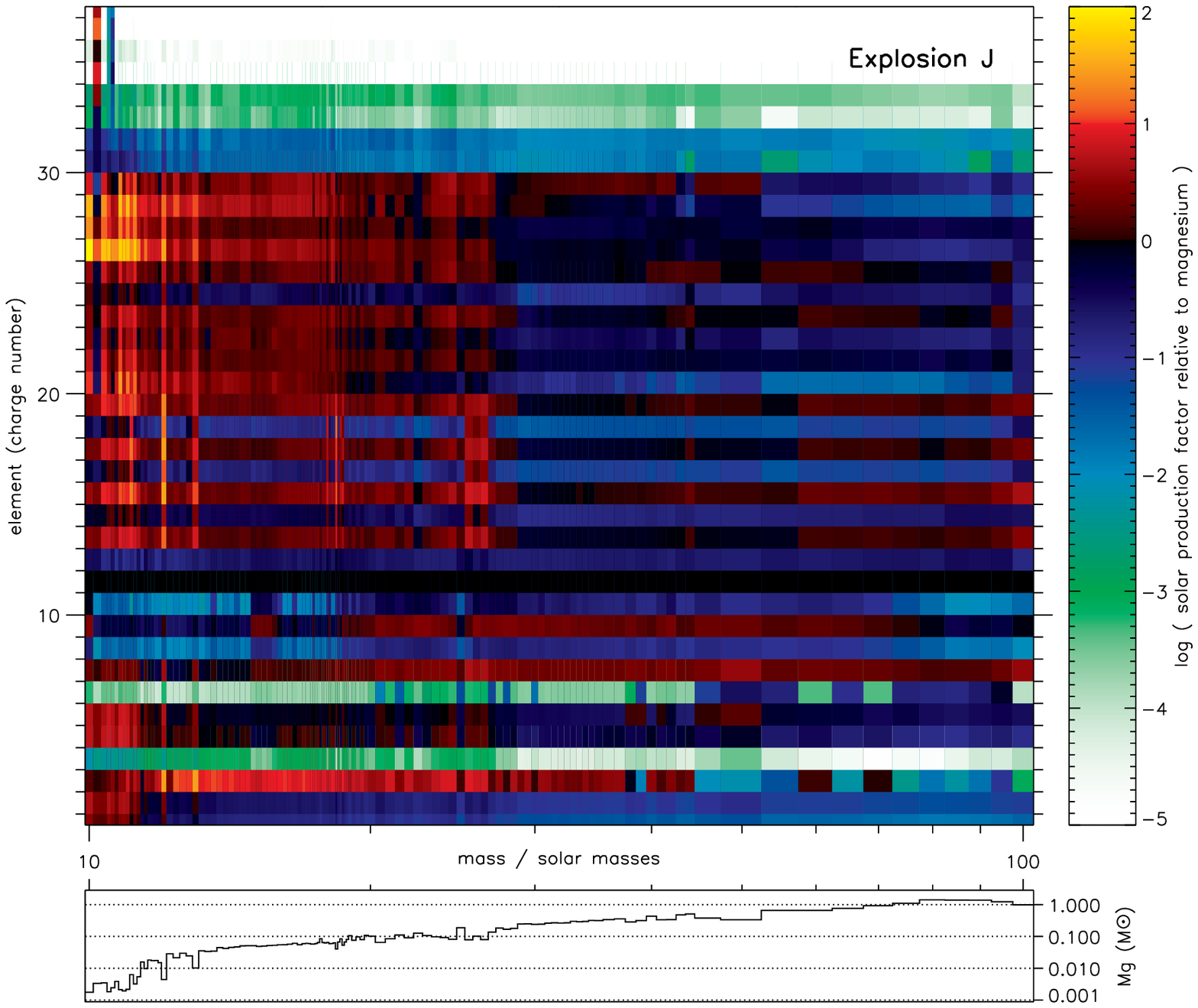}
\vskip 24pt
\includegraphics[width=0.475\textwidth,clip=true]{\figurepath 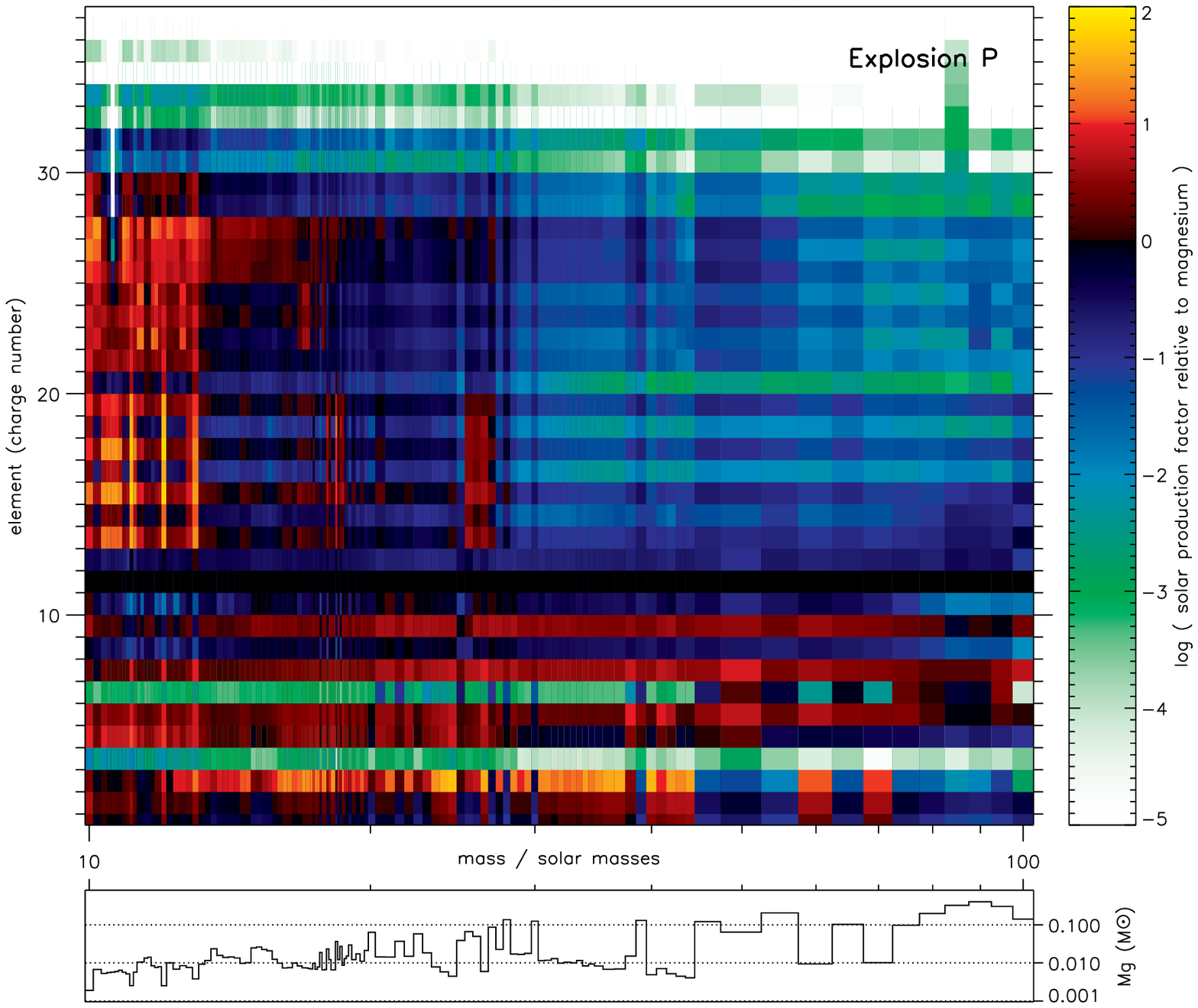}
\hfill
\includegraphics[width=0.475\textwidth,clip=true]{\figurepath 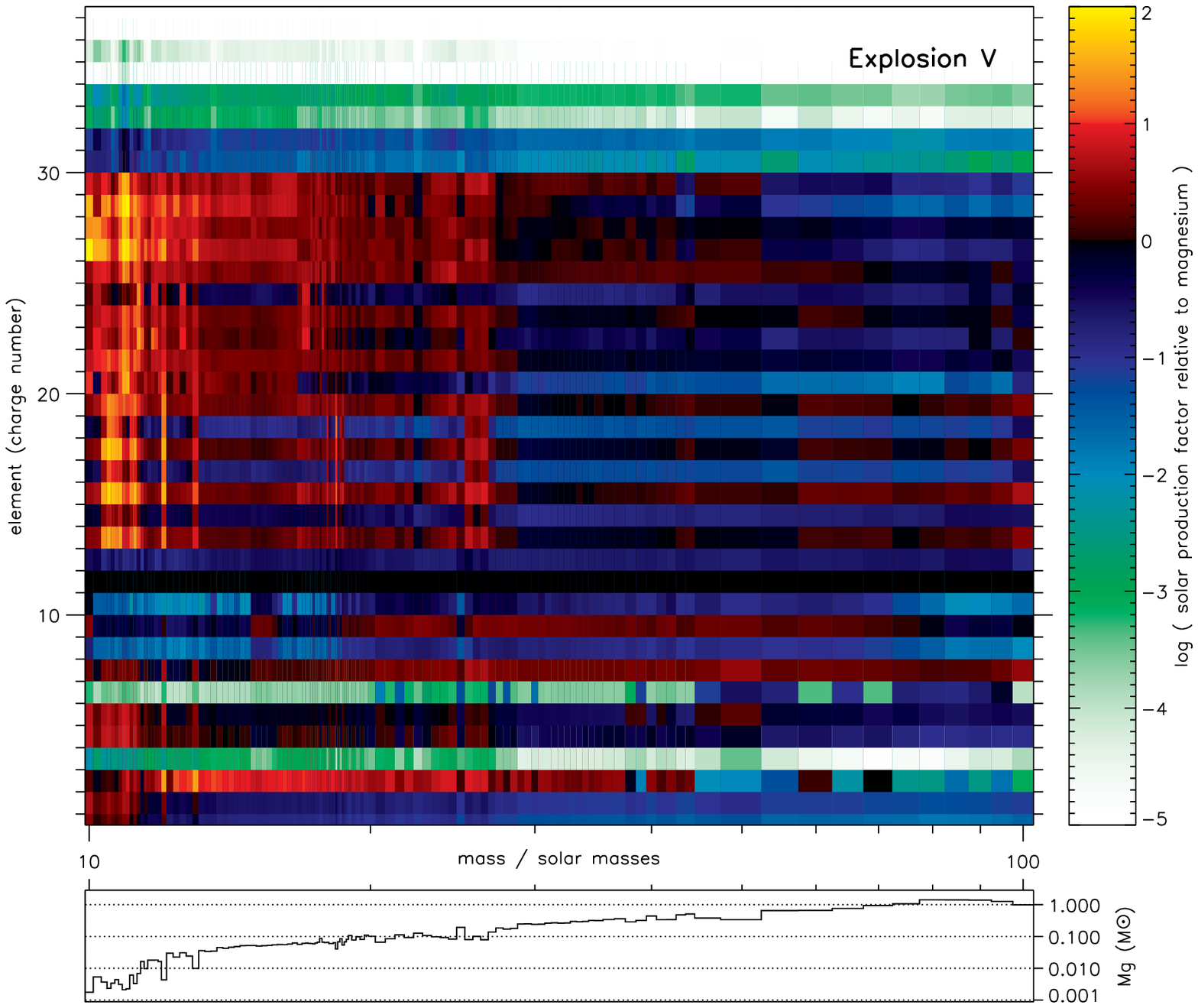}
\caption{ \textsc{(Continued)} Production factors as a function of
  mass for the $2.4$--$10\,\B$ explosions (Explosion Models G--J), and
  the $1.2\,\B$ and $10\,\B$ explosions with pistons located at the
  edge of the deleptonized core (Explosion Models P and
  V). \lFig{pf:G-V}}
\end{figure*}
\clearpage

\begin{figure}
\centering
\includegraphics[width=0.475\textwidth]{\figurepath 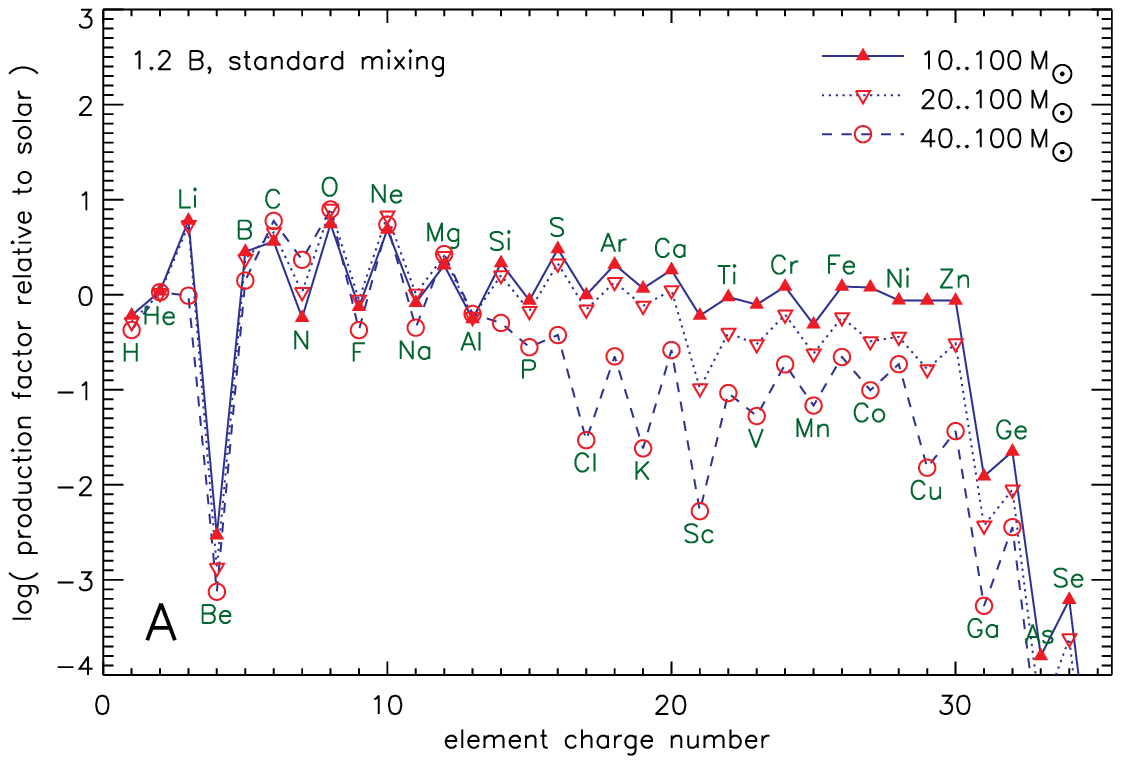}
\hfill                                                                       
\includegraphics[width=0.475\textwidth]{\figurepath 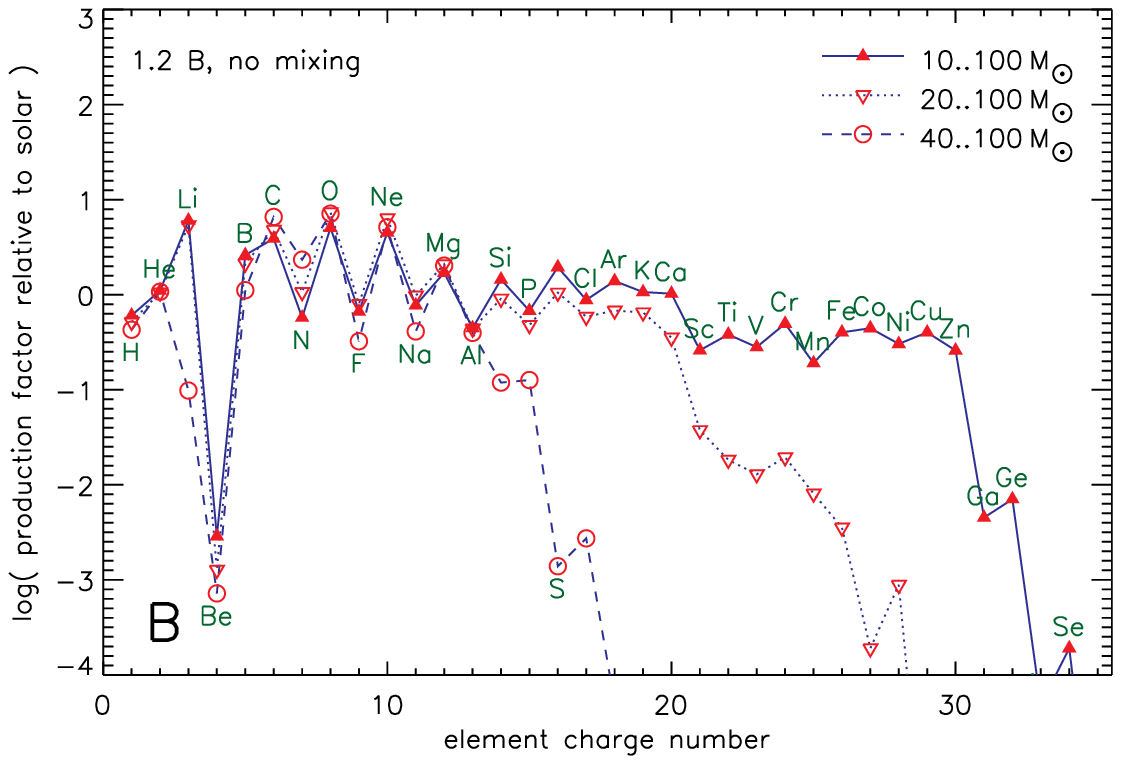}
\\\bigskip                                                                   
\includegraphics[width=0.475\textwidth]{\figurepath 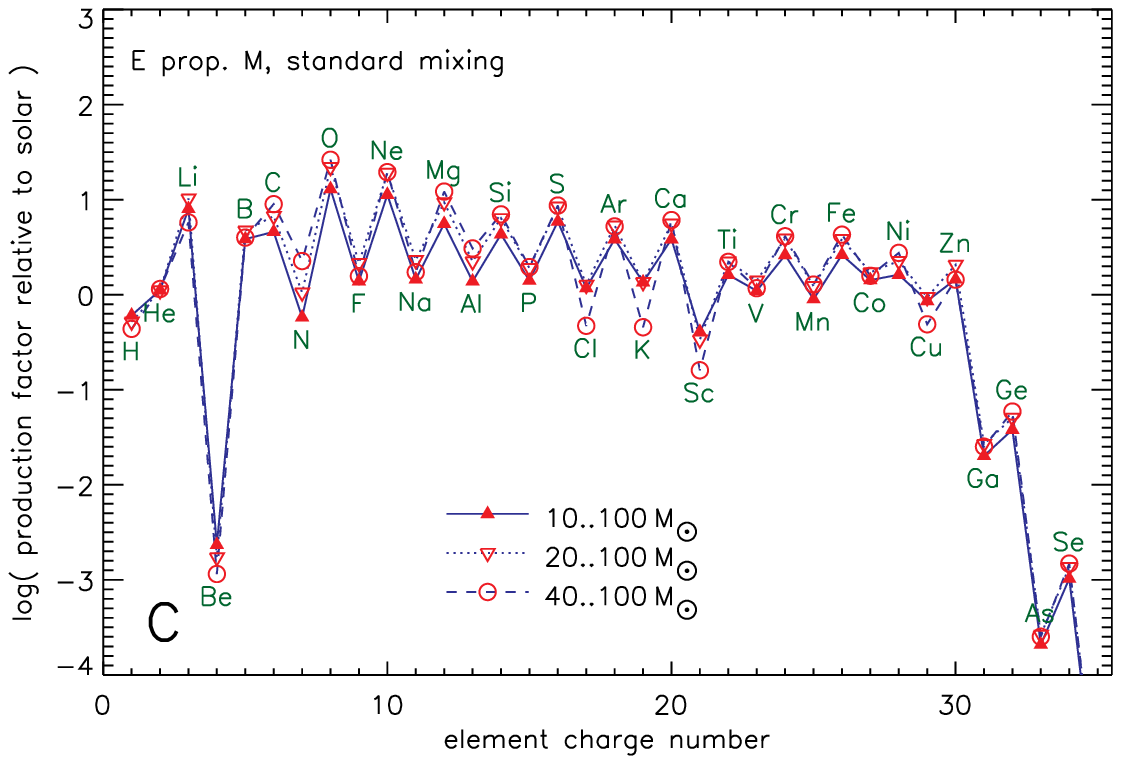}
\hfill                                                                       
\includegraphics[width=0.475\textwidth]{\figurepath 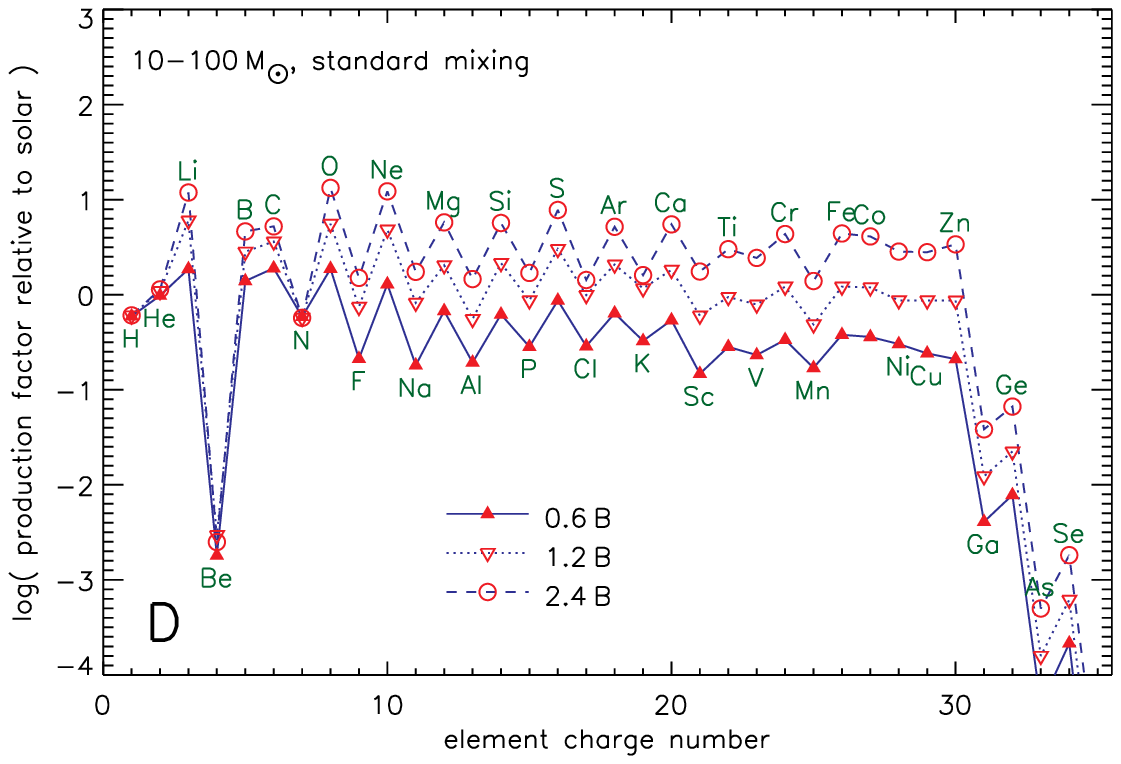}
\\\bigskip                                                                   
\includegraphics[width=0.475\textwidth]{\figurepath 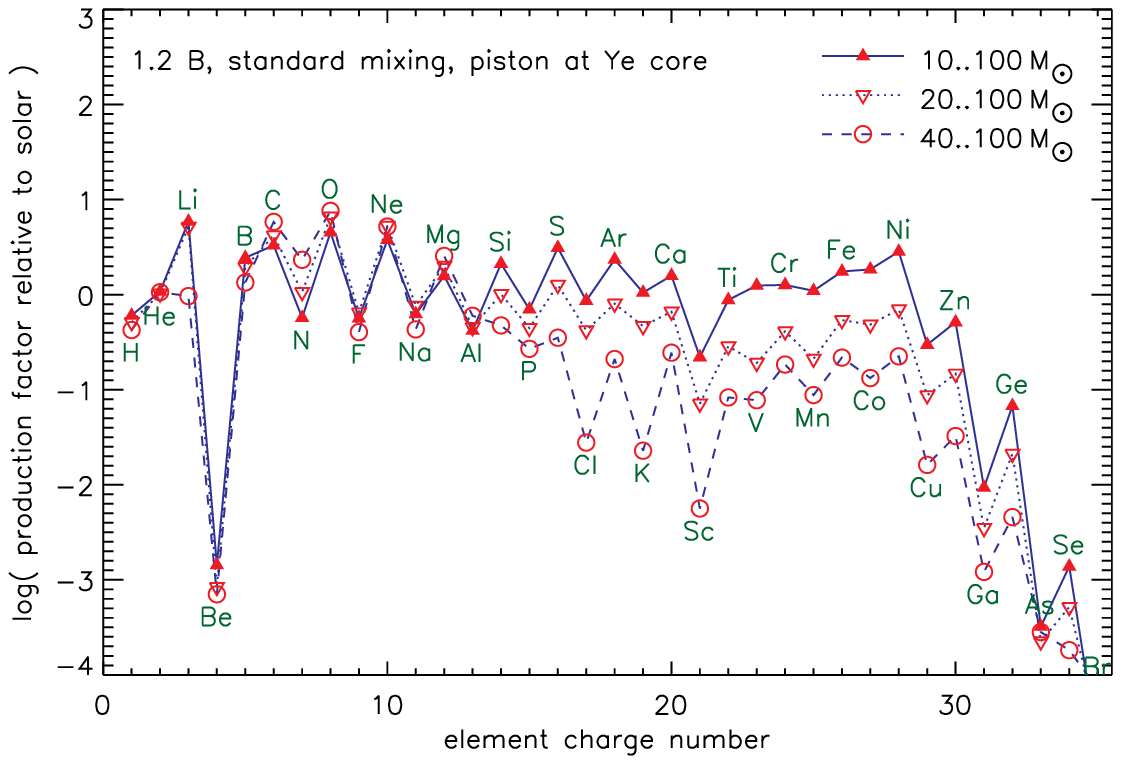}
\hfill                                                                                                   
\includegraphics[width=0.475\textwidth]{\figurepath 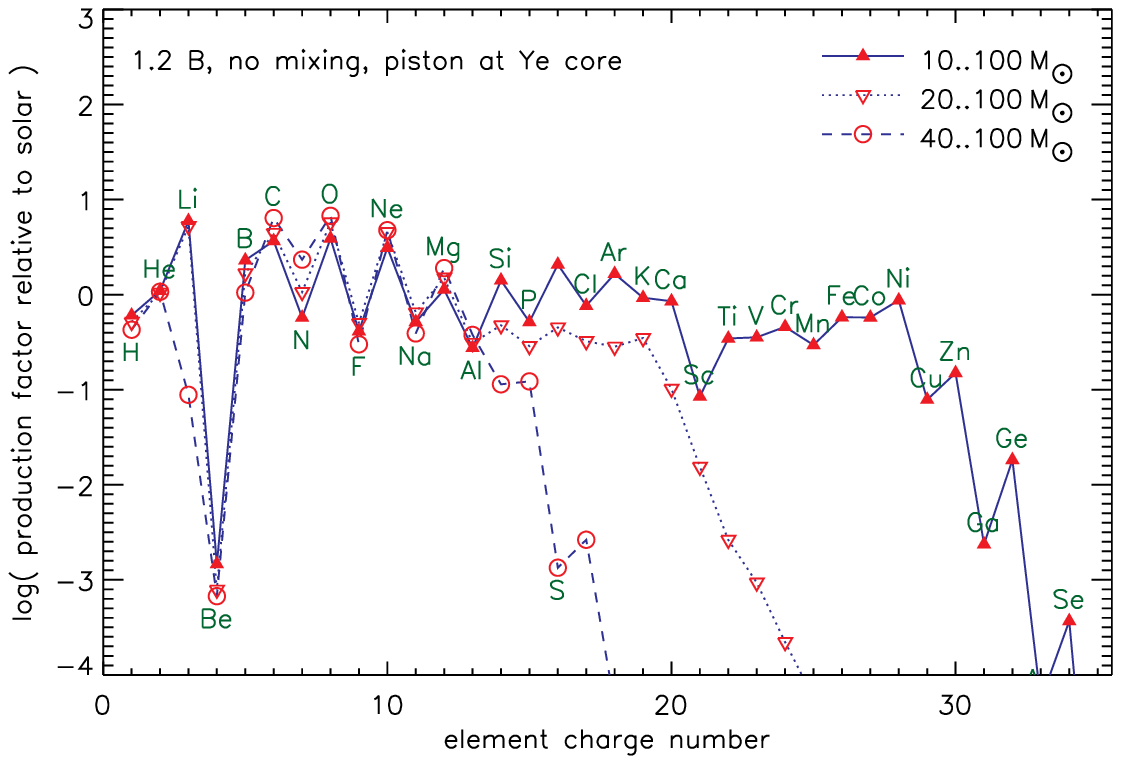}
\caption{IMF integral production factors using the Salpeter IMF with
  slope with $\Gamma=1.35$.  
\textbf{Panel A:} Standard mixing of $0.1$, $1.2\,$B explosions, low-mass
cut-offs for IMF at $10$, $20$, and $40\,\Msun$, $S=4\kB/$baryon.
\textbf{Panel B:} No mixing, $1.2\,$B explosions, low-mass
cut-offs for IMF at $10$, $20$, and $40\,\Msun$, $S=4\kB/$baryon.
\textbf{Panel C:} Standard mixing of $0.1$, low-mass
cut-offs for IMF at $10$, $20$, and $40\,\Msun$, $S=4\kB/$baryon.  
Explosion energy is $1.2\,$B$\,\times M/20\,\Msun$.
\textbf{Panel D:} Standard mixing of $0.1$, low-mass
cut-offs for IMF at $10$--$100\,\Msun$, $S=4\kB/$baryon.  
Constant explosion energies of $0.6$, $1.2$, and $2.4$\,B.
\textbf{Panel E:} Standard mixing of $0.1$, $1.2\,$B explosions, low-mass
cut-offs for IMF at $10$, $20$, and $40\,\Msun$, $\Ye$ core piston.
\textbf{Panel F:} No mixing, $1.2\,$B explosions, low-mass
cut-offs for IMF at $10$, $20$, and $40\,\Msun$, $\Ye$ core piston.
  \lFig{IMFyield} }
\end{figure}

\clearpage

\begin{figure}
\centering
\includegraphics[width=0.475\textwidth]{\figurepath 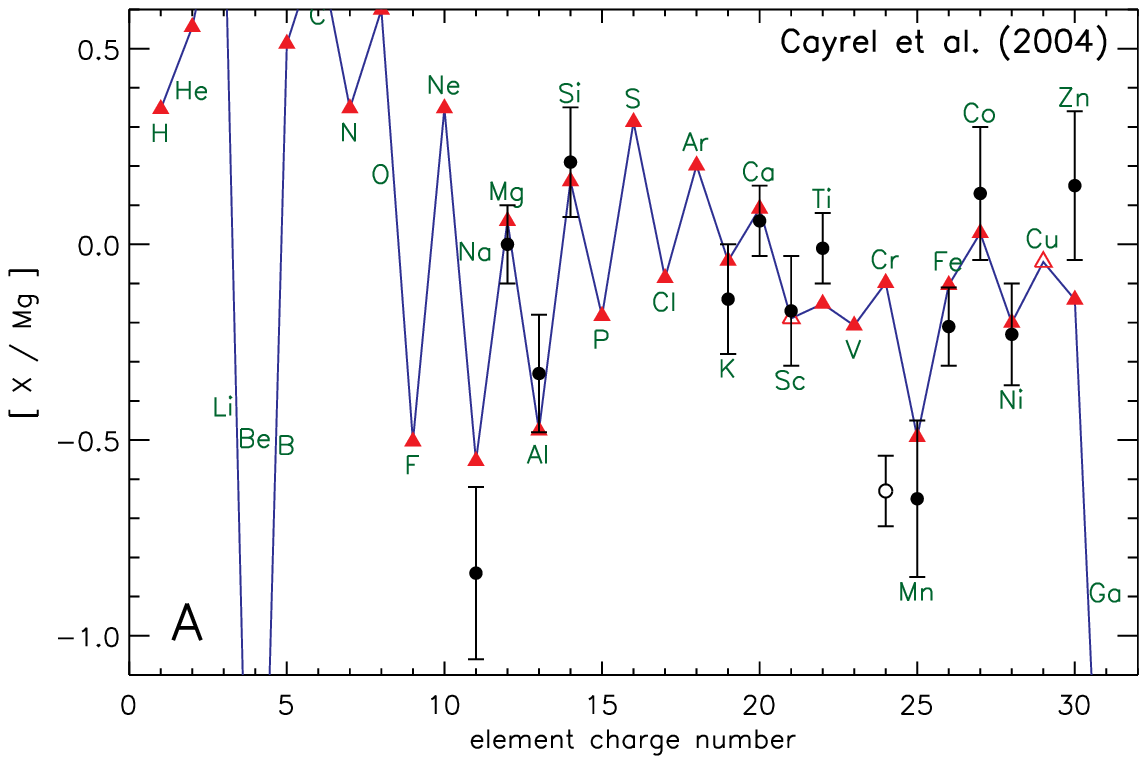}
\hfill
\includegraphics[width=0.475\textwidth]{\figurepath 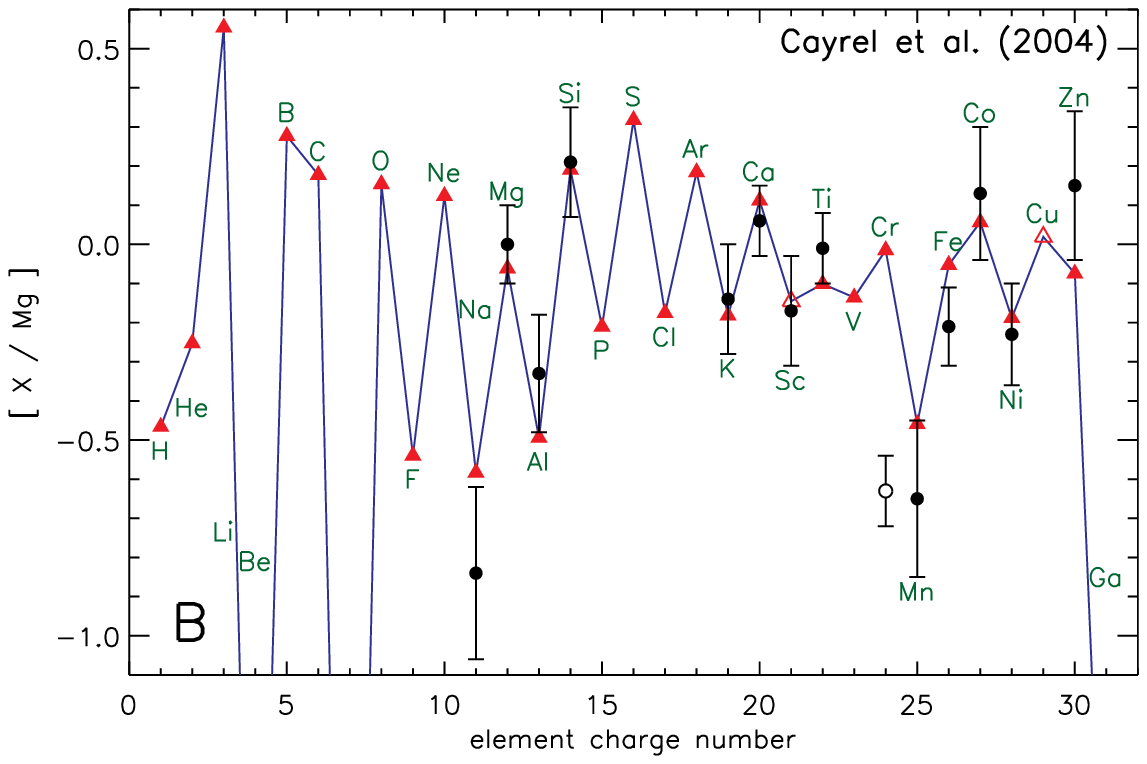}
\\\bigskip
\includegraphics[width=0.475\textwidth]{\figurepath 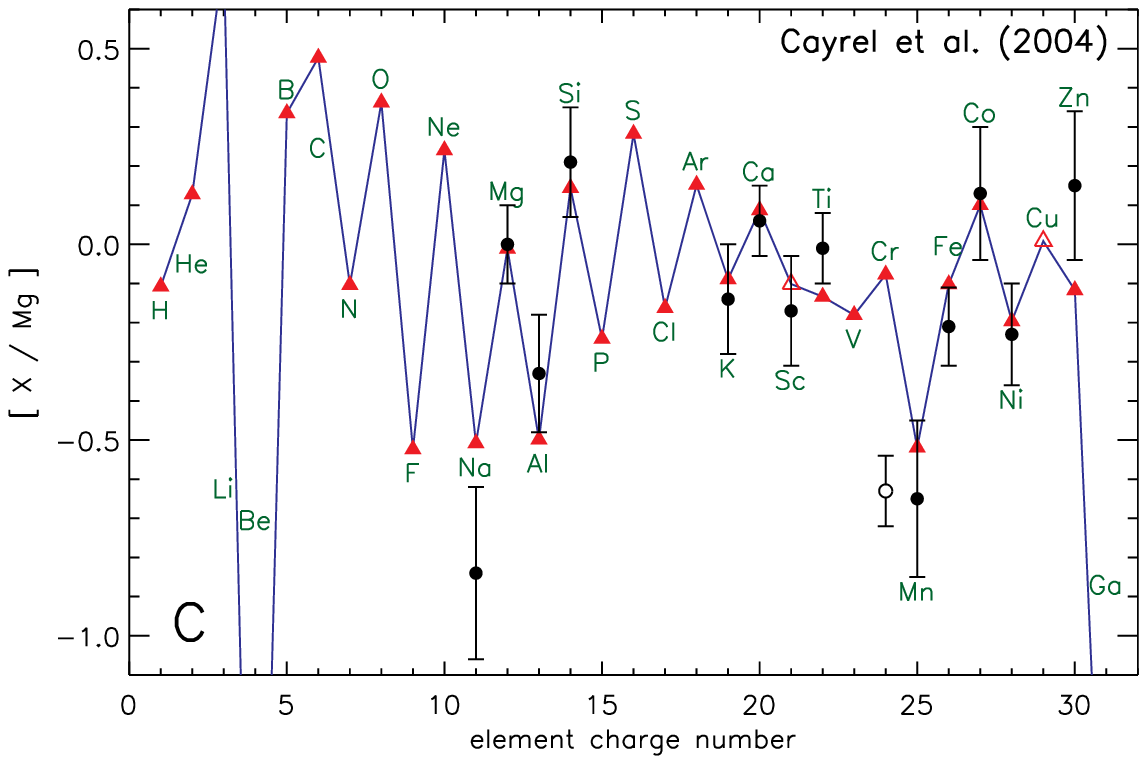}
\hfill
\includegraphics[width=0.475\textwidth]{\figurepath 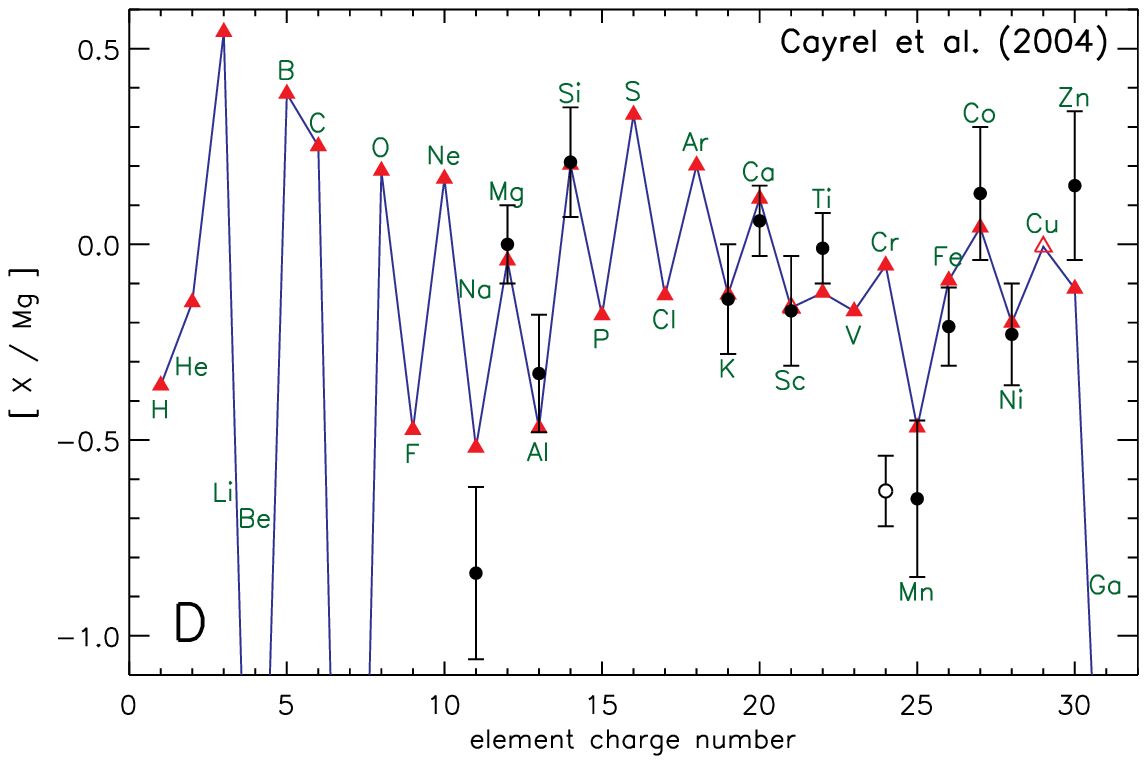}
\\\bigskip
\includegraphics[width=0.475\textwidth]{\figurepath 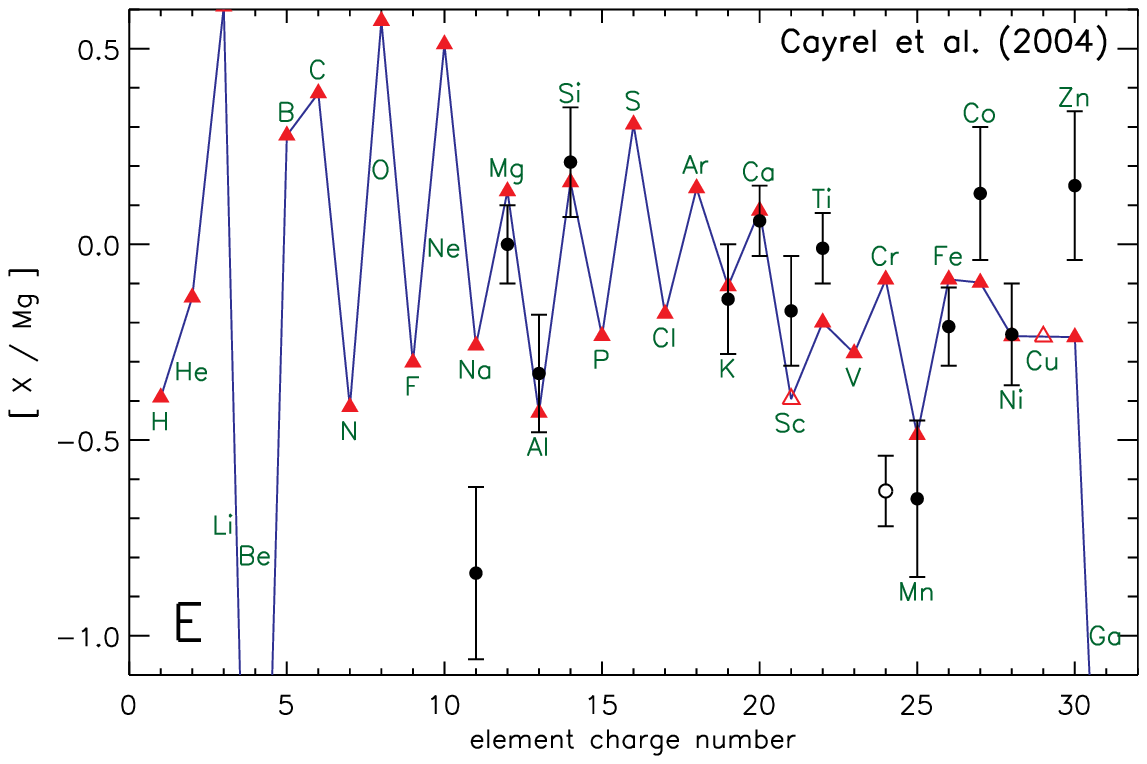}
\hfill
\includegraphics[width=0.475\textwidth]{\figurepath 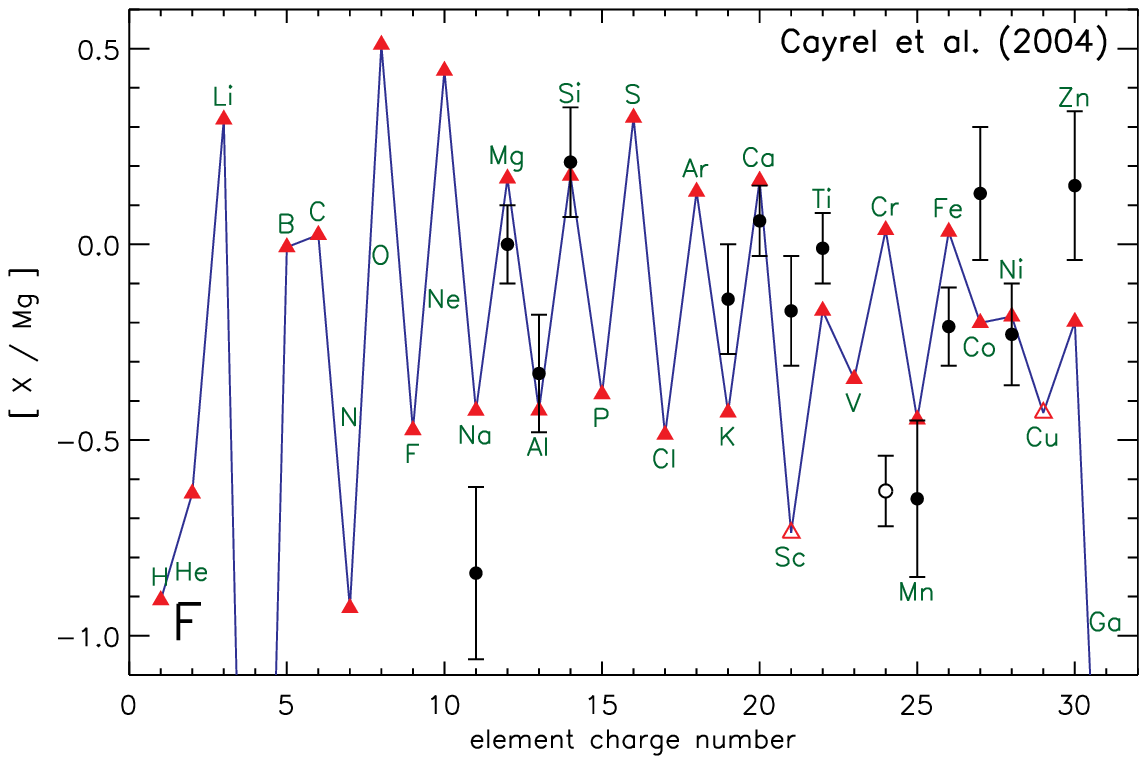}
\caption{Fits to \citet{cay04}.  Cr is ignored (\textsl{hollow
    circle}).  Cu (no data from by \citeauthor{cay04}) and Sc may have
  other nucleosynthesis contributions and we treat them only as
  theoretical lower limits (\textsl{hollow triangles}). 
A list of fits and their properties is given in \Tab{fit_table}.
\textbf{Panel A:} Gauss IMF fit with 
  $M=11.0\,\Msun\pm0.3\,$dex (truncated at $10\,\Msun$), 
  $E=0.6\,\B$, 
  $\EExp=-1$, 
  mixing$=0.0398$, 
  $\chi=0.824$.
\textbf{Panel B:} Best power-law IMF fit with 
  $M=11-15\,\Msun$,
  $\Gamma=-.650$, 
  $E=0.9\,\B$, 
  $\EExp=-1$, 
  mixing$=0.0158$, 
  $\chi=0.732$ (overall best fit).
\textbf{Panel C:} Best power-law IMF fit with fixed mass range
and IMF power-law exponent: 
  $M=10-100\,\Msun$,
  $\Gamma=1.35$, 
  $E=0.9\,\B$, 
  $\EExp=-1$, 
  mixing$=0.0158$, 
  $\chi=0.753$ (overall best fit).
\textbf{Panel D:} Best power-law IMF fit with fixed explosion energy:
  $M=10-15\,\Msun$,
  $\Gamma=-0.65$, 
  $E=1.2\,\B$, 
  $\EExp=0$, 
  mixing$=0.0158$, 
  $\chi=0.740$ (overall best fit).
\textbf{Panel E:} ``Standard'' IMF fit: 
  $M=10-100\,\Msun$,
  $\Gamma=1.350$, 
  $E=1.2\,\B$, 
  $\EExp=0$, 
  mixing$=0.1$, 
  $\chi=1.695$.
\textbf{Panel F:} High explosion energy case; IMF fit with fixed mass-dependent
energy, full mass range and Salpeter IMF (mixing was allowed to float):   
  $M=10-100\,\Msun$, 
  $\Gamma=1.350$,
  $E=1.8\,\B$, 
  $\EExp=1$, 
  mixing$=0.1$, 
  $\chi=2.287$.
  \lFig{Cayrel} }
\end{figure}

\clearpage

\begin{figure}
\centering
\includegraphics[width=0.475\textwidth]{\figurepath 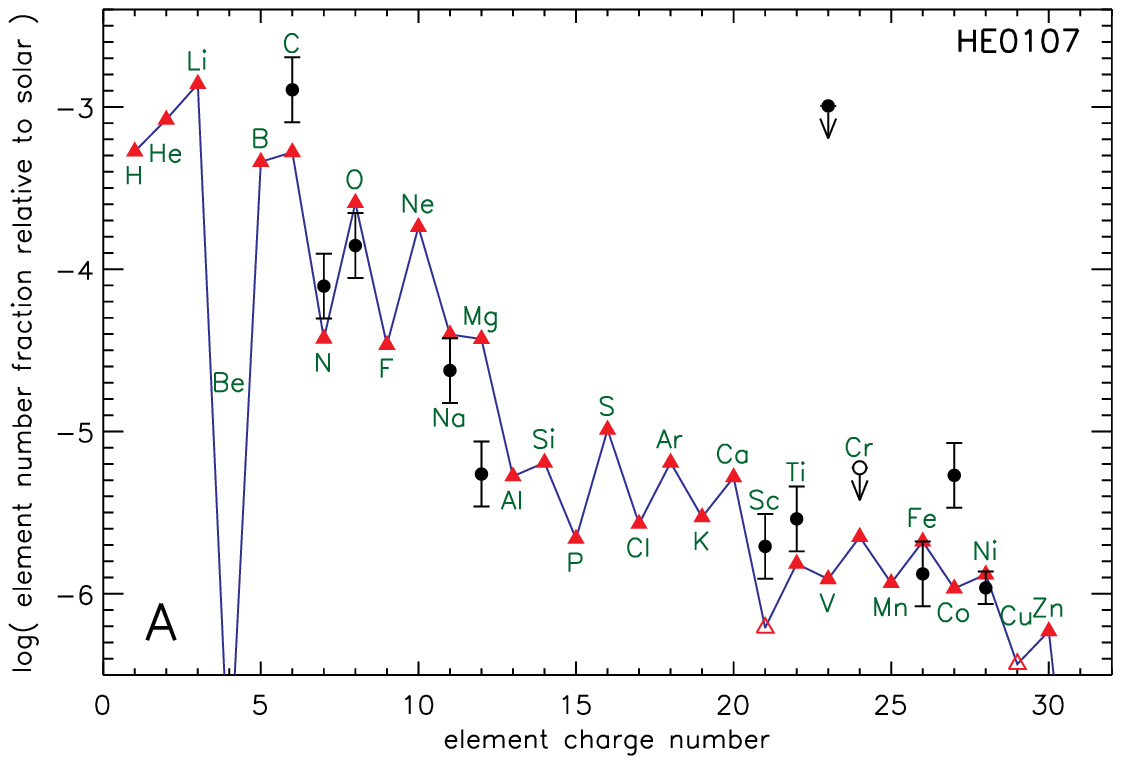}
\hfill
\includegraphics[width=0.475\textwidth]{\figurepath 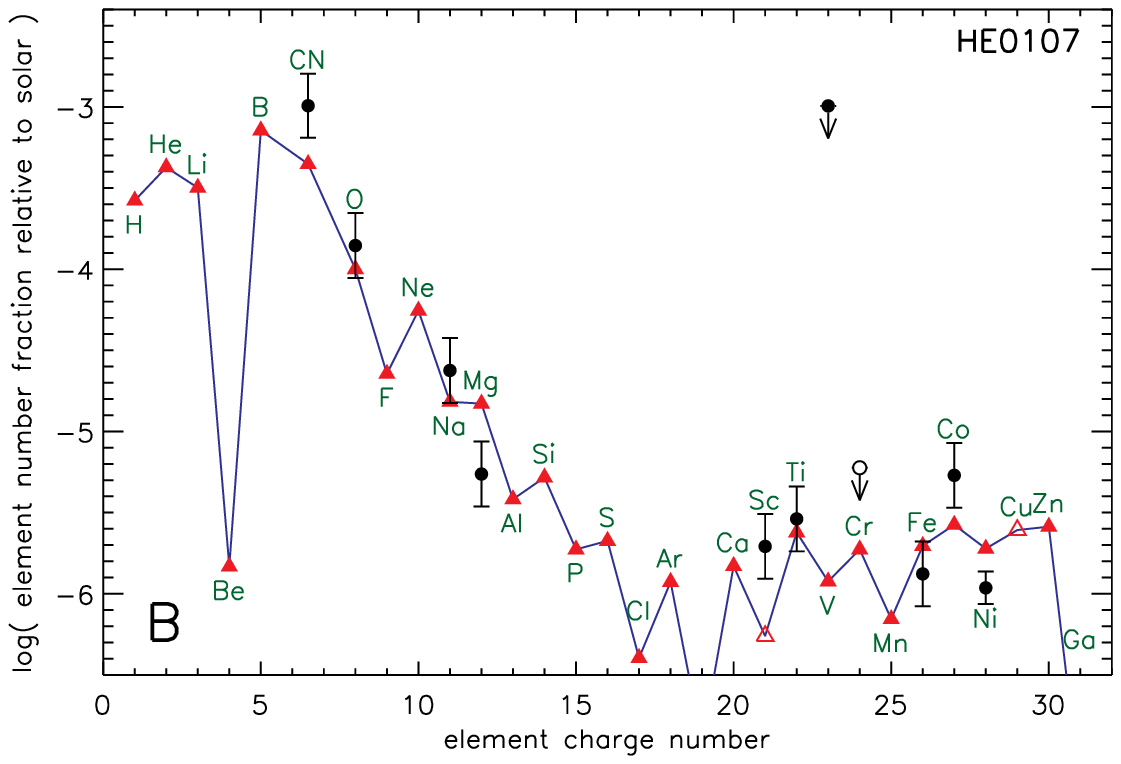}
\\\bigskip
\includegraphics[width=0.475\textwidth]{\figurepath 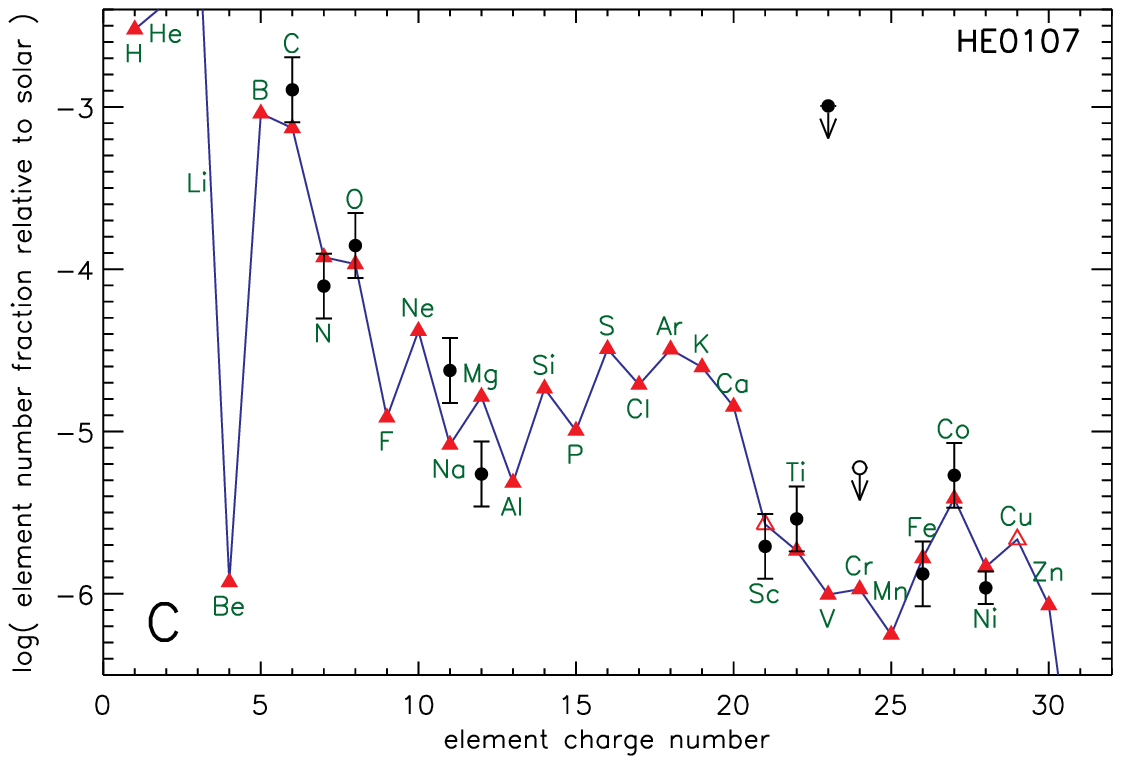}
\hfill
\includegraphics[width=0.475\textwidth]{\figurepath 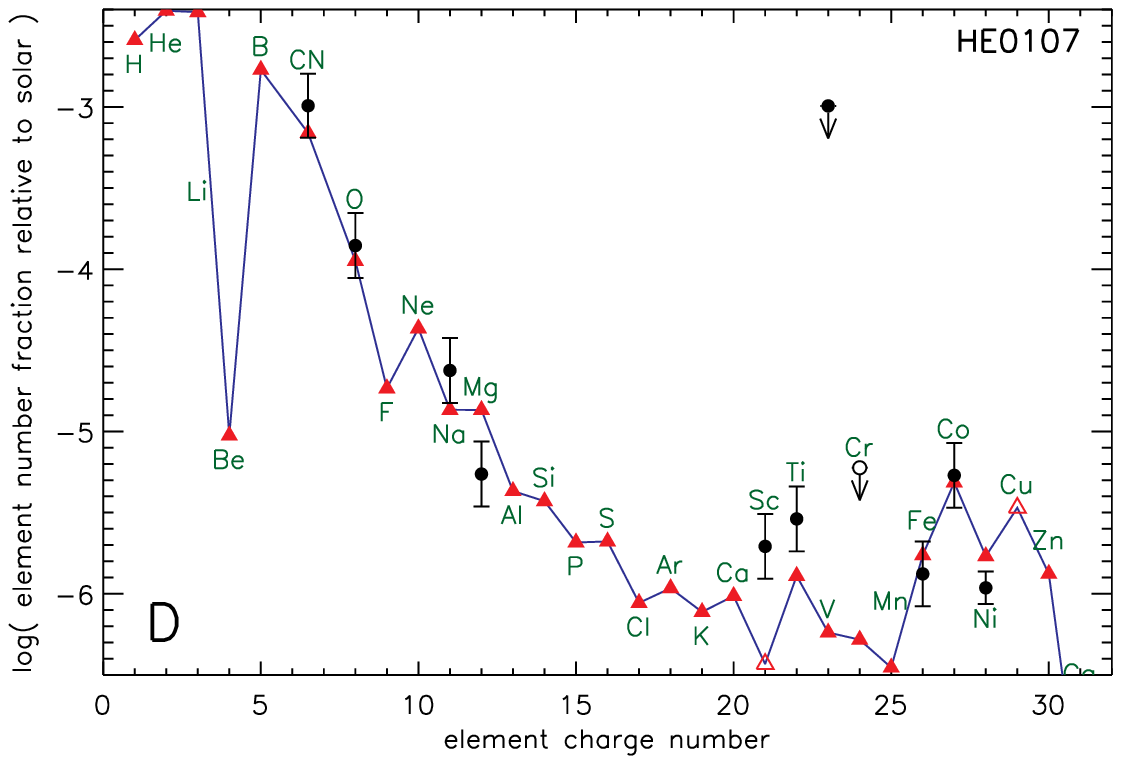}
\\\bigskip
\includegraphics[width=0.475\textwidth]{\figurepath 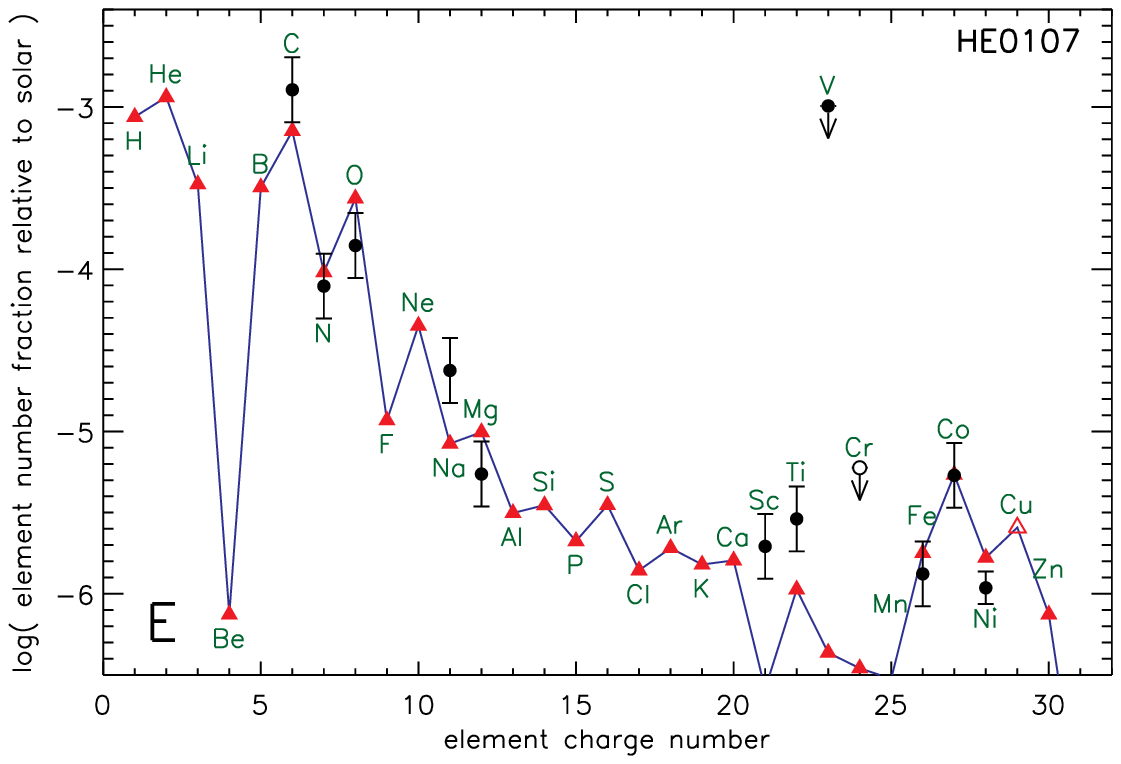}
\hfill
\includegraphics[width=0.475\textwidth]{\figurepath 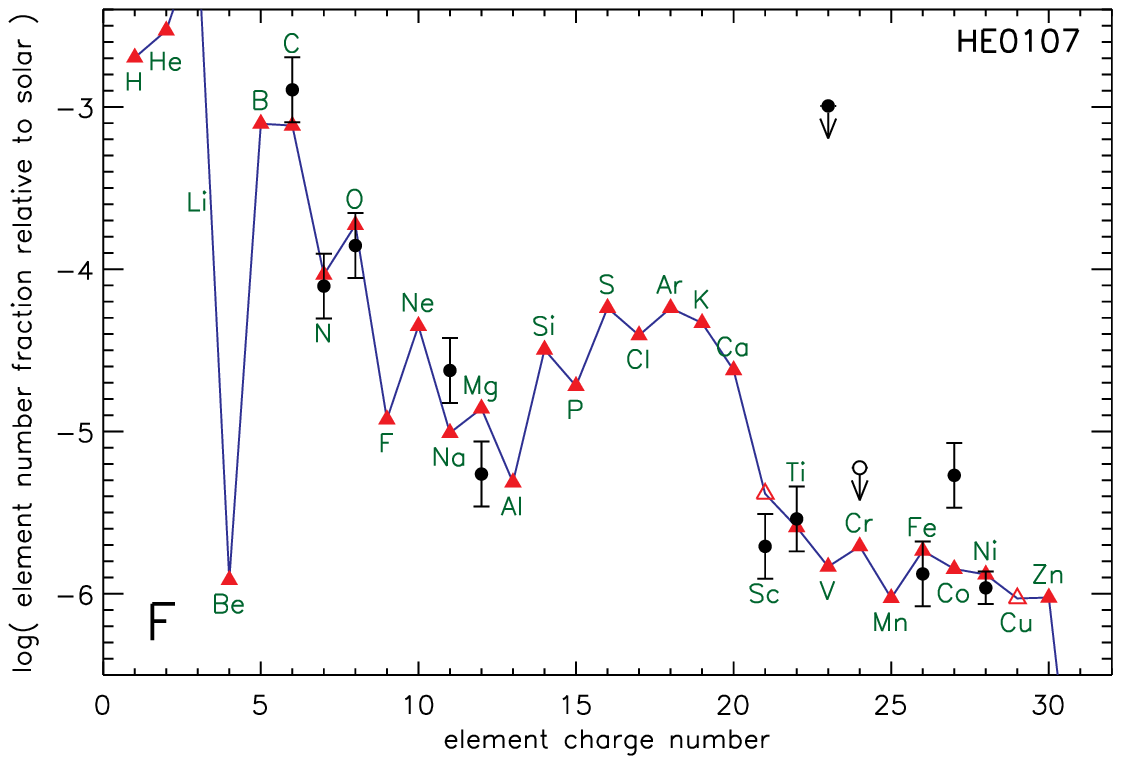}
\caption{Fits to HE0107-5240 \citep{CBE07}.  Cr is ignored
  (\textsl{hollow circle}), but its upper limit is always above the
  fits shown.  Cu (no data from HE0107) and Sc may have other
  nucleosynthesis contributions and we treat them only as theoretical
  lower limits (\textsl{hollow triangles}).  A list of fits and their
  properties is given in \Tab{fit_table}.
\textbf{Panel A:} Best single stars fit: 
  $M=20.5\,\Msun$,
  $E=0.6\,\B$, 
  mixing$=0.0631$, 
  $\chi=3.845$.
\textbf{Panel B:} Best single stars fit combing C+N:  
  $M=10.5\,\Msun$,
  $E=0.3\,\B$, 
  mixing$=0.0063$, 
  $\chi=1.856$. 
  Combining C+N+O does not give a better fit.
\textbf{Panel C:} Gauss IMF fit with 
  $M=17\,\Msun\pm0.1\,\dex$, 
  $E=0.6\,\B$, 
  $\EExp=1.0$, 
  mixing$=0.0025$, 
  $\chi=1.571$.
\textbf{Panel D:} Gauss IMF fit combing C+N with 
  $M=12\,\Msun\pm0.025\,\dex$, 
  $E=0.3\,\B$, 
  $\EExp=0.0$, 
  mixing$=0.0040$, 
  $\chi=1.359$.
  Combining C+N+O does not give a better fit.
\textbf{Panel E:} IMF fit with  
  $M=10-40\,\Msun$, 
  $\Gamma=-0.65$,
  $E=0.3\,\B$, 
  $\EExp=1.0$, 
  mixing$=0.004$, 
  $\chi=1.752$. 
  Combing C+N does not give a better fit, 
  combing C+N+O gives allows a slightly better fit.
\textbf{Panel F:} IMF fit with  
  $M=12-30\,\Msun$, 
  $\Gamma=1.350$ (Salpeter),
  $E=0.6\,\B$, 
  $\EExp=1.0$, 
  mixing$=0.0251$, 
  $\chi=1.974$. 
  \lFig{HE0107} }
\end{figure}

\clearpage

\begin{figure} 
\centering
\includegraphics[width=0.475\textwidth]{\figurepath 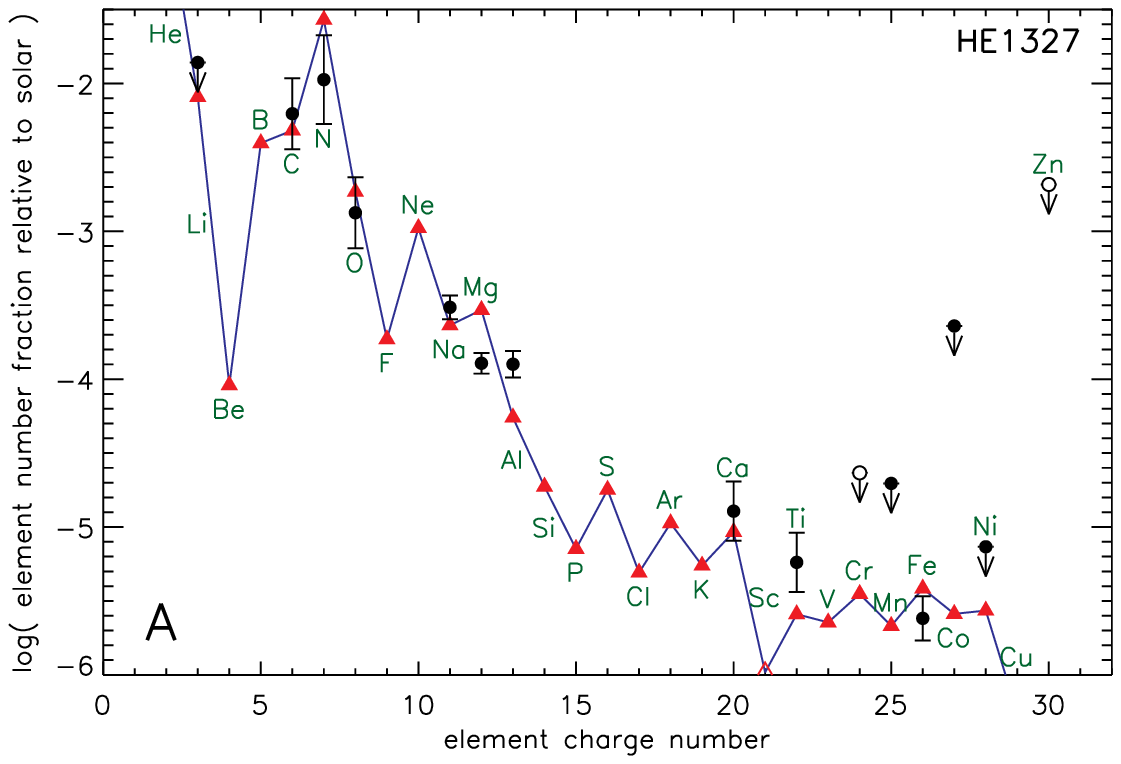}
\hfill
\includegraphics[width=0.475\textwidth]{\figurepath 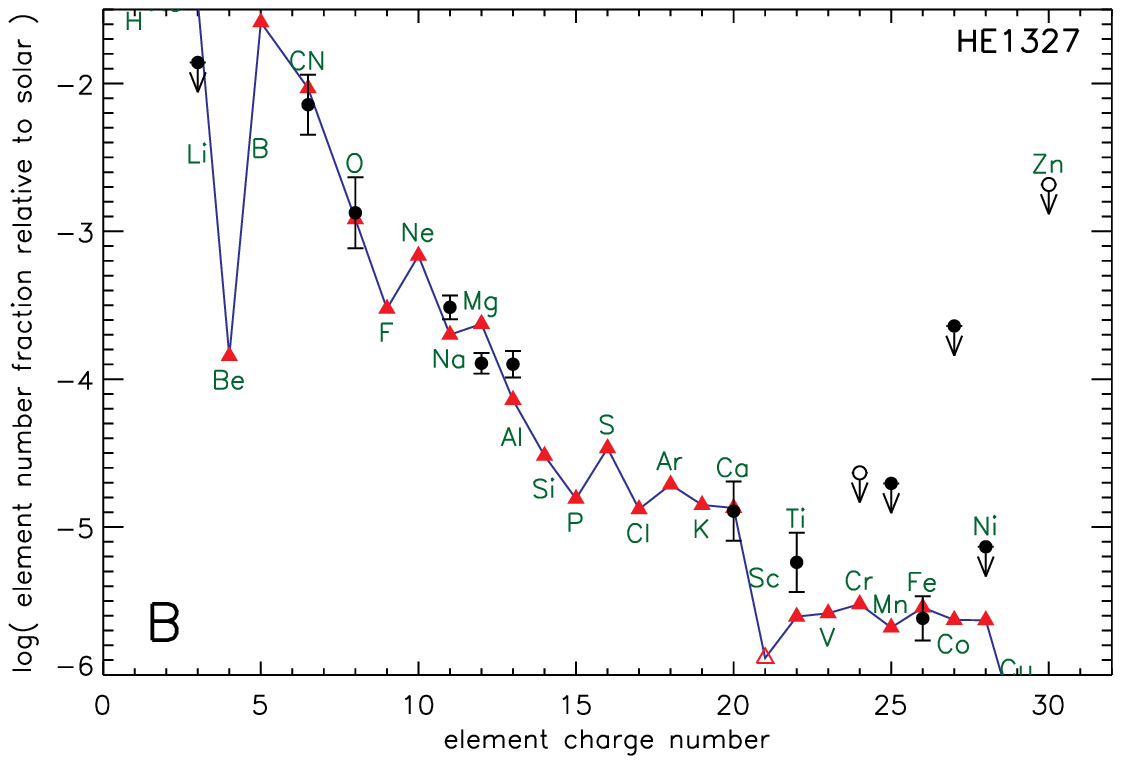}
\\\bigskip
\includegraphics[width=0.475\textwidth]{\figurepath 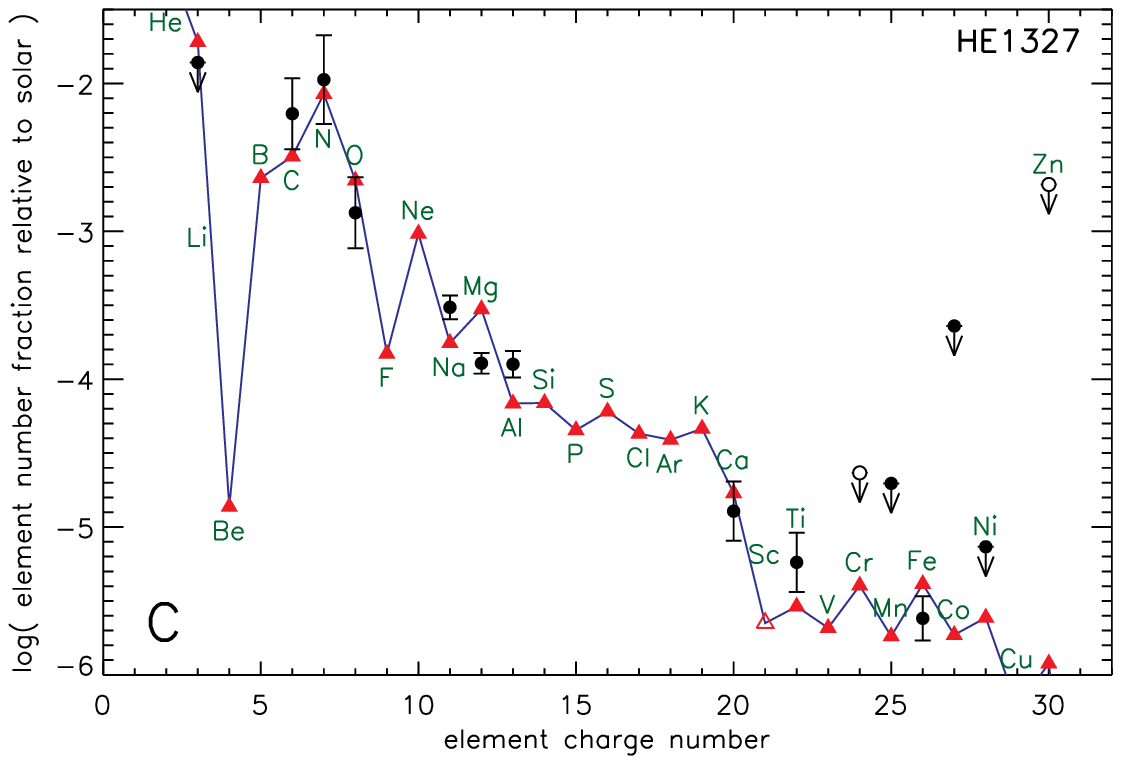}
\hfill
\includegraphics[width=0.475\textwidth]{\figurepath 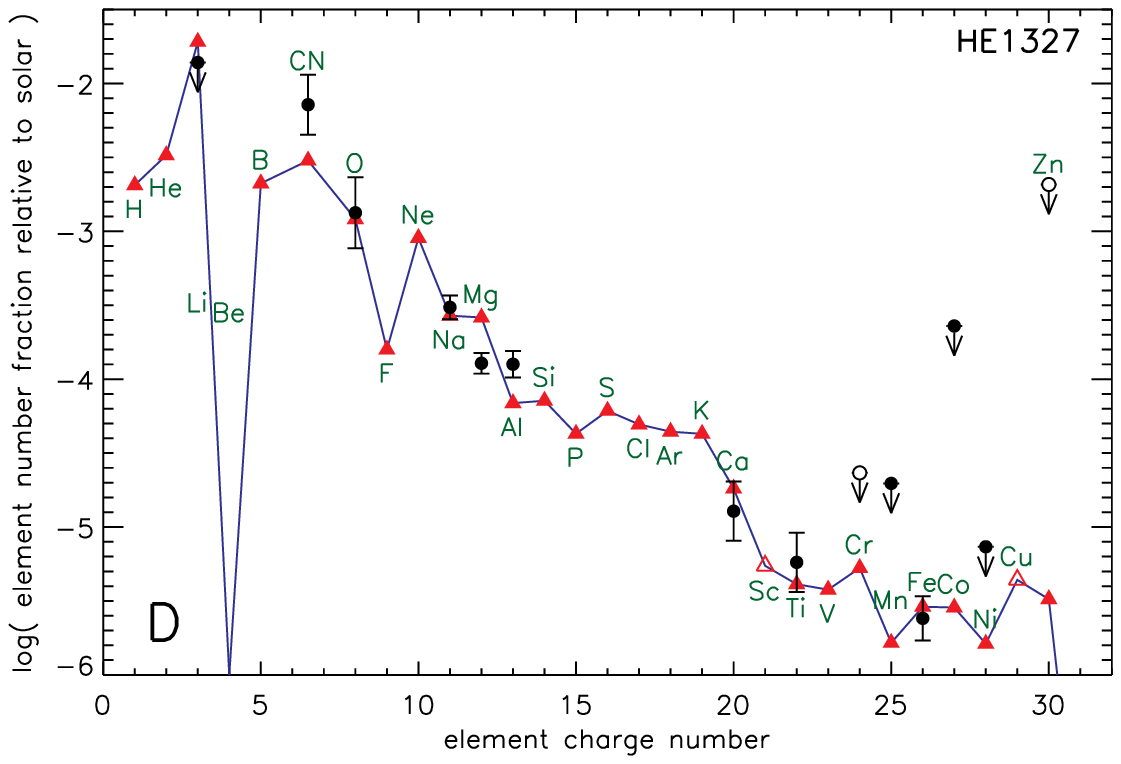}
\\\bigskip
\includegraphics[width=0.475\textwidth]{\figurepath 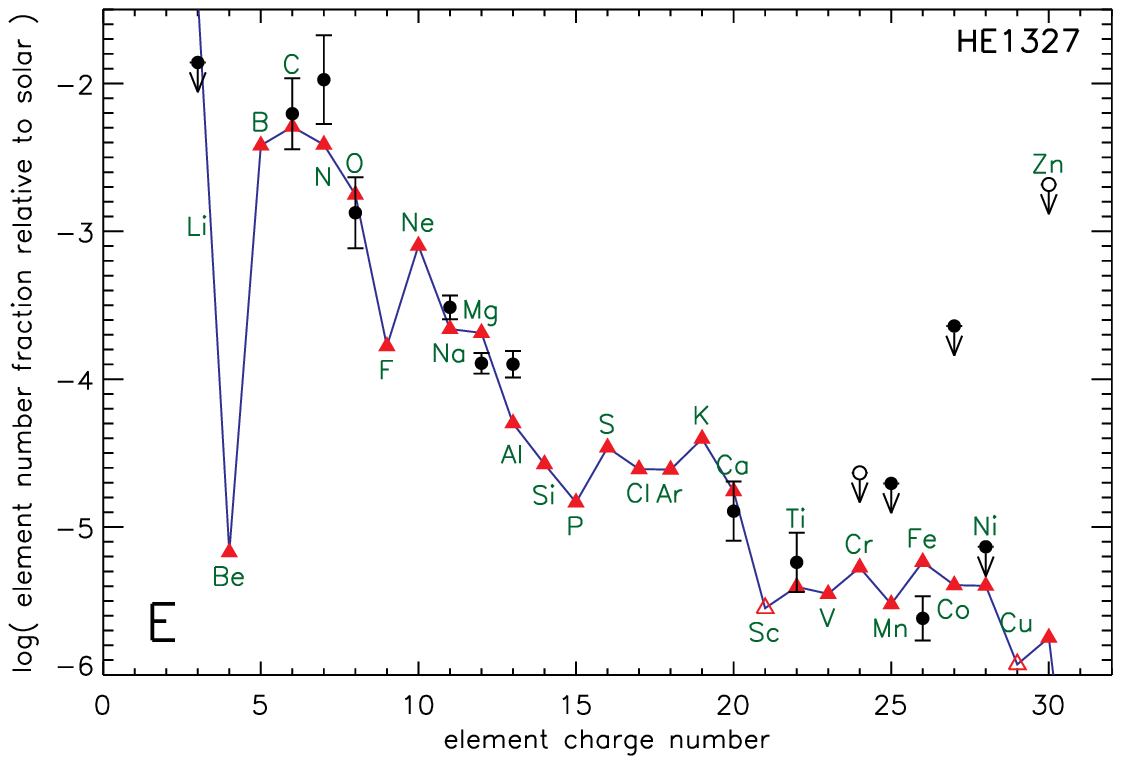}
\hfill
\includegraphics[width=0.475\textwidth]{\figurepath 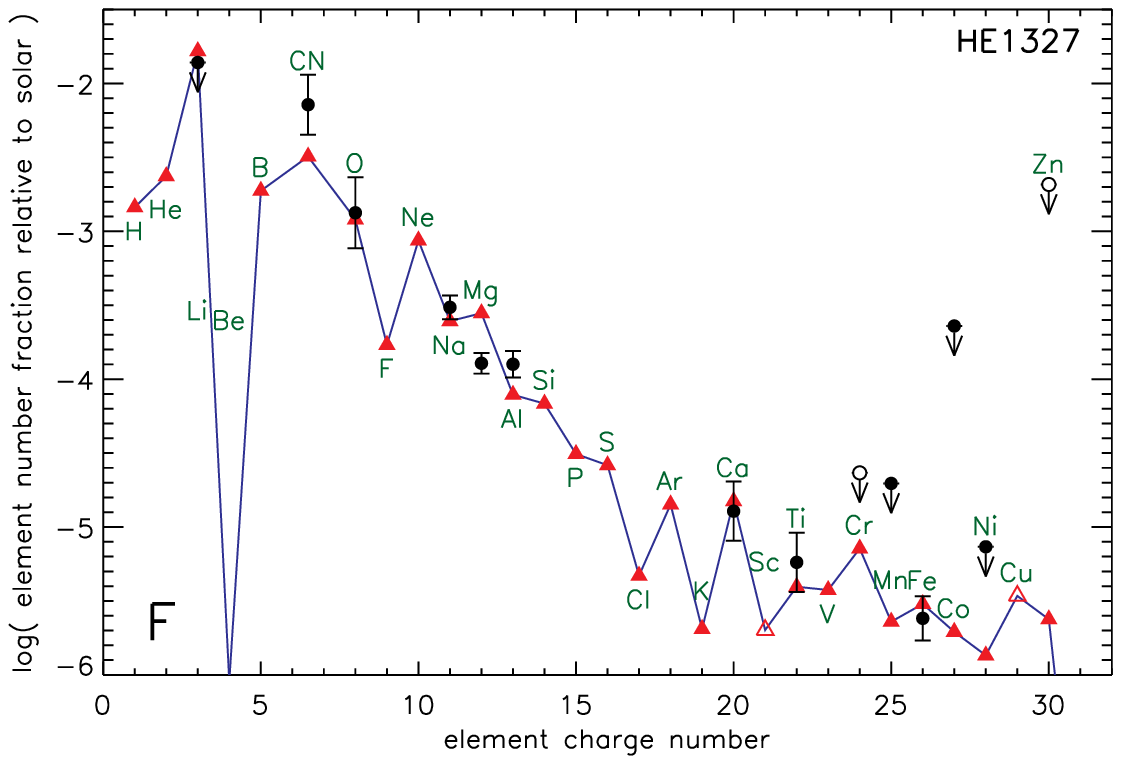}
\caption{Fits to HE1327 \citep{aok06,fre06}.  Cr and Zn are
  ignored, but their upper limit is always above the fits shown.  
  Cu and Sc (both no data from HE1327) may have other
  nucleosynthesis contributions and we treat them only as theoretical
  lower limits (\textsl{hollow triangles}).
A list of fits and their properties is given in \Tab{fit_table}.
\textbf{Panel A:} Best single stars fit: 
  $M=21.5\,\Msun$,
  $E=0.3\,\B$, 
  mixing$=0.0631$, 
  $\chi=4.051$.
\textbf{Panel B:} Best single stars fit combing C and N:  
  $M=11.6\,\Msun$,
  $E=0.3\,\B$, 
  mixing$=0.0158$, 
  $\chi=2.941$.
  Combing C+N+O does not allow a better fit than just combing C+N. 
\textbf{Panel C:} Gauss IMF fit with 
  $M=25.0\,\Msun\pm0.05\,$dex, 
  $E=0.3\,\B$, 
  $\EExp=0.5$, 
  mixing$=0.0631$, 
  $\chi=4.072$.
\textbf{Panel D:} Gauss IMF fit with 
  $M=15.0\,\Msun\pm0.025\,$dex, 
  $E=0.9\,\B$, 
  $\EExp=0.5$, 
  mixing$=0.0063$, 
  $\chi=2.833$.
  Combing C+N+O does not find a better match than just combing C+N. 
\textbf{Panel E:} IMF fit with  
  $M=15-40\,\Msun$, 
  $\Gamma=1.350$,
  $E=0.3\,\B$, 
  $\EExp=-1.0$, 
  mixing$=0.0398$, 
  $\chi=3.620$. 
\textbf{Panel F:} IMF fit with  
  $M=13.5-17\,\Msun$, 
  $\Gamma=1.350$,
  $E=0.9\,\B$, 
  $\EExp=0.0$, 
  mixing$=0.0040$, 
  $\chi=2.892$. 
  Combing C+N+O does not allow a better fit than just combing C+N. 
  \lFig{HE1327} }
\end{figure}

\clearpage

\begin{figure} 
\centering
\includegraphics[width=0.475\textwidth]{\figurepath 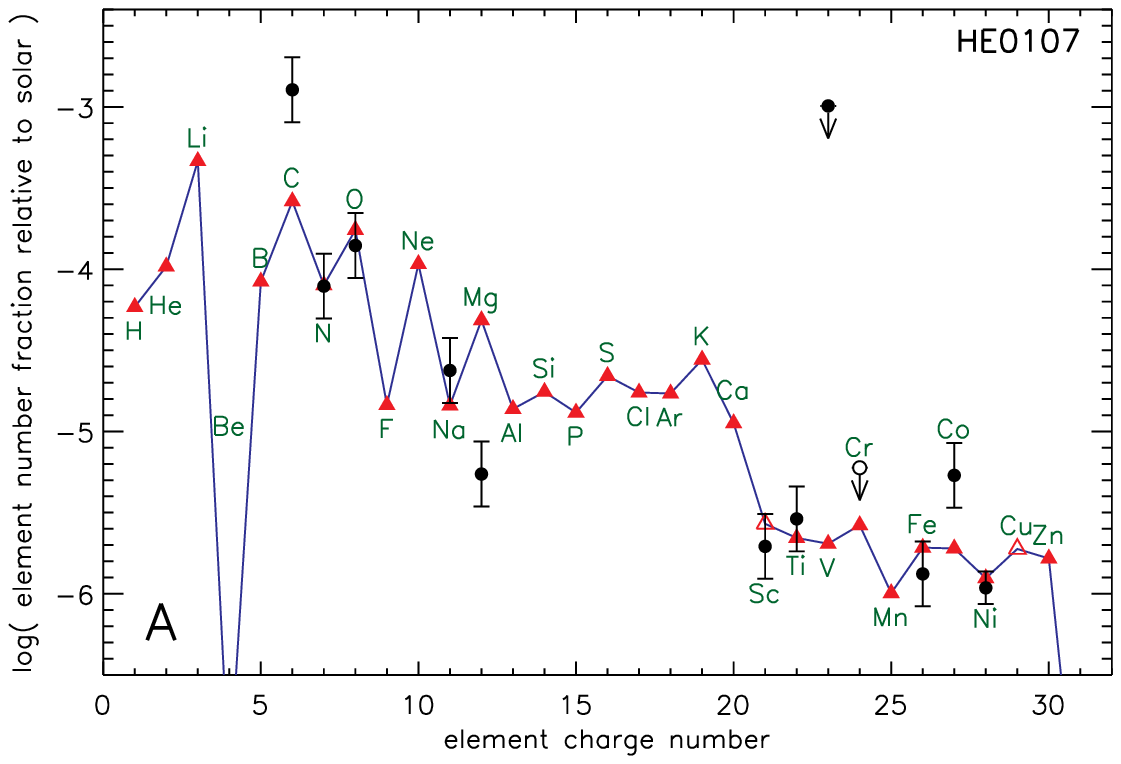}
\hfill                                                                                          
\includegraphics[width=0.475\textwidth]{\figurepath 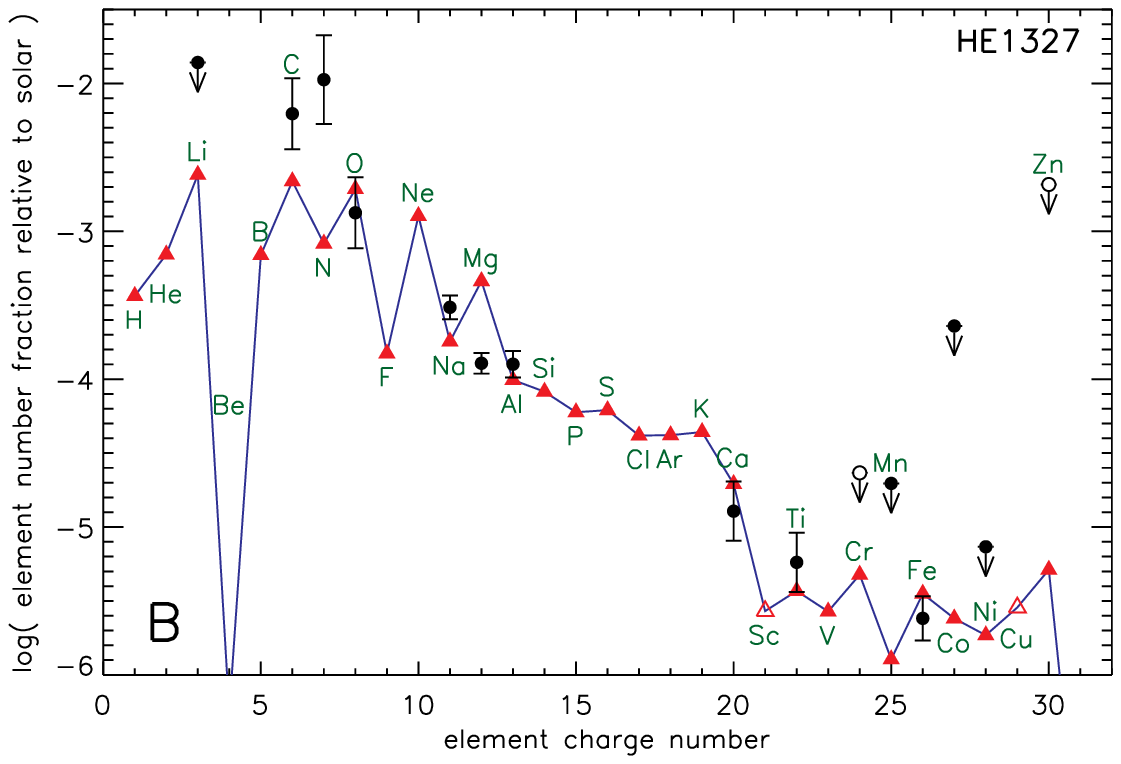}
\\\bigskip                                                                                      
\includegraphics[width=0.475\textwidth]{\figurepath 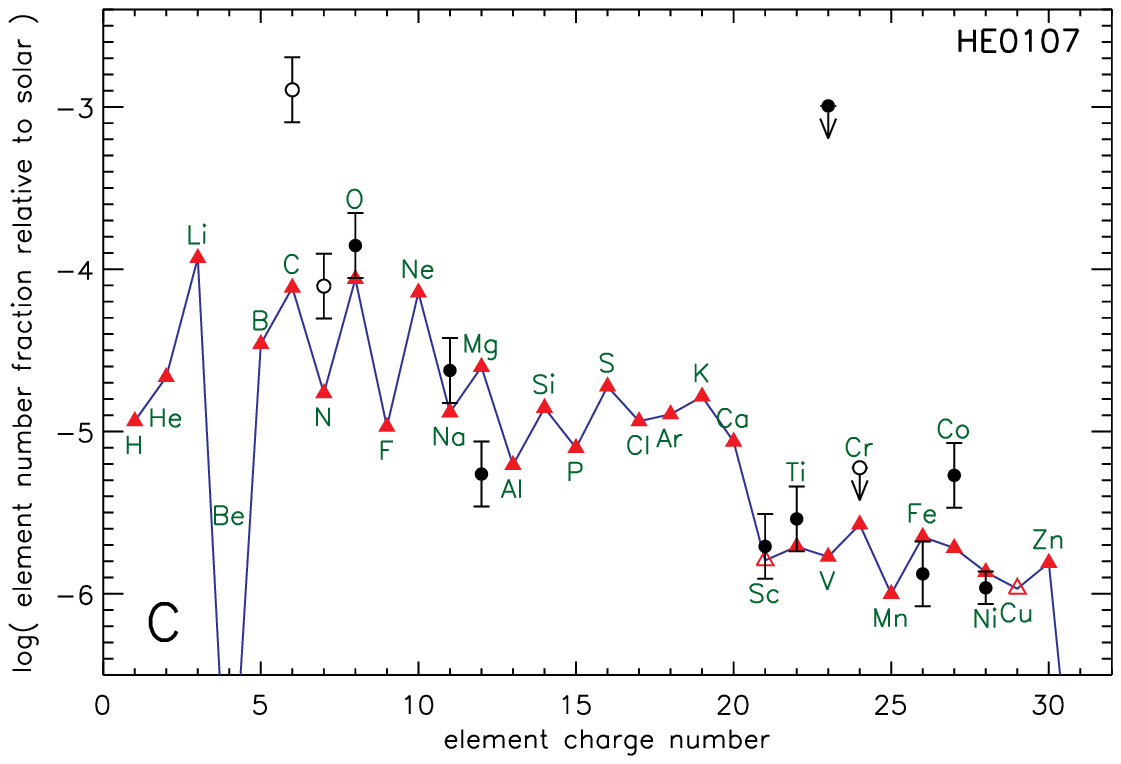}
\hfill                                                                                          
\includegraphics[width=0.475\textwidth]{\figurepath 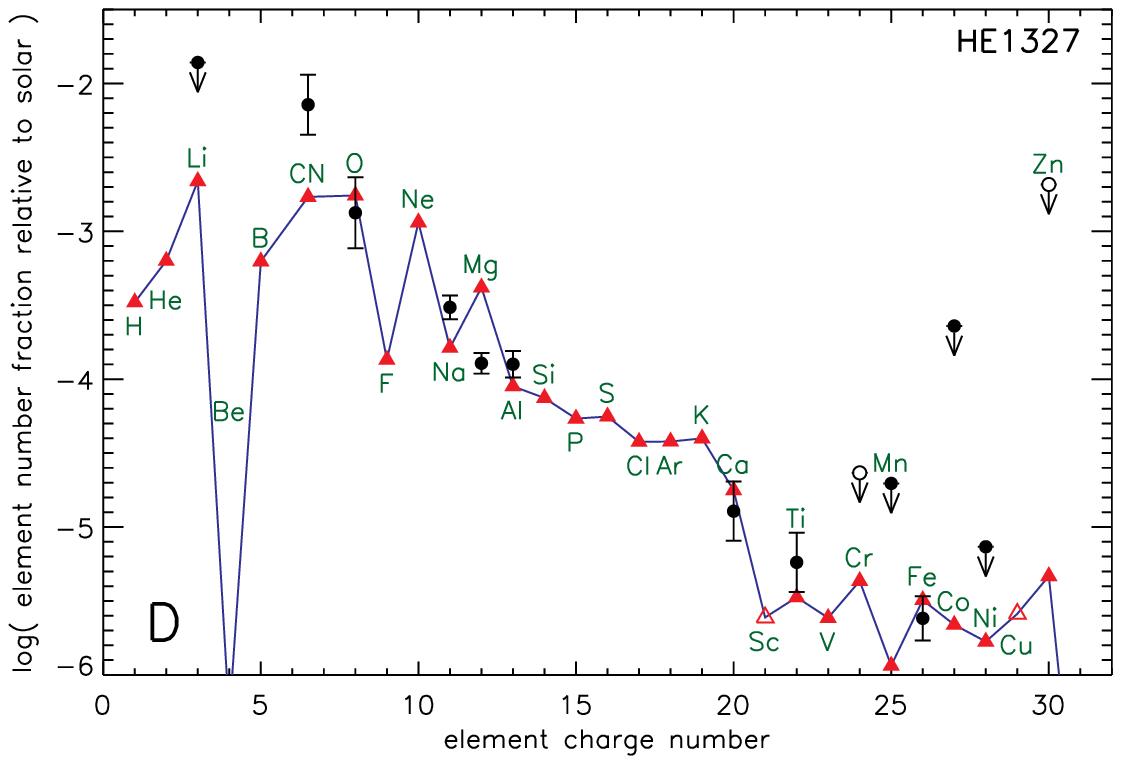}
\\\bigskip                                                                                      
\includegraphics[width=0.475\textwidth]{\figurepath 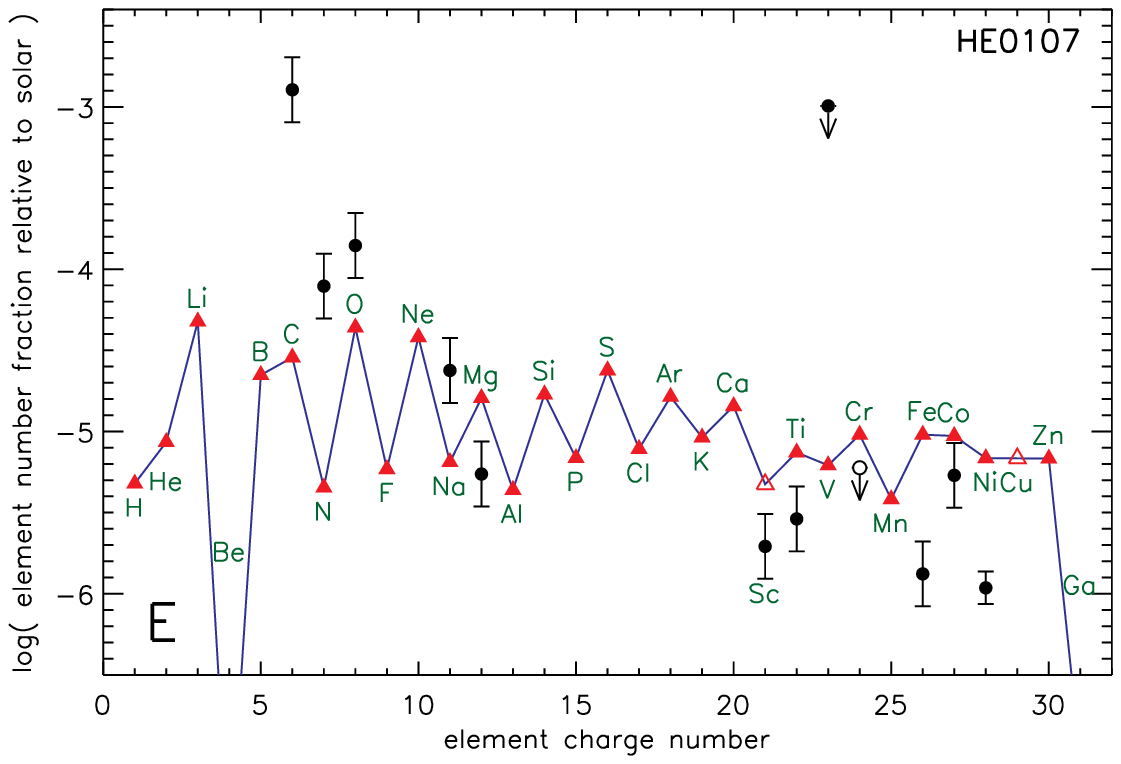}
\hfill                                                                                          
\includegraphics[width=0.475\textwidth]{\figurepath 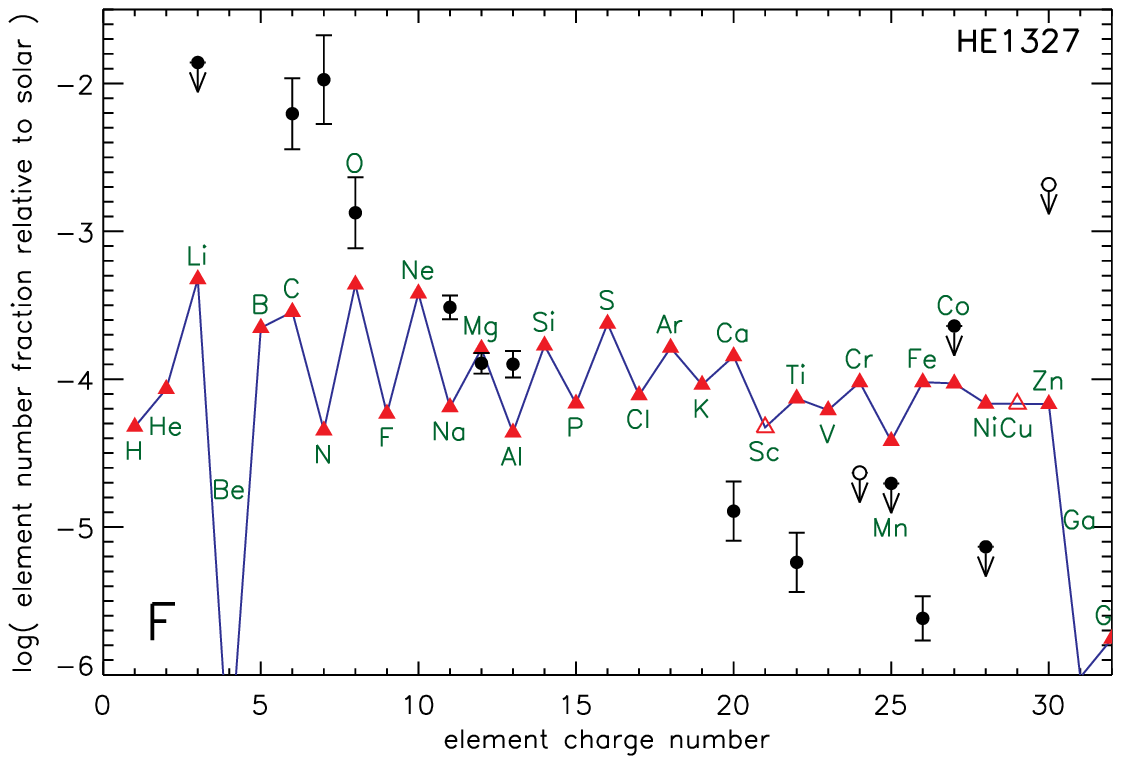}
\caption{Fits with upper mass limit fixed to 100\,\Msun and fixed
  $\Gamma = 1.350$ for HE0107-5240 \citep{CBE07} (\textsl{Panels A, C,
    and E}) and HE1327-2326 \citep{aok06,fre06} (\textsl{Panels B, D,
    and F}).  For comparisons for the ``standard'' IMF, mixing, and
  energies are given in \textsl{Panels E and F} for HE0107 and HE1327
  respectively.  Cr is ignored, but except for the poor ``standard''
  fits its upper limit is always above the fits shown where present.
  Cu and Sc (both no data from HE1327) may have other nucleosynthesis
  contributions and we treat them only as theoretical lower limits
  (\textsl{hollow triangles}).  A list of fits and their properties is
  given in \Tab{fit_table}.
\textbf{Panel A:} HE0107.  IMF fit with  
  $M=13.5-100\,\Msun$, 
  $\Gamma=1.350$,
  $E=0.9\,\B$, 
  $\EExp=-0.5$, 
  mixing$=0.0158$, 
  $\chi=3.870$.  
  Combining C+N or C+N+O does not allow better fits.
\textbf{Panel B:} HE1327.  IMF fit with  
  $M=20-100\,\Msun$, 
  $\Gamma=1.350$,
  $E=1.2\,\B$, 
  $\EExp=-1.0$, 
  mixing$=0.0158$, 
  $\chi=7.189$.  
  Combining C+N or C+N+O gives $\chi\approx0.5$ better fits.
\textbf{Panel C:} HE0107.  Ignoring carbon and nitrogen.  IMF fit with  
  $M=15-100\,\Msun$, 
  $\Gamma=1.350$,
  $E=1.2\,\B$, 
  $\EExp=-0.5$, 
  mixing$=0.01$, 
  $\chi=2.399$.  
\textbf{Panel D:} HE1327.  Combining C+N.  IMF fit with  
  $M=20-100\,\Msun$, 
  $\Gamma=1.350$,
  $E=1.2\,\B$, 
  $\EExp=-1.0$, 
  mixing$=0.0158$, 
  $\chi=6.699$.  
  Combining C+N+O improves the best fit found only slightly.
  \lFig{MLfits} 
\textbf{Panel E:} HE0107.  ``Standard'' IMF and energies:
  $M=10-100\,\Msun$, 
  $\Gamma=1.350$,
  $E=1.2\,\B$, 
  $\EExp=0.0$, 
  mixing$=0.1$, 
  $\chi=19.809$.  
\textbf{Panel F:} HE1327.  ``Standard'' IMF and energies:
  $M=10-100\,\Msun$, 
  $\Gamma=1.350$,
  $E=1.2\,\B$, 
  $\EExp=0.0$, 
  mixing$=0.1$, 
  $\chi=30.341$.  
\lFig{HE1327norm}}
\end{figure}

\end{document}